\documentclass[fleqn,usenatbib,useAMS]{mnras}

\usepackage{graphicx}
\usepackage{amsmath}
\usepackage{amssymb}
\usepackage{multicol}
\usepackage{multirow}
\usepackage{bm}	
\usepackage{nccmath}
\usepackage[T1]{fontenc}
\usepackage{ae,aecompl}
\usepackage{newtxtext,newtxmath}
\usepackage{threeparttable, tablefootnote}
\usepackage{xcolor}

\newcommand{\kms}{\,km\,s$^{-1}$}
\newcommand{\angstrom}{\mbox{\normalfont\AA}}
\newcommand{\mgii}{Mg\hspace{0.5mm}\textsc{ii} }
\newcommand{\feii}{Fe\hspace{0.5mm}\textsc{ii} }
\newcommand{\ovi}{O\hspace{0.5mm}\textsc{vi} }

\newcommand{\mean}[1]{\langle\,#1\,\rangle}
\newcommand{\ewGr}[1]{$EW_{\rm 2796}\,>\,#1\,$\angstrom}
\newcommand{\ewMg}{$EW_{\rm 2796}$ }

\defcitealias{anand21}{Paper I} 

\title[The nature of \mgii absorbers in DESI galaxy clusters]{Cool circumgalactic gas in galaxy clusters: connecting the DESI legacy imaging survey and SDSS DR16 \mgii absorbers}

\author[Anand, Kauffmann \& Nelson]{Abhijeet Anand$^{1}$\thanks{E-mail: abhijeet@mpa-garching.mpg.de},
Guinevere Kauffmann$^{1}$,
Dylan Nelson$^{2}$
\\\\
$^{1}$Max-Planck-Institut f\"{u}r Astrophysik, Karl-Schwarzschild-Str. 1, 85741 Garching, Germany\\
$^{2}$Universit\"{a}t Heidelberg, Zentrum f\"{u}r Astronomie, Institut f\"{u}r theoretische Astrophysik, Albert-Ueberle-Str. 2, 69120 Heidelberg, Germany\\
}

\pubyear{2021}

\begin{document}
\maketitle

\begin{abstract} 
We investigate the cool gas absorption in galaxy clusters by cross-correlating \mgii absorbers detected in quasar spectra from Data Release 16 of the Sloan Digital Sky Survey (SDSS) with galaxy clusters identified in the  Dark Energy Spectroscopic Instrument (DESI) survey. We find significant covering fractions ($1-5\, \%$ within $r_{500}$, depending on the chosen redshift interval), $\sim 4-5$ times higher than around random sightlines. While the covering fraction of cool gas in clusters decreases with increasing mass of the central galaxy, the total \mgii mass within $r_{\rm 500}$ is nonetheless $\sim 10$ times higher than for SDSS luminous red galaxies (LRGs). The \mgii covering fraction versus impact parameter is well described by a power law in the inner regions and an exponential function at larger distances. The characteristic scale of the transition between these two regimes is smaller for large equivalent width absorbers. Cross-correlating \mgii absorption with photo $-z$ selected cluster member galaxies from DESI reveals a statistically significant connection. The median projected distance between \mgii absorbers and the nearest cluster member is $\sim200$ kpc, compared to $\sim500$ kpc in random mocks with the same galaxy density profiles. We do not find a correlation between \mgii strength and the star formation rate of the closest cluster neighbour. This suggests that cool gas in clusters, as traced by \mgii absorption, is: (i) associated with satellite galaxies, (ii) dominated by cold gas clouds in the intracluster medium, rather than by the interstellar medium of galaxies, and (iii) may originate in part from gas stripped from these cluster satellites in the past.
\end{abstract}



\section{Introduction}\label{intro}

Understanding the physical processes that form galaxies and determine their evolution is a central goal of the field. The circumgalactic medium (CGM) that surrounds galaxies is of particular interest, because it is the interface between galaxies and the larger-scale intergalactic medium (IGM). The complex interplay between gas inflows and outflows powered by supernova explosions or accreting black holes is known as the cosmic baryon cycle, and is a key ingredient to understanding galaxy formation and evolution \citep[see][for reviews]{tumlinson17,peroux20a}. Both observations \citep{stocke13, zhu14, lan18, zahedy19, tchernyshyov21} and simulations \citep{oppenheimer18, peeples19, fielding20, nelson20, byrohl21} have revealed that the CGM is multi-phase and hosts gas with a wide range of temperatures and densities.

Thanks to advances in space and ground-based telescopes, the CGM has been studied extensively in past decade. The metal absorption lines detected in distant quasar spectra trace gas at different temperatures and densities in the CGM of foreground galaxies \citep{rudie12, rubin14, chen17, anand21, hasan21, dutta21}. At the same time, modern cosmological simulations that produce reasonably realistic global galaxy population properties, such as the stellar mass function and the evolution of the cosmic star formation density with redshift, include strong galactic-scale outflows, powered by supernovae explosions as well as accreting supermassive black holes \citep{naab17, pillepich18, dave19, nelson19}. These outflows propagate through the halo gas, and observations of the CGM can therefore place strong constraints on the physical models that are included in theoretical simulations.

One of the most studied metal species is \mgii, which is a doublet with wavelengths $\lambda\lambda 2796,\, 2803\, \angstrom$. Given the relatively high abundance of Mg and high oscillation strength of the transition, it can be detected in optical wavelengths at $z\gtrsim 0.4$, both in absorption \citep{churchill03, zhu13a} and in emission \citep{nelson21, burchett21, zabl21}. The low-ionization potential of magnesium, I.E $\approx 7.65\, eV$, \citep[see][for detailed energy diagram]{prochaska11} implies that it predominantly traces cool, $T\sim 10^{4-4.5}\, \rm K$ gas in and around galaxies. 

The nature of cool, low-ionization gas traced by \mgii absorbers in the CGM of galaxies has been studied in great detail as a function of galactic properties and environment \citep{zhu14, zahedy19, lan18, lan20, schroetter21}. In star-forming galaxies \mgii absorption is 2-3 times higher within $\sim 50$ kpc from the central galaxy than passive galaxies \citep{lan18, lan20} and covering fractions within the virial radius correlate positively with the star formation rate (SFR) of the central galaxy \citep{anand21}. Gas metallicity and \mgii equivalent widths are also higher along the minor axes of galaxies, pointing towards a galactic outflow origin of \mgii absorbers in star-forming or emission-line galaxies (ELGs) \citep{bordoloi11, chen17, lan18, peroux20b}. On the other hand, absorber-galaxy cross-correlation studies have also revealed that passive galaxies (SDSS luminous red galaxies, LRGs) host a significant amount of cool low-ionization gas (traced by \mgii absorbers) in their inner ($\lesssim 50\, \rm kpc$) and outer ($\sim 200\, \rm kpc$) CGM \citep{zhu14, perez15, lan18, anand21, zu21}. The presence of \mgii absorbers is likely a result of complex balance between gas accretion, cooling, and ram-pressure stripping of cool gas from satellite galaxies \citep{zhu14, huang16, zahedy19}. Additionally, because massive galaxies can host AGNs, they can still drive outflows with significant amounts of metal-bearing gas at large velocities out to $\sim$ tens of kpc from the central galaxy \citep{circosta19, nelson19}. 

Although the cool CGM (traced by \mgii absorbers) of galaxies has been studied extensively, few studies have explored the nature and origin of \mgii absorbers in galaxy clusters \citep{lopez08, padilla09, lee21} or groups \citep{nielsen18, dutta20}. In this high-mass regime we have the potential to shed light on the role of environment in shaping the CGM of galaxies. Cluster halo gas, known as the intracluster medium (ICM), is heated up to high temperatures $T\sim \rm 10^{7} - 10^{8} \, K$ due to gravitational collapse. In addition, outflows powered by supermassive black holes in the centre of clusters can also heat the gas \citep{zuhone09, yang16}. The hot ICM emits mostly at X-ray wavelengths due to radiation from thermal bremsstrahlung produced in the highly ionized gas \citep[for a review, see][]{bohringer10, kravtsov12}. Although the ICM is hot, cold/cool gas has sometimes been detected in and around clusters. The most frequently observed elements are hydrogen (H$\alpha$, Ly$\alpha$) \citep{yoon17, muzahid17, pradeep19} and metal absorption lines (\mgii, \ovi) \citep{lopez08, burchett18, lee21}, which are detected in the spectra of background quasars. 

Similarly, recent radio observations with the Very Large Array (VLA) have revealed a large amount of neutral hydrogen (H \textsc{i}) in the Virgo cluster, mostly stripped from cluster galaxies due to ram-pressure of hot ICM \citep{vollmer12, bahe13}. In addition to H \textsc{i}, observations have also revealed the presence of cool low-ionization gas (e.g., \mgii) in groups and clusters \citep{fossati12, dutta20, hamanowicz20, dutta21} and  submillimeter observations have also revealed long extended filaments of cold molecular gas in cool-core clusters \citep{edge01}. In a recent paper \citep{olivares19}, the authors analysed ALMA (CO lines) and MUSE (H$\alpha$) data of three clusters: Centaurus, Abell S1101, and RXJ1539.5 and detected long (3-25 kpc) extended molecular gas filaments in the inner part of cluster halo. By cross-correlating SDSS DR3 quasars and clusters from the Red-Sequence Cluster Survey (RCS), \citet{lopez08} and \citet{padilla09} found that strong absorbers (\ewGr{1}) are more abundant within $<1$ Mpc than $<2$ Mpc from the cluster centre, while weak absorbers follow the expected distribution of \mgii absorbers in field. The authors argued that the difference could be explained if the strong absorbers trace the overdensity of galaxies in clusters. In a recent paper, \citet{lee21} performed a similar \mgii - galaxy cluster cross-correlation with SDSS DR14 quasars and redMaPPer clusters and found $\sim3$ times higher detection rate of \mgii absorbers in inner part ($D_{\rm proj}\lesssim r_{\rm vir}$) of clusters than at large distances ($D_{\rm proj}\gtrsim 3-4r_{\rm vir}$). 

On the other hand, simulations have also predicted the formation and survival of cold gas in cluster environments. Recently, \citet{qiu21} reported a radiation-hydrodynamic simulation of AGN feedback in galaxy clusters (particularly for cool-core clusters) and proposed a scenario in which cold gas filaments ($T\sim 10^{4}\rm \, K$) can condense out of warm-hot ($10^{4}$ to $10^{7}$ K) AGN outflows, as they move out of the central galaxy. Idealized simulations of gas in the ICM have shown that cold gas can form in-situ out of hot ICM due to local temperature fluctuations and thermal instabilities (TI), if the ratio of thermal instabilities time-scale to the free-fall time-scale is below some critical threshold ($t_{\rm TI}/t_{\rm ff}\lesssim 10$) \citep{sharma12, mccourt12, voit15, voit17}. Besides the in-situ formation of cold gas in clusters \citep{dutta21a}, other phenomena such as interactions between cluster galaxies \citep{wang93} and ram pressure stripping \citep{gunn72} of cold gas from satellite galaxies \citep{cortese07, mccarthy08, yun19} can also give rise to cold gas in clusters. Therefore, it is important to explore the nature of cool low-ionization gas in cluster environments and compare its incidence and physical properties with cool gas surrounding isolated galaxies in order to understand its origin and impact on galaxy evolution.

In this paper, our goal is to use the latest datasets of \mgii absorbers and galaxy clusters to perform a high S/N absorber-cluster cross-correlation study to shed light on the nature and origin of cool metal gas ($T\sim 10^{4}$ K, traced by \mgii absorbers) in cluster environments. We aim to detect \mgii absorption and understand how it traces the underlying halo, as well as to connect cool gas with the physical properties of clusters and their member galaxies. We also explore the nature of \mgii absorbers as a function of equivalent width, and study their properties in the inner and outer parts of clusters to shed light on physical processes that might impact the observed properties of cool gas in dense environments. As far as we are aware, this is the largest statistical study of \mgii absorbers in galaxy clusters to date, over a broad range of redshift, $0.4<z<1$ and halo mass, $\rm 10^{13.8}\, M_{\odot}\,<\,M_{500}\,<\, 10^{14.8}\, M_{\odot}$. 

The outline of the paper is as follows: Section~\ref{data} introduces the samples of \mgii absorbers, galaxy clusters and galaxies. We describe the methods in Section~\ref{method} and present our main results in Section~\ref{results}. Finally, we discuss the possible interpretations of our findings and conclusions in Section~\ref{discuss}, followed by a detailed summary of our results in Section~\ref{summary}. We adopt a Planck-consistent cosmology \citep{planck20} for our analysis, i.e. $\Omega_{\rm m,\,0}=0.307,\, H_{0}=67.7$\kms\,$\rm Mpc^{-1}$ and $\Omega_{\rm \Lambda,\, 0} = 1-\Omega_{\rm m,\,0}$ throughout the paper.


\section{Data}\label{data}

\subsection{\mgii absorber catalogue}

\begin{figure*}
	\includegraphics[width=0.45\linewidth]{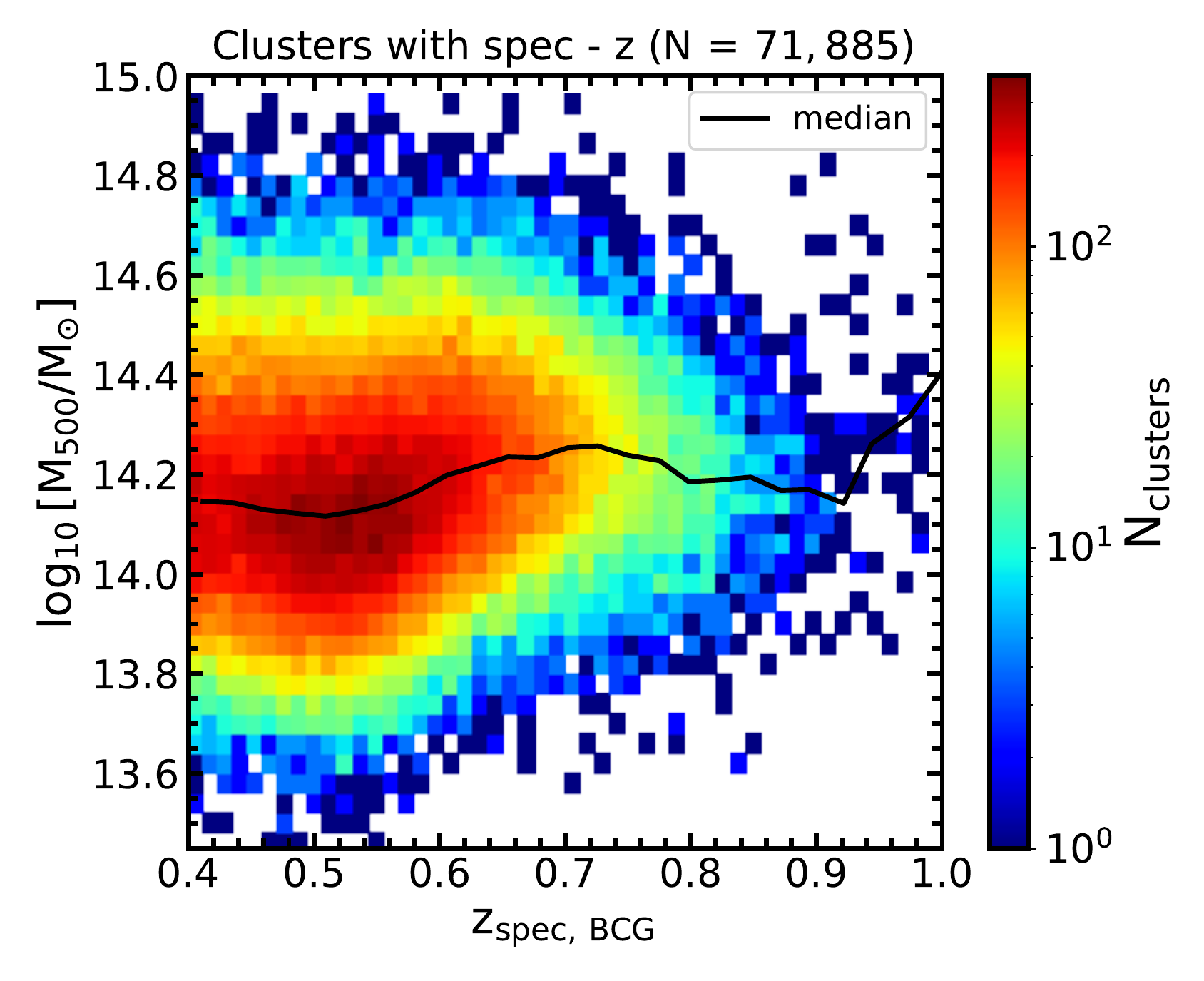}
	\includegraphics[width=0.45\linewidth]{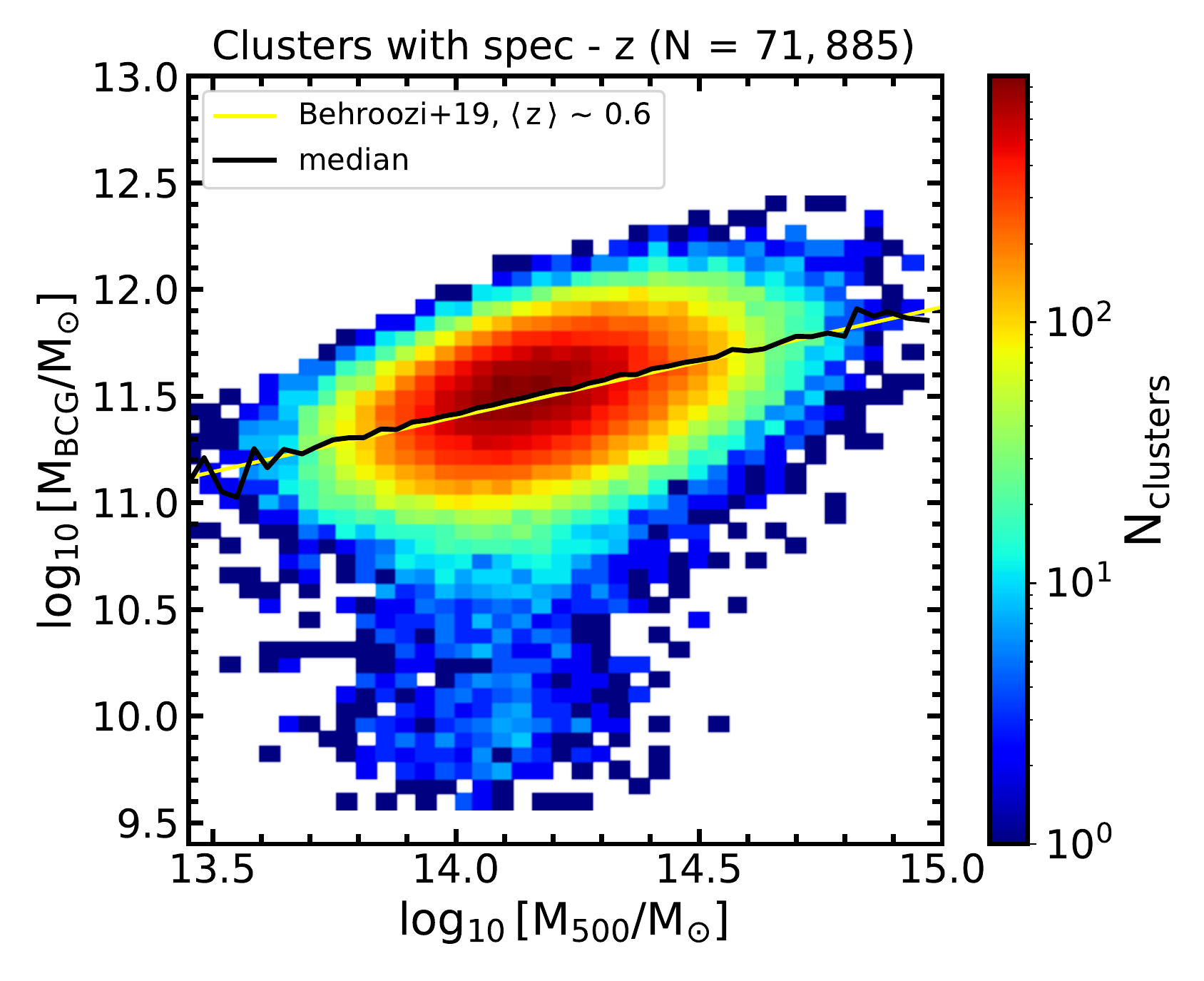}
    \caption{\textbf{Left: }2D histogram of DESI galaxy cluster total mass ($\rm M_{500}$) and spectroscopic redshifts of BCGs. \textbf{Right: }2D distribution of DESI galaxy cluster's BCG stellar mass and total halo mass ($\rm M_{500}$) of clusters. The yellow solid line shows the analytical stellar mass-halo mass (SMHM) model at $z\sim 0.6$ from \citet{behroozi19}. In both panels, the black solid lines show the median values in square bins.}
    \label{fig:mass_z_hist}
\end{figure*}

We use our recent \mgii/\feii absorber catalogue\footnote{publicly available at \url{www.mpa-garching.mpg.de/SDSS/MgII/}} (MPA-SDSS catalogue, \cite{anand21}, hereafter \citetalias{anand21}) based on $\sim 1$ million quasars from SDSS data release (DR16) \citep{ahumada20, lyke2020}. The catalogue includes $\sim 160,000$ \mgii systems and $\sim 70,000$ \feii confirmed systems (see Table \ref{tab:sample}). The completeness and purity both are very high for the catalogue as shown in \citetalias{anand21}, where the pipeline to model quasar continuum and detect absorbers is described in detail. 

To summarize, the detection of absorption lines uses a dimensionality reduction technique, called Non-negative Matrix Factorization \citep[NMF, see also][]{lee99, zhu13a}, to estimate the quasar continuum, followed by a matched kernel convolution technique and adaptive S/N criteria to automatically detect the \mgii/\feii doublets in spectra. In this paper, we focus our analysis on systems with \ewGr{0.4} (denotes the equivalent width of \mgii $\lambda 2796$ line) and redshifts $z>0.4$, which constitutes $\sim \rm 95\%$ of the absorbers in the parent catalogue for which the completeness estimates are determined to be high ($\gtrsim40-50\, \%$, and goes up to $\gtrsim 90\,\%$ for strong systems). The sample with \ewGr{0.4} is fairly complete ($\gtrsim30-40\, \%$) and results obtained for this sample should be considered as most robust to uncertainties resulting from observational selection effects.

\par In the current analysis, we also divide absorbers into two \ewMg bins, namely $0.4\, \angstrom\,<EW_{2796}<\,1\,\angstrom$ (\textit{weak} absorbers) and \ewGr{1} (\textit{strong} absorbers) to characterize the absorber properties in cluster environments as a function of absorber strength. In addition, we also show some results where we combine these two bins (denoted by \ewGr{0.4}) to investigate a few global trends and compare with previous studies. Note that we take only those absorbers that are associated with quasars having $\rm S/N_{QSO}>2$, as both the completeness and the detection purity
of absorbers are high for these quasars \citepalias[see section 3.3 in][for more discussion]{anand21}.

\subsection{Galaxy Clusters}\label{desi_clusters}

We use the galaxy clusters identified in Data Release 8 (DR8) of the legacy imaging survey \citep{dey19} of the Dark Energy Spectroscopic Instrument (DESI). The full galaxy cluster catalogue was compiled by \cite{zou21}. 
It includes $540,432$ clusters with photometric redshifts, $z_{\rm photo}\lesssim1$, for which $122,390$ systems also have the spectroscopic redshifts\footnote{Mostly taken from SDSS, in private communication with Hu Zou.} for their brightest cluster galaxies (BCG) (given as `SZ\_BCG' in the catalogue). In this paper, we use the clusters with spectroscopic redshifts for their BCGs in order to perform a robust cross-correlation study in both projected and redshift space. Since only $z>0.4$ \mgii absorbers can be detected in SDSS spectra due to wavelength coverage, we apply the same redshift cut $z_{\rm BCG}>0.4$ for selecting the final cluster sample with which we will perform the absorber-cluster correlation study. After applying this redshift cut, the final sample includes $N= 71,885$ galaxy clusters. \citet{zou21} do not provide the completeness limits for the stellar mass of BCGs and cluster halo mass as a function of redshift. However, as described in \citet{zou21} the completeness of photo $-z$ galaxy catalogue \citep{zou19} is higher than $\gtrsim90 \%$ for galaxies with $r<23$ ($r-$ band magnitude), which is the magnitude cut for the photo $-z$ galaxies used for cluster identification. 

We summarize the sample sizes with different cuts in Table \ref{tab:sample}. The mean redshift of BCGs in the sample and their corresponding errors are $\mean{z_{\rm BCG}}\approx 0.55$ and $\sigma_{z_{\rm BCG}}\lesssim 60$ \kms, respectively. The mean cluster mass $M_{\rm 500}$, defined as the total mass within the radius $r_{\rm 500}$, where $r_{\rm 500}$ is the radius at which the total mass density is estimated to be 500 times the critical density of the universe\footnote{We derive $r_{\rm 200}$ from $r_{\rm 500}$ as: $r_{\rm 200} \approx\, 1.55\, r_{\rm 500}$ \citep{reiprich13}.} is $ \mean{M_{\rm 500}}\sim 10^{14.2}\, \rm M_{\odot}$. The cluster halo masses $M_{\rm 500}$ are derived with an empirical scaling relation between luminosity ($L_{\rm 1\, Mpc}$) and mass \citep[see also][]{gao20}. The empirical relation was derived based on a calibrated sample of galaxy clusters detected in X-ray and Sunyaev–Zeldovich (SZ) surveys \citep{wen15} that have reliable estimates of $M_{\rm 500}$ and $R_{\rm 500}$. The typical $r_{\rm 500}$ of clusters in our sample is $\sim 700$ kpc. The mean stellar mass of the cluster BCGs is $\mean{M_{\rm BCG}}\sim 10^{11.5}\, \rm M_{\odot}$. We show the distributions of $M_{\rm 500}$ vs redshift in the left panel of Figure~\ref{fig:mass_z_hist} and $M_{\rm 500}$ vs $M_{\rm BCG}$ in the right panel of Figure~\ref{fig:mass_z_hist}. We also plot the median values (solid black line) in each bin in both panels. In the $M_{\rm BCG}$ vs $M_{\rm 500}$ panel, we contrast the median values with the analytical stellar mass-halo mass (SMHM) relation from \citet{behroozi19} (solid yellow line) and they are consistent. There is considerable  scatter around this median line. Part of this scatter arises because clusters are distributed over a broad range of redshifts, $0.4<z_{\rm BCG}<0.8$, while the rest is intrinsic scatter.

\par To better understand the statistical significance of our cluster-absorber correlation, we also estimate the expected average absorption signals for random sky sightlines. For this purpose, following the recipe in \citetalias{anand21}, we define 100 \textit{random galaxy cluster samples}, each equal in size to our parent cluster sample. To do so, we shuffle both the true sky positions and redshifts. The shuffling de-correlates the positions and redshifts of the clusters while preserving the original sky coverage and redshift distribution of the DESI legacy imaging surveys. We compute all observational measurements for both true and random samples and show the results. 

\subsection{Photometric Galaxy Sample}\label{desi_galaxies}

To investigate the association of \mgii absorbers in clusters with their galaxy members, we use the photometric sample of galaxies detected in same legacy imaging surveys of DESI. \citet{zou19} constructed the photo$-z$ galaxy catalogue which is publicly available.
It contains more than $\sim 300$ million galaxies with $z_{\rm photo} < 1$ out of which $\sim 244$ million galaxies have photometric redshifts $z_{\rm photo}>0.4$ and stellar masses in the range $\rm 8.5\,<\, log\, [M_{\star}/M_{\odot}]\,<\, 11.8$ (see Table \ref{tab:sample}). The stellar mass cut is applied to retain high completeness independent of galaxy redshift. As described in the previous section, the overall $r-$ band magnitude completeness of galaxies used in the current analysis is $\gtrsim 90\%$ as all galaxies have $r<23$ in the final sample. The photometric $z$ is estimated with a linear regression method developed in \citet{beck16}. The stellar mass and other galaxy parameters such as star formation rate (SFR) were derived using the Le Phare code \citep{ilbert09} which adopts \citet{bruzal03} stellar population models and the \citet{chabrier03} initial mass function (IMF). 

\begin{figure}
	\includegraphics[width=0.9\linewidth]{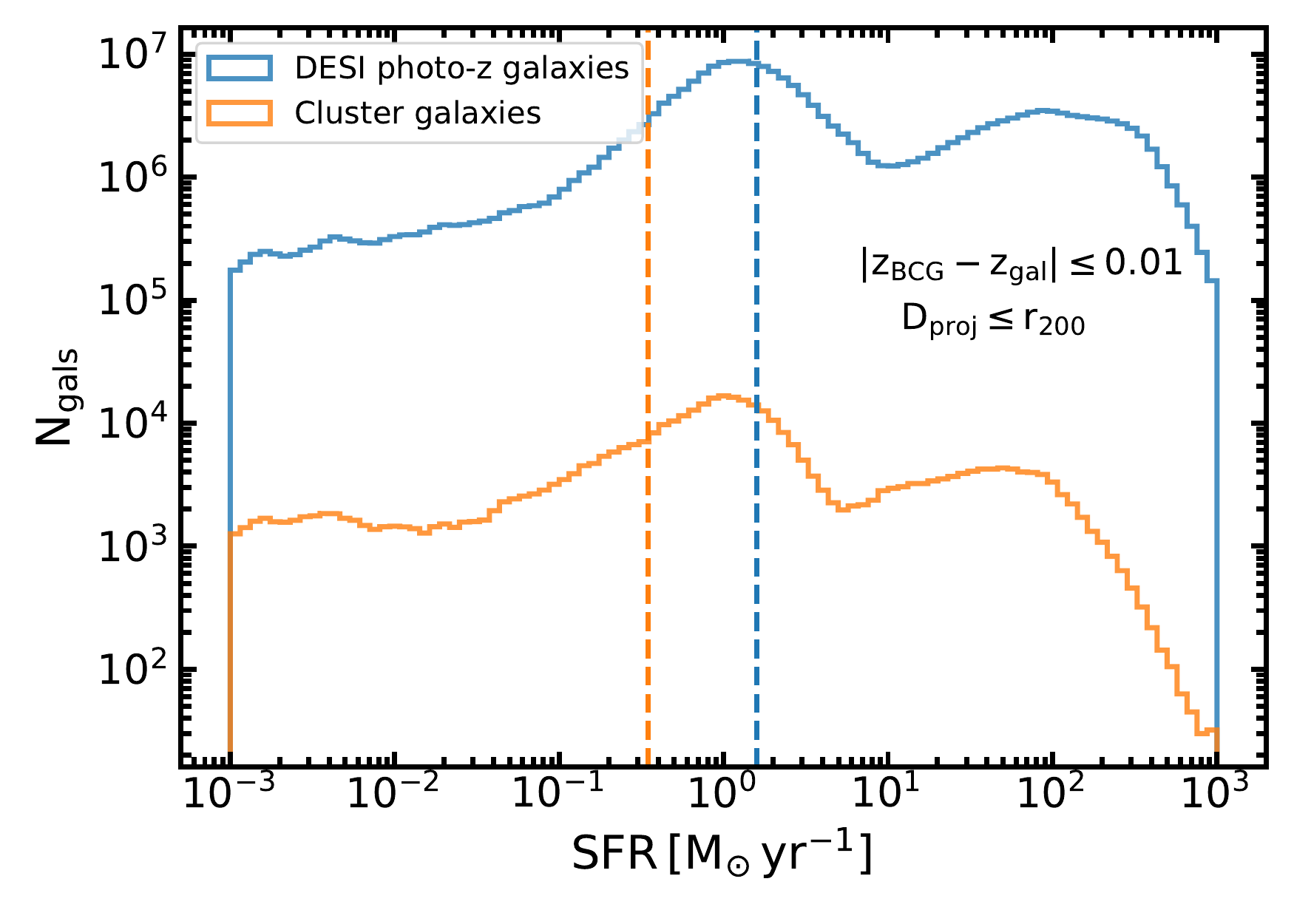}
    \caption{1D distribution of the star formation rate (SFR) of DESI photo$-z$ main galaxy sample (blue) and galaxies that lie within $r_{\rm 200}$ of clusters (orange) as defined in section~\ref{gal_clusters}. We see that typical SFR (shown in dashed vertical orange line) of cluster galaxies is $\sim 6-7$ times lower than main galaxy sample (shown in dashed vertical blue line).}
    \label{fig:photo_z_gals}
\end{figure}

The median photo$-z$ redshift accuracy of the galaxies in our sample is $z_{\rm photo}\approx 0.66$ ($\sigma_{\Delta z_{\rm norm}}\approx 0.017$).\footnote{Defined as the standard deviation of $\Delta z_{\rm norm}=\frac{z_{\rm phot}-z_{\rm spec}}{1+z_{\rm spec}}$ \citep{zou19}.} The median stellar mass of
the galaxies in the sample is $M_{\rm \star}\sim \rm 10^{10.3}\, M_{\odot}$ (with $0.1$ dex uncertainty). 
Similarly the median SFR and specific SFR (sSFR) are $\approx 2\, \rm M_{\odot}\, yr^{-1}$ and $\rm log[sSFR/yr^{-1}]\approx -9.92$, respectively. We show the SFR distribution of the DESI legacy photo$-z$ galaxies (blue) in Figure~\ref{fig:photo_z_gals}.

\subsection{Connecting Galaxies with Clusters}\label{gal_clusters}

In order to find the members of a given cluster, we apply cuts on projected distance and redshift separation between a cluster BCG and nearby galaxies. To this end, we take all the galaxies ($g_{i}$) within the projected distance, $D_{\rm proj}\lesssim 2.5$ Mpc\footnote{This corresponds to $3-4\,r_{500}$ of our clusters.} from the BCG of the cluster with redshift separation, $|\Delta z| \,=\, |z_{\rm BCG} - z_{\rm g_{i}}|\,\leq\,0.01$\footnote{The $|\Delta z|$ condition corresponds to $|\Delta v|\,=\Big |\,\frac{c\Delta z}{1+{z_{\rm BCG}}}\Big |\,\lesssim\, 1900\,$ \kms $\lesssim 2\,v_{\rm 500},\,$ (for $M_{\rm 500}\sim 10^{14.2}\, \rm M_{\odot}$) where $c\, = 3\times10^{5}$ \kms\, is the speed of light.}. Note that we are using photo$-z$ for our calculations which have typically large uncertainties, and hence we are applying a conservative cut. This means we will intentionally exclude some of the cluster members, which should be kept in mind when interpreting the results. In practice, we experiment with a number of cuts on $|\Delta z|$ to make sure that our main conclusions are not sensitive to this choice. 

We show the SFR distribution (in orange) for the cluster galaxies that are found within the $r_{\rm 200}$ of the clusters in Figure~\ref{fig:photo_z_gals}. The typical $r_{\rm 200}$ is $\sim 1$ Mpc for our clusters. Note that the typical SFR of cluster galaxies is $\approx 0.3\, \rm M_{\odot}\, yr^{-1}$ which is $\sim 6-7$ times lower than the SFR of galaxies in the full photo$-z$ sample (see section~\ref{desi_galaxies}, blue curve in Figure~\ref{fig:photo_z_gals}). On the other hand, the typical stellar mass of the full DESI legacy photo-$z$ galaxy sample is roughly the same as for cluster member galaxies. This implies that galaxies that lie in cluster environments have lower SFR at a given stellar mass, as is expected in cluster environments because most of the cluster galaxies have lost their cool gas due to environmental effects \citep[see also,][]{cen14}.

Moreover, we also want to explore the connection between \mgii absorption in clusters and the stellar activity of its member galaxies. To this end, we divide our clusters into two subsamples: a) clusters with predominantly star-forming galaxies, i.e. the star-forming fraction is larger than half ($f_{\rm star}>0.5$), or b) clusters with more passive galaxies. In both cases, we consider galaxies within $r_{500}$, (or $r_{200}$) of the BCG. For this purpose, we use the specific star-formation rate (sSFR) of the member galaxies from our photo$-z$ galaxy catalogue to estimate the fraction of star-forming and passive galaxies in cluster halo. A value of $\rm log(sSFR) = -11$ often defines the boundary (at $z=0$) below which galaxies are classified as passive, while galaxies above this limit are considered star-forming. A similar cut was applied by \citet{werk12} in COS-Halos survey, though other studies may use slightly different cuts for this purpose \citep{donnari19}, e.g, a higher cut, $\rm log(sSFR) = -10$ was applied by \citet{tchernyshyov21} in CGM$^{2}$ survey galaxies. 

We adopt the $\rm log(sSFR) = -11$ threshold, we find that this procedure divides the sample roughly equally. The average fraction of star-forming galaxies within $r_{\rm 500}$ is $f_{\rm star}\approx 0.35$ in our clusters. As we move outwards from the cluster centre, the fraction of star-forming galaxies increases and galaxies in the outskirts of cluster are more star-forming \citep{cen14, butsky19}.

\begin{table}
  \centering
  \caption{Sample statistics of \mgii absorbers, clusters and galaxies in this paper. References are: [a] \citetalias{anand21}; [b] \citet{zou21}; [c] \citet{zou19}}
  \begin{tabular}{||cc||}
    \hline
    Objects & N \\
    \hline
    SDSS DR16 \mgii Absorbers ($\rm S/N_{QSO}>2$) &158,508$\rm ^{a}$\\
    \mgii Absorbers (\ewGr{0.4} and $z>0.4$) & 155,141$\rm ^{a}$\\
    &\\
    DESI DR8 photo$-z$ Clusters & 540,432$\rm ^{b}$\\
    Clusters (with $z_{\rm spec}$) & 122,490$\rm ^{b}$\\
    Clusters ($z_{\rm photo}>0.4$) & 383,085$\rm ^{b}$\\
    Clusters ($z_{\rm spec}>0.4$) & 71,885$\rm ^{b}$\\
    &\\
    DESI Galaxies (with photo$-z$) & 303,379,640$\rm ^{c}$\\
    DESI Galaxies ($z_{\rm photo}>0.4$) & 244,428,600$\rm ^{c}$\\
    \hline
  \end{tabular}
  \label{tab:sample}
\end{table}

\begin{table*}
  \centering
  \caption{Sample statistics comparison versus previous samples. Notes: [1] Number of Mg \textsc{ii} absorbers - cluster pairs ($N_{\rm pair}$) within $D_{\rm proj}<2h^{-1}$ Mpc from the cluster centre; [2] Number of Mg \textsc{ii} absorbers found in 82,000 quasar sightlines that are within $D_{\rm proj}/r_{\rm vir}<5$ from the cluster centre; [3] $N_{\rm pair}$ within $D_{\rm proj}<2h^{-1}\rm Mpc$ from the BCG in our sample; [4] $N_{\rm pair}$ within $D_{\rm proj}/r_{\rm 200}<5$ from the BCG in our study. Our sample sizes are larger by $5-8$ times. [$\star$] $\rm S/N_{QSO}>2$, i.e. signal-to-noise of quasar spectra.}
  
  \begin{tabular}{||cccccc||}
    \hline
    Study& \ewMg limit&QSOs & Clusters & Mg \textsc{ii} Pairs &  References\\
    \hline
    \citet{lopez08}& $>0.3$ \angstrom & $46,420^{\rm a}$ & $\lesssim 500^{\rm b}$ & $421^{\rm 1}$& [a] \citet{schneider05}; [b] \citet{gladders05}\\
    \citet{lee21}&$>0.3$ \angstrom &$526,356^{\rm c}$ & $26,000^{\rm d}$ & $\sim 200^{\rm 2}$& [c] \citet{paris18}; [d] \citet{rykoff14, rykoff16}\\
    Current study & $>0.4$ \angstrom & $773,594$ $\rm ^{e, f, \star}$ & $71,885$& $1337^{\rm 3}$& [e] \citetalias{anand21}; [f] \citet{ahumada20}\\
     Current study & $>0.4$ \angstrom & $773,594$ & $71,885$& $2768^{\rm 4}\, $& \\
    \hline
  \end{tabular}
  \label{tab:sample_comp}
\end{table*}

Previous studies of \mgii absorption aiming specifically to characterize the abundance of cool gas in clusters have had significantly smaller sample sizes ($\approx 3-8$) than the current study -- a comparison is shown in Table \ref{tab:sample_comp}. Recently, \citet{mishra22} performed a stacking analysis of \mgii absorption around \citet{wen15} SDSS clusters with SDSS DR16 quasars \citep{lyke2020} and found a significant reservoir of metal-rich gas in the outskirts of clusters.


\section{Methods}\label{method}

\subsection{\mgii Covering Fraction}\label{cov_frac}

A key value that quantifies the incidence rate of \mgii absorbers in a given environment is the covering fraction ($f_{c}$). Similar to our previous analysis \citepalias{anand21}, we define the covering fraction as the fraction of quasar sightlines in a given projected distance bin $\Delta R$ (optionally normalized by $r_{\rm 500}$ of the cluster), having one or more absorbers with \ewMg larger than a given \ewMg threshold. 

More precisely, we define the following quantities in the same manner as in \citetalias{anand21}. First, $N_{\rm BCG,\, j}^{\rm abs}|_{\rm \Delta R}^{\Delta z}$ is the number of \mgii absorbers (corrected for their completeness, see below) detected around the $j^{\rm th}$ cluster (defined by its BCG) within an annulus $\rm \Delta R$ satisfying a maximum $\Delta z$ separation. We also correct the absorbers by their completeness factor, $c({\rm{EW}_{\rm 2796}},z)$, that we estimated in \citetalias{anand21} and the `effective number' is given as $1/c({\rm{EW}_{\rm 2796}},z)$. As the completeness factor, $c({\rm{EW}_{\rm 2796}},z)\leq1$, the corrected number of absorbers for a given $EW_{\rm 2796},\, z$ is, $N_{\rm abs}\geq 1$.

Analogously, $N_{\rm BCG,\, j}^{\rm QSO}|_{\Delta R}$ is the corresponding number of quasars, i.e. the number of sightlines in that annular bin. The subscript $\Delta R$ denotes an annulus with projected inner and outer radii $\rm R_{1}$ and $\rm R_{2}$, and $|\Delta z| = |z_{\rm BCG} - z_{\rm \mgii}|$ is the redshift separation between BCG and the \mgii; we adopt $|\Delta z|\leq\rm 0.01$\footnote{The $|\Delta z|$ condition corresponds to $|\Delta v|\,=\Big |\,\frac{c\Delta z}{1+{z_{\rm BCG}}}\Big |\,\lesssim\, 1900\,$ \kms $\lesssim 2\,v_{\rm 500},\,$ (for $M_{\rm 500}\sim 10^{14.2}\, \rm M_{\odot}$) where $c\, = 3\times10^{5}$ \kms\, is the speed of light.} in our analysis same as \citetalias{anand21}. Moreover, we also show a comparison for the $|\Delta z|\leq 0.003$\footnote{This corresponds to $|\Delta v|\leq 600$ \kms\, which is smaller than the typical $v_{\rm 500}$ of cluster halo.} case in Appendix~\ref{dz_compare}, to understand if our conclusions change significantly with this choice. We find that decreasing $|\Delta z|$ to such small values does not change our overall conclusions, particularly at small distances from the BCG (see Figures~\ref{fig:dz003_diff_fc} and \ref{fig:dz003_ew_mgii}).

Given a cluster BCG with angular positions and redshift $(\alpha,\delta,z)$, we derive the covering fraction (defined above) of absorbers around clusters. 
\begin{ceqn}
\begin{equation} \label{eqn:fc_def}
  f_{\rm c}|^{\Delta R} = \rm \frac{N(sightlines\, with\, absorbers)}{N(QSO\, sightlines)}
  = \frac{ \sum_{j} N_{BCG,\,j}^{abs}|_{\Delta R}^{\Delta z} }{ \sum_{j} N_{BCG,\,j}^{QSO}|_{\Delta R} } .
\end{equation}
\end{ceqn}
\noindent where $j$ varies over all clusters in given galaxy cluster sample. If multiple absorbers satisfy the given \ewMg threshold, they are counted as multiple sightlines as we are counting absorbers over all clusters (the sum is over all $j$) in a given radial bin. As described in \citetalias{anand21} and \citet{zhu13a}, the typical width of a single \mgii absorption system can vary from $4-8$ pixels that correspond to $280-500$ \kms in SDSS quasar spectra\footnote{Each wavelength pixel in SDSS spectra corresponds to $70$ \kms in SDSS spectra.}. 

We estimate errors by bootstrapping the \mgii- cluster cross-correlation 100 times, as described in \citetalias{anand21}. 

\subsection{\mgii Absorption and Surface Mass Density} \label{surf_mass}

Another physical quantity that characterizes the strength and overall distribution of \mgii absorbers in a given space around cluster or galaxies, is the mean \mgii absorption strength. To this end, we first estimate the total $\mean{EW_{2796}}$ by summing \ewMg, contributed by all absorbers with \ewGr{0.4} in a given radial bin $\Delta R$ around clusters. Then we divide this total \ewMg by the number of absorbers in that radial bin. Next, to account for the incidence rate of absorbers, we also multiply this statistical mean by the covering fraction, to measure the mean $\mean{EW_{2796}}$ of \mgii absorbers per sightline. We combine both weak and strong absorbers here because we want to measure the total \mgii absorption. Mathematically,

\begin{equation}
    \langle EW_{2796} \rangle ^{\Delta R} = \frac{\sum_{i}EW_{\rm2796,\, i}^{\Delta R}}{N_{\rm abs}}\, f_{c}^{\Delta R}
    \label{eqn:mg_ew}
\end{equation}

where $i$ varies over all the absorbers in a given radial bin, $N_{\rm abs}$ is the number of absorbers (i.e. the sum over all $i$) and $f_{c}^{\Delta R}$ is the covering fraction in that annular bin ($\Delta R$). Note that these mean values should be close to the mean absorption that we would have obtained by stacking the quasar spectra directly, i.e. accounting for rare (weak on average) absorption in the spectra. We test this hypothesis for SDSS LRGs from our previous analysis \citepalias{anand21} and compare the results with \citet{zhu14, lan18}\footnote{For more details see also section 2.2 of \citet{lan18}.}, the values are consistent within error bars (see Figure~\ref{fig:sdss_lrg_anand21_zhu14}).

Once we have the mean equivalent widths ($\mean{EW_{2796}}^{\Delta R}$) we use the curve of growth for \mgii assuming we are in the linear regime (Eqn.~\ref{eqn:mg_col}), following the derivation in \citet{draine11}, to estimate the column density \citep[see also][Eqn. 11]{zhu14}. Assuming a thermal broadening factor of $b\sim 4$ \kms $\implies\, T\sim 25,000\, \rm K$, the stronger \mgii line (i.e. $\lambda\,2796$) starts saturating at \ewMg$ \gtrsim 0.15-0.2\, \angstrom$ \citep{churchill2000, zhu14}. As a result, only a lower limit on $\rm N_{\mgii}$ can be estimated for the strongest absorbers, if we use the linear part of curve of growth. For very weak absorbers in optically thin ($\tau \ll 1$, i.e. unsaturated), regions $\rm \mean{N_{\mgii}}^{\Delta R}$ can be estimated as 

\begin{equation} 
  \langle N_{\rm \mgii}\rangle^{\Delta R} = 1.13\, \times 10^{20}\, \frac{\mean{EW_{2796}}^{\Delta R}}{f_{\rm 2796} \lambda_{2796}^{2}}\,\,\rm cm^{-2}
  \label{eqn:mg_col}
\end{equation}

where, \ewMg and $\lambda_{\rm 2796}$ are in \angstrom\, and $f_{\rm 2796}$ is the oscillator strength\footnote{$f_{\rm 2796} = 0.608, \, \lambda_{2796}\,=\, 2796.35\,\angstrom$.} of \mgii $\lambda\,2796$ line. Note that this will only give lower limits on column densities for saturated systems, and we must use the full curve of growth in such cases. Next we estimate the \mgii surface mass density, which is defined as the mass of \mgii absorbers per unit surface area, by multiplying the column densities by the atomic mass of \mgii\footnote{$\rm m_{\mgii} = 24.305\, amu$.}. We show the results in units of $\rm M_{\odot}\,kpc^{-2}$.

\begin{equation} 
  \sum\nolimits_{\rm \mgii}^{\Delta R} = \rm \mean{N_{\mgii}}^{\Delta R}\,m_{\mgii}
  \label{eqn:surf_def}
\end{equation}

In the section~\ref{mgii_ew_cluster}, we will further illustrate that most of the absorbers have small $\mean{EW_{2796}}$ in clusters and therefore our assumption that they are unsaturated and lie on the linear part of the curve of growth is fairly well justified.

\par Next, we also estimate the total mass of \mgii gas in clusters and SDSS LRGs within a given radius, using our surface mass density estimates. We describe the methods in Section \ref{mgii_mass}.

\subsection{Connecting \mgii Absorbers with Cluster Members}\label{mgii_members}

To further investigate the connection between \mgii absorbers (detected in clusters) and cluster members (described in section~\ref{desi_galaxies}), we also connect \mgii absorbers with cluster galaxies in projected and redshift space. In particular, for each cluster we first find all the absorbers within a given $D_{\rm proj}$ (e.g., $\lesssim r_{\rm 200}$). Then we find the nearest member galaxy to each absorber -- the results are discussed in Section~\ref{mgii_gals_connection}.

\par Furthermore, we also want to test the hypothesis that absorbers are not related to the locations of cluster member galaxies. To do so we construct a pseudo-random \mgii - galaxy cross-correlation sample. First, for each cluster we take all the absorbers within a maximum aperture ($D_{\rm proj}\lesssim r_{\rm 200}$) and distribute them uniformly in angle, keeping their distance to the BCG fixed.\footnote{To distribute absorbers randomly in angle (both polar and azimuthal) on a spherical shell (centred at BCG with coordinates $x_{\rm o},\, y_{\rm o}, \, z_{\rm o}$) with radius $R$, where $R$ is the true distance between BCG and the absorber, we use the following recipe: Generate azimuthal and polar angles with uniform distributions, $\rm \phi = U\, [0, 2\pi], \, cos\theta = U\,[-1, 1]$, respectively. The coordinates in 3D space for new absorber ($x_{\rm a},\, y_{\rm a}, \, z_{\rm a}$) would then be: $x_{\rm a} = x_{\rm o}+\rm R\,cos\phi\,sin\theta$, $y_{\rm a} = y_{\rm o}+\rm R\,sin\phi\,sin\theta$, $z_{\rm a} = z_{\rm o}+\rm R\,cos\theta$.} While distributing absorbers randomly we keep the number of absorbers fixed to preserve their number density profile. As above, we then find the nearest member galaxy to each random absorber -- the results of this exercise are discussed in Section~\ref{mgii_gals_connection}.

\subsection{Galaxy Samples for a Comparative Study}

The measurements from previous studies of \mgii absorption around star-forming and passive galaxies (described in section~\ref{intro}) allow us to compare the properties of \mgii absorption in clusters with measurements around field galaxies. To perform such a comparative study, we contrast our \mgii measurements in clusters with cool gas properties around emission-line galaxies (ELGs, \citet{raichoor21}) and luminous red galaxies (LRGs, \citet{chen12}) that we measured in \citetalias{anand21}. As ELGs, LRGs and BCGs trace a wide range of halo masses from around $10^{12}\, \rm M_{\odot}$ to $10^{15}\, \rm M_{\odot}$, this comparison will help understand how \mgii absorber properties vary as a function of halo mass. We summarize the basic properties (redshift, halo mass, stellar mass and virial radii) of these objects in Table~\ref{tab:gal_comp} for a detailed comparison.

\begin{table}
  \centering
  \caption{Summary of galaxy properties of ELGs, LRGs and BCGs.}
  \begin{tabular}{||cccccc||}
    \hline
    Objects & $\langle\, z\,\rangle$& $\langle\,M_{\rm \star}\,\rangle$& $\langle\,M_{\rm 500}\,\rangle$& $\langle\,R_{\rm 500}\,\rangle$&$\langle\,R_{\rm 200}\,\rangle$\\
    && [$\rm M_{\odot}$] & [$\rm M_{\odot}$] & [$\rm kpc$]&[$\rm kpc$]\\
    \hline
    ELGs & 0.84& $\rm 10^{10.5}$ & $\rm 10^{12.2}$ & $120$&$200$\\
    LRGs & 0.54 & $\rm 10^{11.3}$ & $\rm 10^{13.5}$ & $350$&$550$\\
    BCGs & 0.55 & $\rm 10^{11.5}$ & $\rm 10^{14.2}$ & $700$&$1100$\\
    \hline
  \end{tabular}
  \label{tab:gal_comp}
\end{table}


\begin{figure*}
	\includegraphics[width=0.45\linewidth]{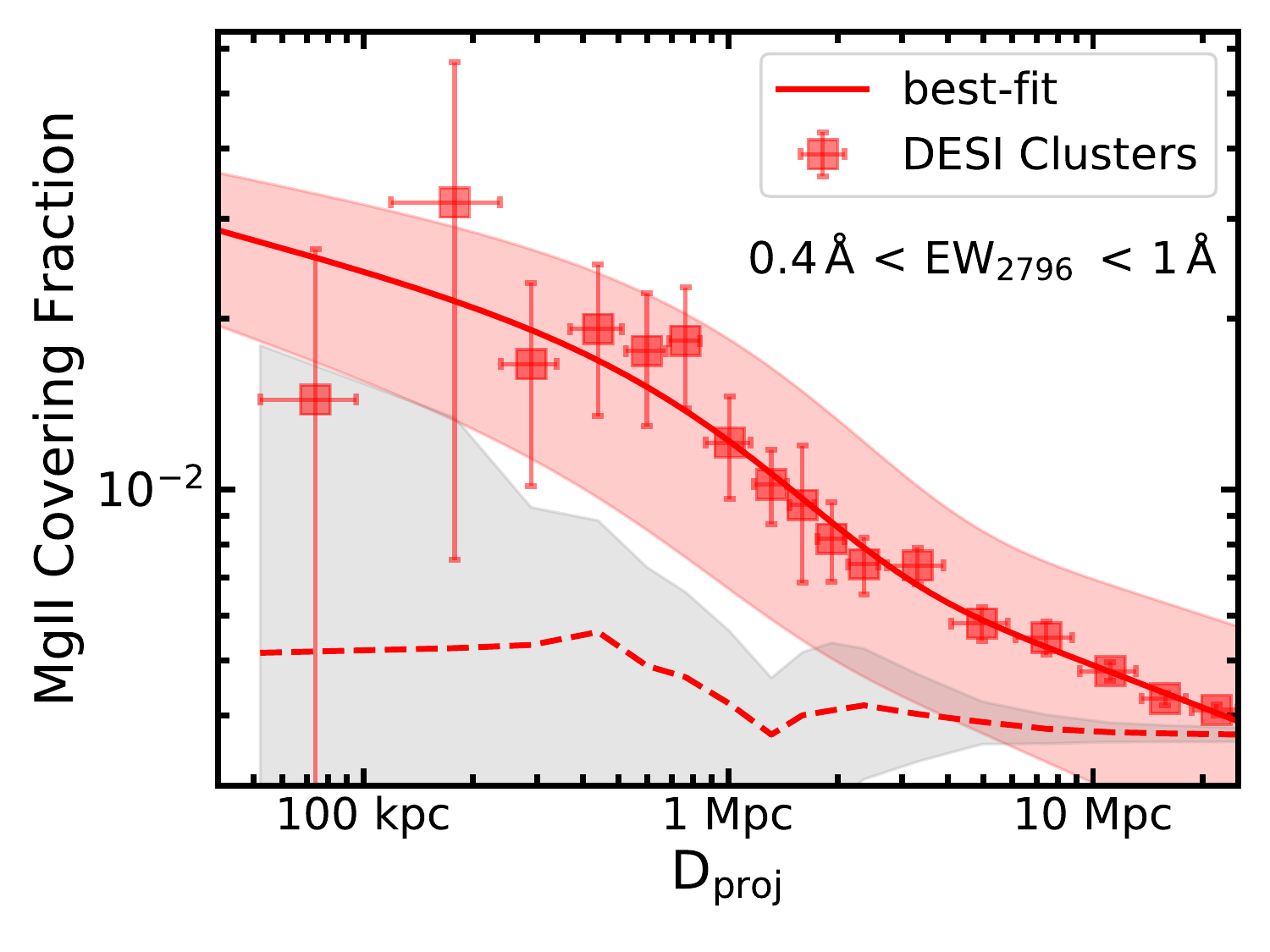}
	\includegraphics[width=0.45\linewidth]{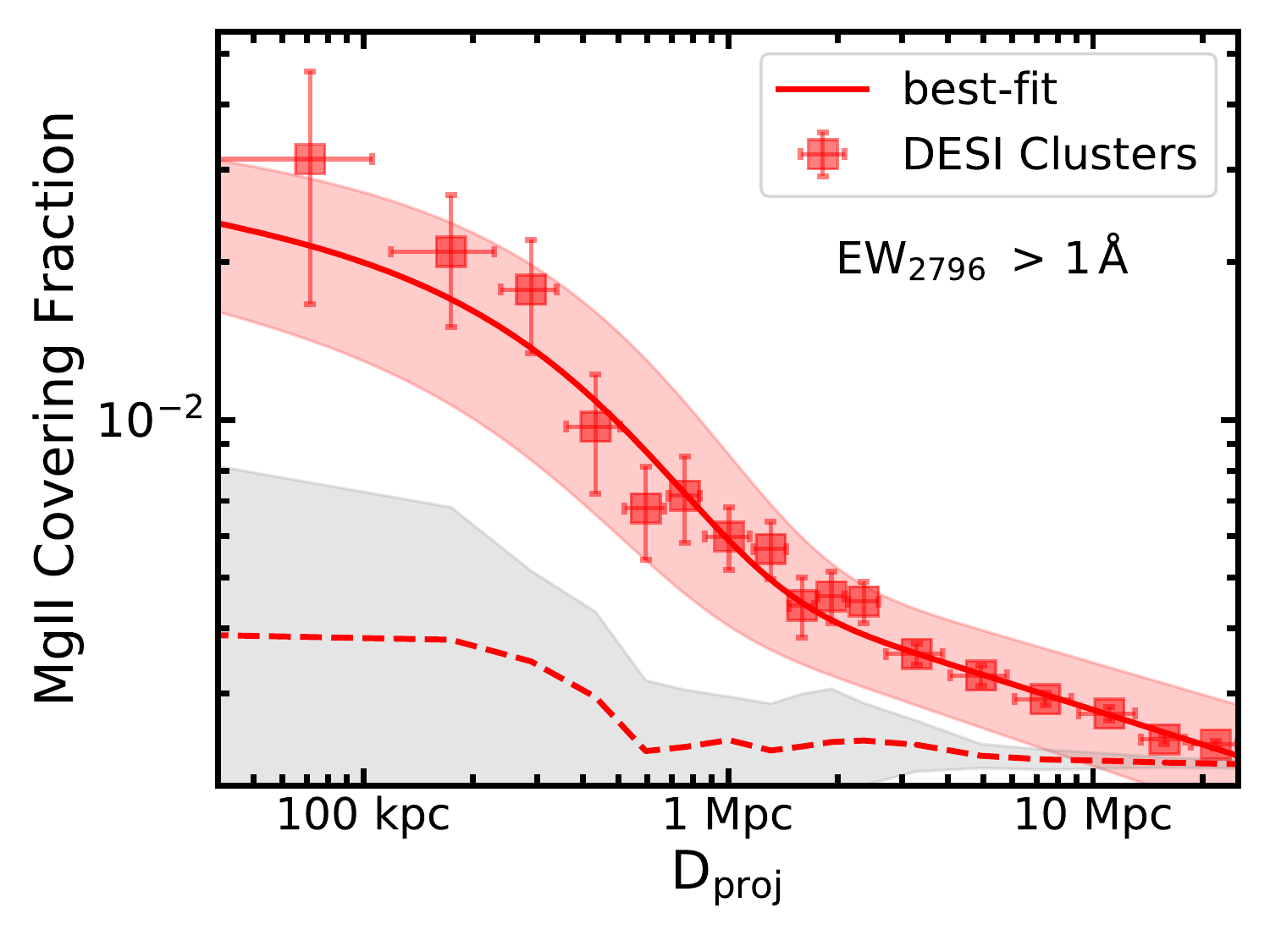}\\
	\includegraphics[width=0.45\linewidth]{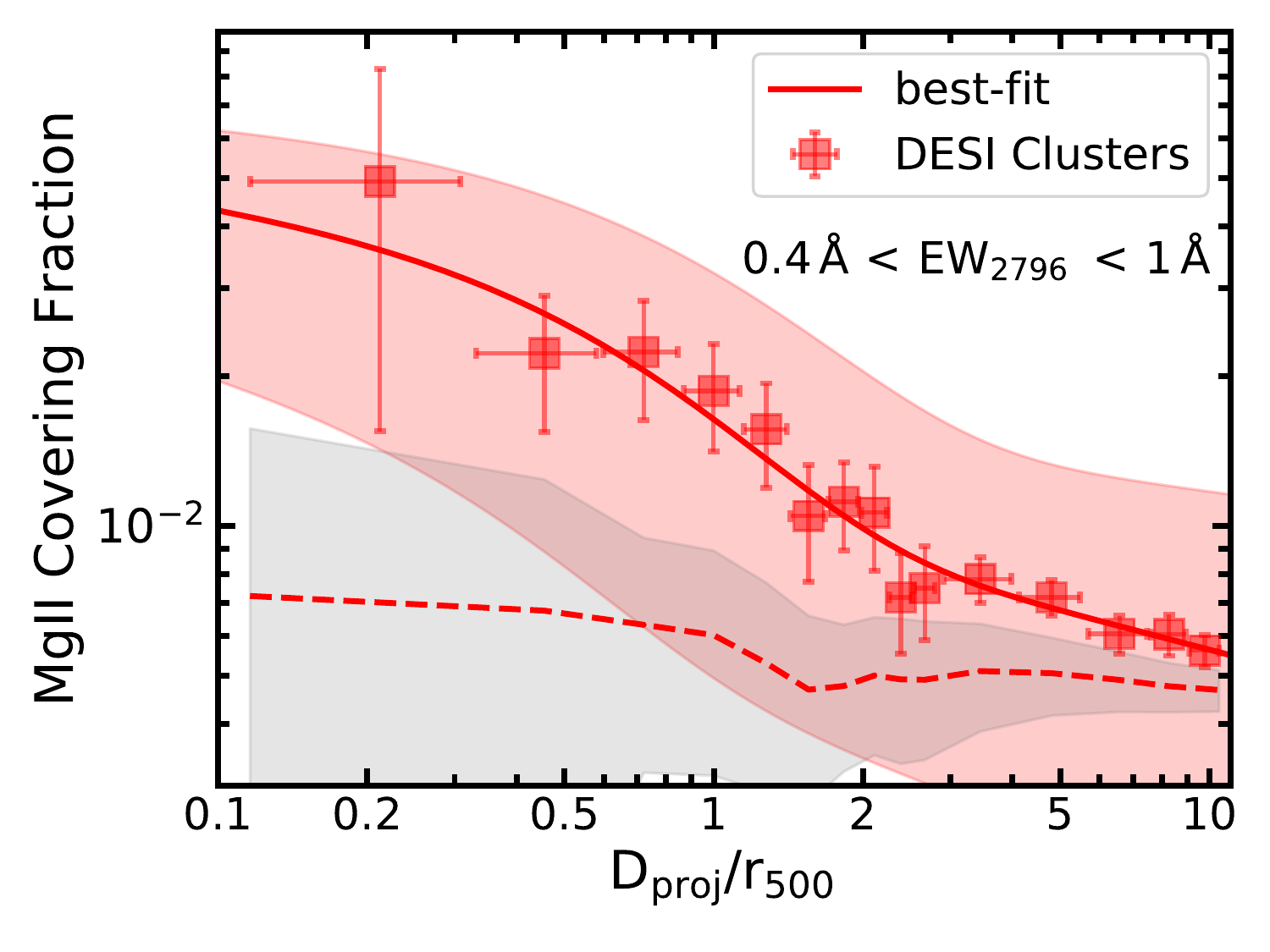}
    \includegraphics[width=0.45\linewidth]{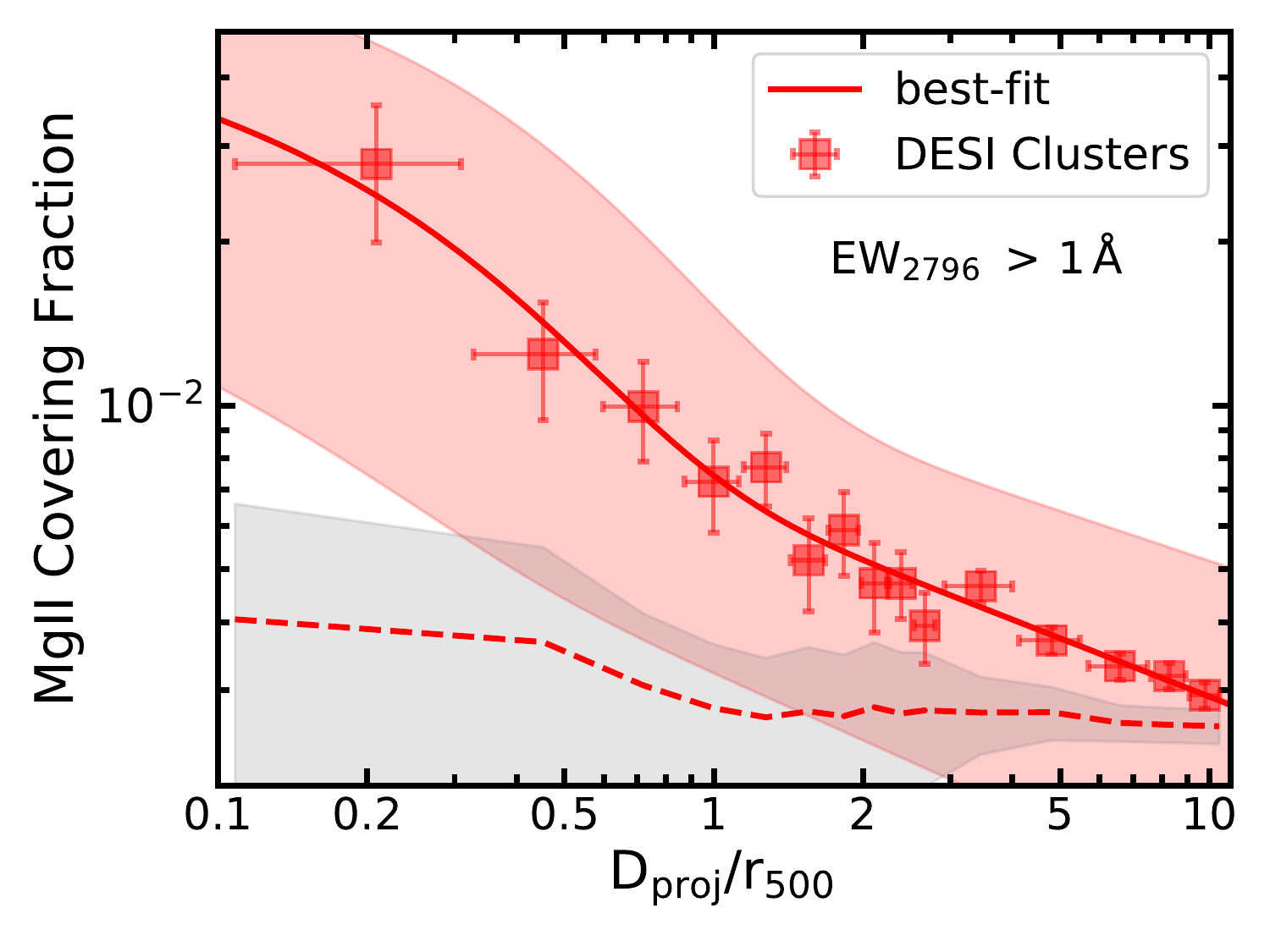}\\
    
    \caption{\textbf{Top:} Differential covering fraction of weak and strong \mgii absorbers around DESI galaxy clusters  as a function of projected distance from the BCG. \textbf{Bottom:} \mgii covering fractions of weak and strong systems as a function of projected distance normalized by $r_{\rm 500}$ of cluster. The corresponding dashed lines show the values measured for the random clusters. We clearly see a decreasing trend with both distance and absorber strength. The solid lines show the best-fitting profiles described in Section~\ref{best_fit}, and the shaded regions show the corresponding $1\sigma$ uncertainty intervals.}
    \label{fig:diff_fc}
\end{figure*}

\section{Results}\label{results}

\subsection{Galaxy Cluster - \mgii Correlation} \label{correlation}

\subsubsection{\mgii Covering Fractions in Galaxy Clusters}\label{mgii_cov_frac}

We start by cross-correlating clusters with \mgii absorbers in redshift and projected distance space. First we derive the \mgii covering fraction ($f_{\rm c}$), described in Section \ref{cov_frac}, as a function of projected distance ($D_{\rm proj}$) from the BCG of the clusters for both \textit{weak ($0.4\,\angstrom\,<\, EW_{\rm 2796}<\,1\, \angstrom$)} and \textit{strong (\ewGr{1})} absorbers. We show the covering fractions ($ d_{\rm 1}<D_{\rm proj}<d_{\rm 2}$) for weak absorbers in top left panel of Figure~\ref{fig:diff_fc} and for strong absorbers in top right panel of Figure~\ref{fig:diff_fc}. The corresponding dashed lines show the measurements for the random mocks.

We see that on smaller scales ($D_{\rm proj}\lesssim 1$ Mpc), the true measurements are $3-5$ times higher than random expectations, indicating the statistical significance of our measurements. The covering fraction decreases with the projected distance. We see that for the weak absorbers, the $f_{\rm c}$ varies from $\sim 2-3$ per cent (similar for $|\Delta z|\leq 0.003$, upper left panel of Figure~\ref{fig:dz003_diff_fc}) within $D_{\rm proj}\lesssim$ 200 kpc to $\lesssim 1$ per cent ($\lesssim 0.5$ per cent for $|\Delta z|\leq0.003$) at $D_{\rm proj}\sim 2$ Mpc. For \ewGr{1} absorbers the \mgii covering fraction varies from $\sim 2-3$ per cent ($\sim 1-2$ per cent for $|\Delta z|\leq 0.003$, upper right panel of Figure~\ref{fig:dz003_diff_fc}) within $D_{\rm proj}\lesssim 200$ kpc to $\lesssim 0.5$ per cent ($\lesssim 0.3$ per cent for $|\Delta z|\leq 0.003$) at $D_{\rm proj}\sim 2$ Mpc. This implies that the covering fraction is also a function of absorber strength, particularly at larger distances. The covering fractions converge to the random expectation at $D_{\rm}\gtrsim 10$ Mpc in each EW bin. 

Next, we normalize the projected distance by dividing by the  $r_{\rm 500}$ of the clusters. Results are
shown in the bottom panels of Figure~\ref{fig:diff_fc}. The covering fraction of weak absorbers is $\sim 2-5$ per cent ($1-4$ per cent for $|\Delta z|\leq 0.003$, lower left panel of Figure~\ref{fig:dz003_diff_fc}) at $D_{\rm proj}\lesssim 0.5r_{\rm 500}$, while it is $\lesssim 3$ per cent ($\lesssim 2$ per cent for $|\Delta z|\leq 0.003$, lower right panel of Figure~\ref{fig:dz003_diff_fc}) for strong absorbers at similar scales. In the outer parts of cluster halo $D_{\rm proj}\gtrsim r_{\rm 500}$, it is about $\sim 1.5$ per cent (similar for $|\Delta z|\leq 0.003$) for weak absorbers and $\lesssim 0.7$ per cent ($\lesssim 0.4-0.5$ per cent for $|\Delta z|\leq 0.003$) for strong absorbers, similar to projected distance trends. The values converge to the random expectations at large distances.

In comparison to SDSS LRGs \citep{lan20, anand21} the covering fractions are slightly larger for DESI legacy clusters, likely due to the bigger halo masses and denser environments. Recently, \citet{nielsen18} performed a similar cross-correlation study of \mgii absorption in groups based on a heterogeneous literature sample and \citet{dutta21} did another study based on MUSE and HST observations and found similar higher covering fractions of \mgii absorbers in groups relative to isolated environments.

\subsubsection{Mean Equivalent Width of \mgii Absorbers}\label{mgii_ew_cluster}

\begin{figure*}
	\includegraphics[width=0.475\linewidth]{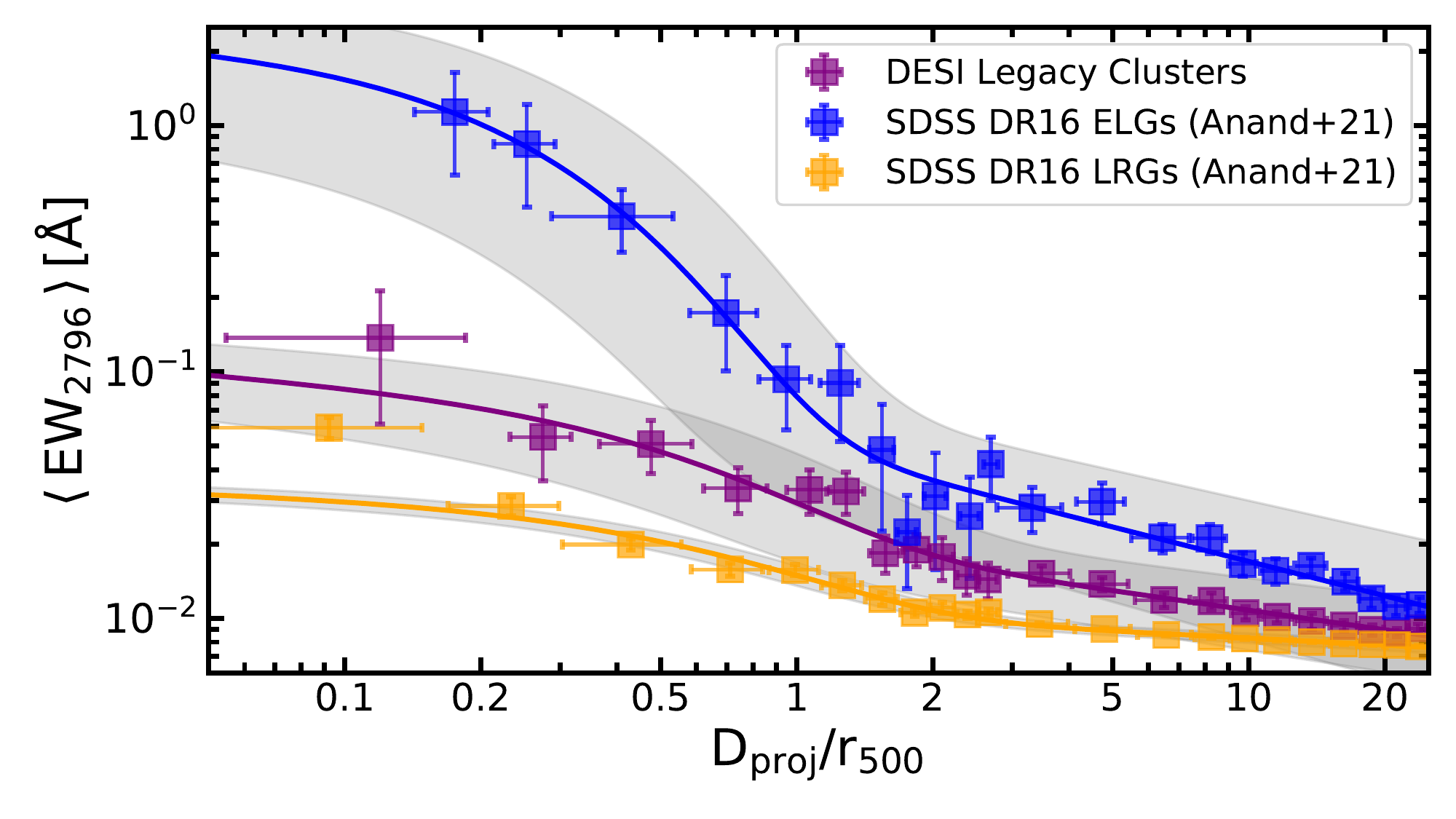}
	\includegraphics[width=0.475\linewidth]{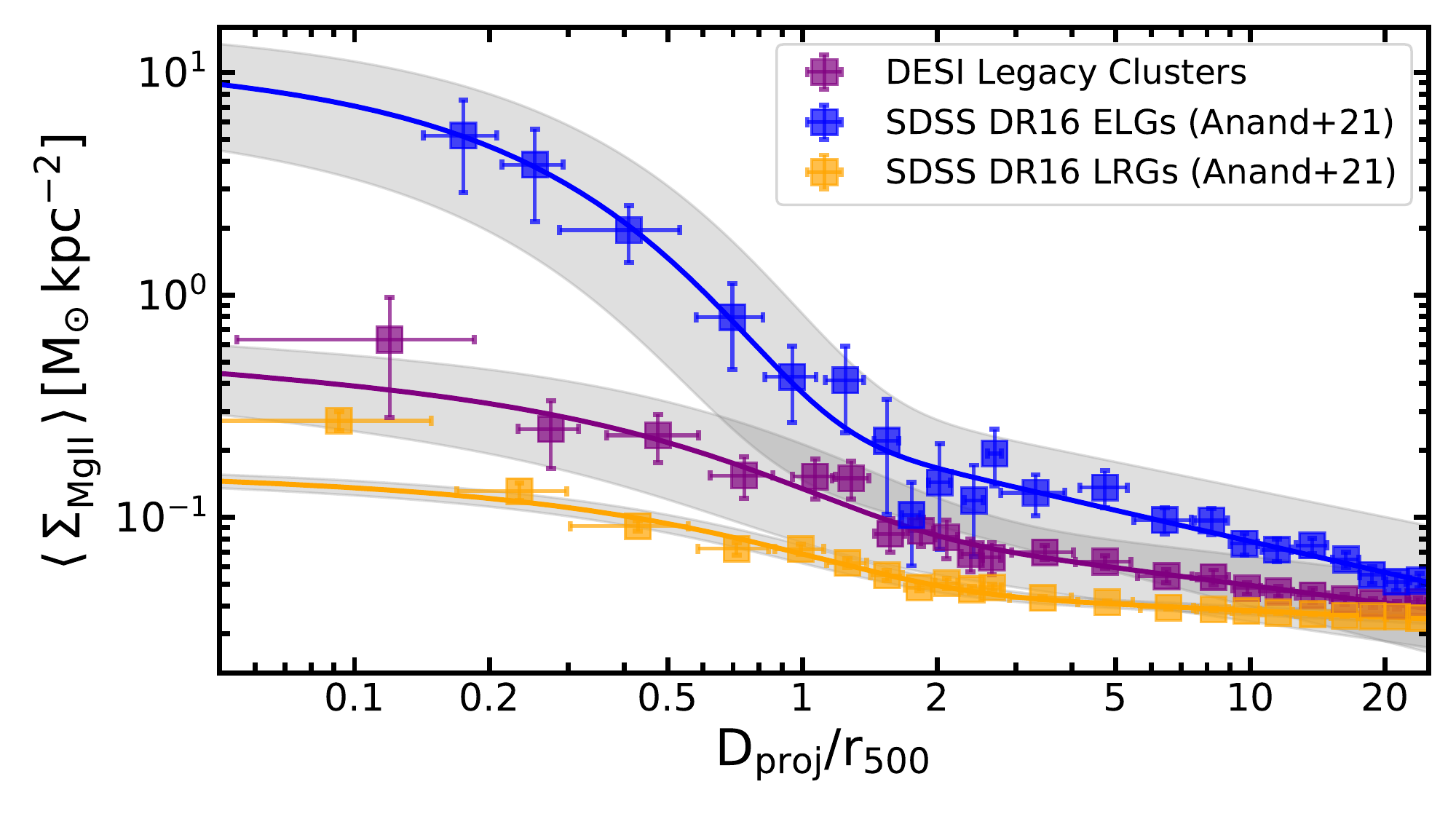}
    \caption{\textbf{Left:} The mean $\mean{EW_{2796}}$ in and around DESI legacy galaxy clusters (purple), SDSS DR16 ELGs (blue) and SDSS DR16 LRGs (orange) as a function of projected distance normalized by $r_{\rm 500}$ of the corresponding haloes. \textbf{Right:} Differential average surface mass density of \mgii absorbers in and around DESI legacy galaxy clusters (purple), SDSS DR16 ELGs (blue) and SDSS LRGs (orange) as a function of normalized distances. The solid lines are the best-fitting curves described in section~\ref{best_fit} and the best-fitting parameters are summarized in Table~\ref{tab:params_fc}. The shaded regions show the corresponding $1\sigma$ uncertainty intervals. For all three samples: SDSS ELGs, SDSS LRGs and DESI legacy clusters, the values converge at large distances.}
    \label{fig:surf_proj}
\end{figure*}

We next estimate the mean equivalent width $\mean{EW_{2796}}$ in galaxy clusters as described in Eqn~\ref{eqn:mg_ew}. We show the results in the left panel of Figure~\ref{fig:surf_proj}, where $\mean{EW_{2796}}$ is shown in purple squares as a function of projected distance ($D_{\rm proj}$) from the BCG, normalized by $r_{\rm 500}$. We also contrast our results with SDSS DR16 LRGs (orange squares) SDSS DR16 ELGs (blue) \citep{anand21} where the projected distance is likewise normalized by $r_{\rm 500}$.

We find that the average \mgii absorption strength is systematically higher in clusters than SDSS LRGs at a given normalized distance from the central galaxy, up to $D_{\rm proj}\lesssim 10\,r_{\rm 500}$, however, the mean $\mean{EW_{\rm 2796}}$ around ELGs is $5-10$ times higher than both clusters and LRGs. It is not surprising, as previous studies \citep{lan14, lan18, anand21} have shown that high SFR of ELGs contributes (via stellar outflows) significantly to the overall \mgii content in the inner part of halo ($\lesssim 50\, \rm kpc$). All three measurements converge at large distances ($D_{\rm proj}\gtrsim 15\,r_{\rm 500}$, $\mean{EW_{\rm 2796}}<0.01\, \angstrom$). The mean $\mean{EW_{\rm 2796}}$ varies from $\approx 0.15-0.05\rm \angstrom$ (similar values for $|\Delta z|\leq 0.003$ case, left panel of Figure~\ref{fig:dz003_ew_mgii}) within $D_{\rm proj}\lesssim r_{\rm 500}$ in DESI legacy clusters while on similar scales it varies from $\approx 0.05-0.02\, \rm \angstrom$ for LRGs and from $\approx 1-0.1\, \rm \angstrom$ for ELGs, respectively. Note that $D_{\rm proj} \approx 10 \, r_{\rm 500}$ will correspond to $2-10$ Mpc on projected scales for ELGs, LRGs and clusters because of their different halo sizes. 

\subsubsection{\mgii Surface Mass Density}\label{surf_dens_mgii}

We proceed to estimate the average column density, $\mean{N_{\rm \mgii}}$ and surface mass density, $\mean{\sum\nolimits_{\rm \mgii}}$ (see Eqn~\ref{eqn:surf_def}) of \mgii absorbers as a function of projected distance ($D_{\rm proj}$) normalized by $r_{\rm 500}$ of the clusters. We use the linear curve of growth (Eqn~\ref{eqn:mg_col}), as justified by the relatively weak equivalent widths measured, particularly for clusters and SDSS LRGs. However, we point out that our measurements are only a lower limit on the total column densities and surface mass densities (particularly for SDSS ELGs, as most of the \mgii absorbers are saturated ($EW_{\rm 2796}\gtrsim 0.2$ \angstrom) around them), as our method is based on the detection of individual absorbers and thus insensitive to very weak \mgii sightlines. A more complete estimate would be possible by measuring both $\mean{EW_{2796}}$ and $\mean{\sum\nolimits_{\rm \mgii}}$ in stacked spectra in order to account for the contribution from the weakest systems \citep[][]{zhu14, perez15, zu21}. 

We show our results in Figure~\ref{fig:surf_proj} (right panel) as a function of $D_{\rm proj}/r_{\rm 500}$. The average surface mass density of \mgii absorbers in galaxy clusters (purple squares) decreases with cluster-centric distance. It varies from $\mean{\sum\nolimits_{\rm \mgii}}\approx 0.6\, \rm M_{\odot}\, kpc^{-2}$ at $D_{\rm proj}\lesssim 0.2\,r_{500}$ to $\mean{\sum\nolimits_{\rm \mgii}}\lesssim 0.2\, \rm M_{\odot}\, kpc^{-2}$ at $D_{\rm proj}\lesssim r_{500}$. At large distances, $D_{\rm proj} \gtrsim 10\,r_{\rm 500}$ the mean $\mean{\sum\nolimits_{\rm \mgii}}\lesssim 0.05\, \rm M_{\odot}\, kpc^{-2}$, $\approx 5$ times lower than inner halo. We also note that distribution becomes flat, possibly indicating that we have started tracing the IGM. We also performed the analysis for $|\Delta z|\leq 0.003$ case (right panel of Figure~\ref{fig:dz003_ew_mgii}) and we do not find any difference in the trend and measurements.

We contrast the cluster measurements with the surface mass density of \mgii absorbers around SDSS LRGs (orange squares) SDSS ELGs (blue squares) \citep{anand21}. For SDSS LRGs, the measurement varies from $\lesssim 0.2-0.3\, \rm M_{\odot}\,kpc^{-2}$ at $D_{\rm proj}\lesssim 0.2\,r_{500}$ to $\lesssim 0.1\, \rm M_{\odot}\,kpc^{-2}$ at $D_{\rm proj}\lesssim r_{500}$, while for SDSS ELGs it varies from $\lesssim 5-6\, \rm M_{\odot}\,kpc^{-2}$ at $D_{\rm proj}\lesssim 0.2\,r_{500}$ to $\lesssim 0.6\, \rm M_{\odot}\,kpc^{-2}$ at $D_{\rm proj}\lesssim r_{500}$ (note that these values are just the lower limits as described above). This indicates that in the inner part ($D_{\rm proj}\lesssim r_{500}$) of cluster haloes, the surface mass density is $\sim 3-4$ times higher than LRGs, but $\sim 10$ times lower than ELGs, while at large distances ($D_{\rm proj}\gtrsim 2-3\,r_{500}$) the corresponding ratio is $\sim 2-3$, indicating a slower decline of \mgii surface mass density in the CGM of LRGs. On the other hand, at such distances the corresponding ratio between ELGs and cluster is $\lesssim 2-3$, indicating a much faster decline of \mgii surface mass density in the CGM of ELGs.

In summary, our analysis shows that the DESI clusters have higher (lower) \mgii surface mass densities out
to $D_{\rm proj}\lesssim 10\,r_{\rm 500}$ relative to SDSS LRGs (ELGs), while all three converge at large distances ($D_{\rm proj}\gtrsim 15\,r_{\rm 500}$, $\mean{\sum\nolimits_{\rm \mgii}}\lesssim 0.01-0.02\, \rm M_{\odot}\, kpc^{-2}$). It is worth pointing out that $r_{\rm 500}$ of clusters are $2-3$ ($5-6$) times larger than LRGs (ELGs). 

\subsection{Characterizing Scale Dependence of \mgii Properties}\label{best_fit}

\begin{table*}
\centering
\caption{Best-fitting parameters for weak and strong \mgii absorber covering fractions ($f_{\rm c}$), mean equivalent widths of \mgii absorbers ($\mean{EW_{\rm 2796}}$) and \mgii surface mass density ($\mean{\sum\nolimits_{\rm \mgii}}$) around DESI legacy clusters, SDSS DR16 ELGs and SDSS DR16 LRGs \citep{anand21}. Note that, we show the best-fitting parameters for both $|\Delta z =z_{\rm BCG}-z_{\rm \mgii}|\leq0.01, \, 0.003$ choices for DESI legacy clusters.}
\begin{tabular}{||cccccccc||}
\hline
\mgii absorbers  & $|\Delta z|$                                          & Quantity &$x$            & $f_{\rm a}$ & $x_{\rm o}$ & $f_{\rm b}$ & $\alpha$ \\ \hline
\multirow{2}{*}{\textbf{Weak}: $0.4\,\angstrom\,<\,EW_{\rm 2796}\,<1\,\angstrom$} &\multirow{2}{*} {$0.01$}& \multirow{2}{*} {$f_{\rm c}$} & $D_{\rm proj}$ &   $0.0087\pm0.0017$ &    $980\pm421$ kpc &   $0.0099\pm0.0049$ & $-0.25\pm0.04$     \\
& &&$D_{\rm proj}/r_{\rm 500}$ & $0.01\pm0.004$ &    $0.67\pm0.34$ &   $0.027\pm0.015$ & $-0.26\pm0.13$ \\ \\

\multirow{2}{*}{\textbf{Strong}: $EW_{2796}\,>\, 1\,\angstrom$} & \multirow{2}{*} {$0.01$}&\multirow{2}{*}{$f_{\rm c}$} &  $D_{\rm proj}$         &   $0.0057\pm0.0006$ &    $395\pm112$ kpc &   $0.0157\pm0.0067$ & $-0.22\pm0.018$  \\
 & &&$D_{\rm proj}/r_{\rm 500}$ &   $0.01\pm0.003$ &    $0.293\pm0.154$ &   $0.025\pm0.019$ & $-0.35\pm0.05$ \\ \\
 
 \multirow{2}{*}{\textbf{Weak}: $0.4\,\angstrom\,<\,EW_{\rm 2796}\,<1\,\angstrom$} &\multirow{2}{*} {$0.003$}& \multirow{2}{*} {$f_{\rm c}$} & $D_{\rm proj}$ &   $0.006\pm0.0026$ &    $1165\pm820$ kpc &   $0.0028\pm0.023$ & $-0.54\pm0.048$     \\
& &&$D_{\rm proj}/r_{\rm 500}$ & $0.007\pm0.003$ &    $1.21\pm0.78$ &   $0.0062\pm0.004$ & $-0.523\pm0.08$ \\ \\

 \multirow{2}{*}{\textbf{Strong}: $EW_{2796}\,>\, 1\,\angstrom$} & \multirow{2}{*} {$0.003$}&\multirow{2}{*}{$f_{\rm c}$} &  $D_{\rm proj}$         &   $0.003\pm0.0007$ &    $765\pm270$ kpc &   $0.0046\pm0.002$ & $-0.42\pm0.037$  \\
 & &&$D_{\rm proj}/r_{\rm 500}$ &   $0.007\pm0.003$ &    $0.393\pm0.257$ &   $0.0094\pm0.009$ & $-0.54\pm0.04$ \\ \\ \hline

\multirow{2}{*}{\mgii absorbers (DESI):} &\multirow{2}{*} {$0.01$}& \multirow{2}{*} {$\mean{EW_{2796}}$} &$D_{\rm proj}/r_{\rm 500}$ & $0.023\pm0.0036$ &    $0.55\pm0.18$ &   $0.06\pm0.026$ & $-0.264\pm0.03$ \\  \\

 \multirow{2}{*}{\mgii absorbers (DESI):} &\multirow{2}{*} {$0.003$}& \multirow{2}{*} {$\mean{EW_{2796}}$} &$D_{\rm proj}/r_{\rm 500}$ & $0.023\pm0.002$ &    $0.49\pm0.12$ &   $0.075\pm0.028$ & $-0.262\pm0.022$ \\  \\
 
 \multirow{2}{*}{\mgii absorbers (LRGs):} &\multirow{2}{*} {$0.01$}& \multirow{2}{*} {$\mean{EW_{2796}}$} &$D_{\rm proj}/r_{\rm 500}$ & $0.010\pm0.0002$ &    $0.69\pm0.06$ &   $0.019\pm0.002$ & $-0.09\pm0.005$
 \\ \\
 \multirow{2}{*}{\mgii absorbers (ELGs):} &\multirow{2}{*} {$0.01$}& \multirow{2}{*} {$\mean{EW_{2796}}$} &$D_{\rm proj}/r_{\rm 500}$ & $0.097\pm0.0287$ &    $0.23\pm0.070$ &   $2.12\pm1.32$ & $-0.46\pm0.053$ \\ \\ 
 \hline

 \multirow{2}{*}{\mgii absorbers (DESI):} &\multirow{2}{*} {$0.01$}& \multirow{2}{*} {$\mean{\sum\nolimits_{\rm \mgii}}$} &$D_{\rm proj}/r_{\rm 500}$ & $0.106\pm0.017$ &    $0.55\pm0.18$ &   $0.26\pm0.12$ & $-0.264\pm0.03$ \\ \\
 
 \multirow{2}{*}{\mgii absorbers (DESI):} &\multirow{2}{*} {$0.003$}& \multirow{2}{*} {$\mean{\sum\nolimits_{\rm \mgii}}$} &$D_{\rm proj}/r_{\rm 500}$ & $0.107\pm0.013$ &    $0.49\pm0.13$ &   $0.34\pm0.13$ & $-0.262\pm0.022$ \\ \\ 
 
\multirow{2}{*}{\mgii absorbers (LRGs):} &\multirow{2}{*} {$0.01$}& \multirow{2}{*} {$\mean{\sum\nolimits_{\rm \mgii}}$} &$D_{\rm proj}/r_{\rm 500}$ & $0.048\pm0.001$ &    $0.69\pm0.06$ &   $0.09\pm0.01$ & $-0.09\pm0.005$ \\ \\

 \multirow{2}{*}{\mgii absorbers (ELGs):} &\multirow{2}{*} {$0.01$}& \multirow{2}{*} {$\mean{\sum\nolimits_{\rm \mgii}}$} &$D_{\rm proj}/r_{\rm 500}$ & $0.44\pm0.12$ &    $0.23\pm0.070$ &   $9.84\pm4.71$ & $-0.46\pm0.053$ \\ \\
\hline
\end{tabular}
\label{tab:params_fc}
\end{table*}

\subsubsection{Covering Fractions}\label{fit_fc}

As visible by eye, the scale-dependence of the \mgii covering fraction, $\mean{EW_{\rm 2796}}$ and the surface mass density cannot be described by a simple power-law function. We therefore fit the observed distributions with a superposition of power-law and exponential functions to account for the steep radial dependence of
\mgii absorption in the inner regions and the flat behaviour at large distances. Specifically, we fit the following function:

\begin{equation}
    f(x) = f_{\rm a}\,  \Big(\frac{x}{x_{\rm o}}\Big)^{\alpha}\,+\, f_{\rm b}\, e^{-(x/x_{\rm o})}
    \label{eqn:fitting}
\end{equation}

\noindent where $x = D_{\rm proj}/r_{\rm 500} \text{ or }D_{\rm proj}$. The function has four free parameters namely $f_{\rm a},\, f_{\rm b}, \, x_{\rm o}$ and $\alpha$. The $x_{\rm o}$ characterizes the scale where the slope (defined by $\alpha$) changes and both functions are comparable and contribute significantly to the observed distributions. Note that similar functional form (including redshift dependence, with 5-6 parameters) has been used to characterize \mgii and \ovi absorber properties around passive and star-forming galaxies from the DESI legacy imaging survey \citep{lan20} and COS-Halos and COS-Dwarfs surveys \citep{tchernyshyov21}.

We fit the stacked profile of each sample with a standard Levenberg-Marquardt minimization, and show the best fitting parameters and their uncertainties (measured with bootstrap approach) in Table~\ref{tab:params_fc}. The best-fitting curves are shown by solid lines in corresponding figures (see Figures~\ref{fig:diff_fc}, ~\ref{fig:surf_proj}).

We observe that the characteristic scale is smaller, $x_{\rm o}\approx 400$ kpc and $x_{\rm o}\approx 0.3$ (normalized) for strong absorbers compared to $x_{\rm o}\approx 1$ Mpc  and $x_{\rm o}\approx 0.7$ (normalized) for weak absorbers. Furthermore, we note that the covering fraction declines faster (characterized by slope of power law, $\alpha$) on normalized scales for strong absorbers ($\alpha\approx -0.35$) than weak systems ($\alpha\approx -0.26$). This implies that strong absorbers are destroyed rapidly in the inner part of the halo, while weak absorbers survive up to large distances.  

We also compile the best-fitting parameters of \mgii covering fractions corresponding to $|\Delta z|\leq 0.003$ choice in Table~\ref{tab:params_fc}. There is one trend that is of particular interest: the characteristic scales ($x_{\rm o}$) are slightly larger (albeit large error bars) for both weak and strong absorbers on both projected and normalized scales. On the other hand, slopes ($\alpha$) are significantly larger ($\sim2$ times). It implies that covering fraction declines more rapidly than $|\Delta z|\leq 0.01$ for both strong and weak absorbers, and the \mgii scales shift to larger values. It is reassuring that the overall trend and conclusions (as described above) remain the same.

\subsubsection{\mgii Mass in Clusters and Galaxies}\label{mgii_mass}

We also fit the mean $\mean{EW_{2796}}$ and \mgii surface mass density profiles (solid lines in Figure~\ref{fig:surf_proj}) in clusters, SDSS ELGs and SDSS LRGs with the same fitting function (Eqn.~\ref{eqn:fitting}). The best-fitting parameters with their errors are also reported in Table~\ref{tab:params_fc}. We see that radial profiles of both quantities decline more steeply in DESI clusters (slope $\alpha\approx -0.26$, same for $|\Delta z|\leq 0.003$ case, see Table~\ref{tab:params_fc}), than SDSS LRGs ($\alpha\approx-0.1$), while the decline is much faster in SDSS ELGs ($\alpha\approx-0.45$). \mgii absorbers in cluster environments seem to be concentrated in the inner parts of halo ($D_{\rm proj}\lesssim r_{\rm 500}$). At large distances from the cluster centre ($D_{\rm proj}\gtrsim 2-3r_{\rm 500}$), the probability of detecting absorbers is much lower. The \ewMg in the inner haloes of clusters is also higher than in SDSS LRGs, but much lower than SDSS ELGs. 

To estimate the total mass of \mgii gas within $r_{\rm 500}$ of the cluster halo, SDSS ELGs and SDSS LRGs, we integrate the surface mass density profile up to $D_{\rm proj}\lesssim r_{\rm 500}$ in all haloes. 

\begin{equation}
    M_{\rm \mgii}\, (x<1) = 2\pi r_{\rm 500}^{2}\int_{0}^{1}\Big\{f_{\rm a}\Big(\frac{x}{x_{\rm o}}\Big)^{\alpha}+f_{\rm b}e^{-x/x_{\rm o}}\Big\}x\,dx
    \label{eqn:surf_mass_fit}
\end{equation}

\noindent where $x = D_{\rm proj}/r_{\rm 500}$ and $r_{\rm 500}$ is in kpc. The best-fitting values for $f_{\rm a},\, f_{\rm b}, \, x_{\rm o}$ and $\alpha$ are compiled in Table~\ref{tab:params_fc}.

Using this approach, we find that the total \mgii mass within $r_{\rm 500}$ (see Table~\ref{tab:gal_comp}) of the clusters is $\rm M_{\mgii,\, cluster}(D_{proj}<r_{500}) \approx (3.26\pm1.58)\times 10^{5}\, M_{\odot}$ ($(3.43\pm1.44)\times 10^{5}\, \rm M_{\odot}$, for $|\Delta z|\leq 0.003$ case), while for LRGs within their $r_{\rm 500}$ it is $\rm M_{\mgii,\, LRG}(D_{proj}<r_{500})\approx (3.68\pm0.34)\times 10^{4}\, M_{\odot}$. This implies that $\rm M_{\mgii,\, LRG}/M_{\mgii,\, cluster}\approx 0.11\pm 0.05$ such that there is more \mgii gas ($\approx 10$ times) in cluster haloes. Similarly, the total \mgii mass within $r_{\rm 500}$ (see Table~\ref{tab:gal_comp}) of ELGs is $\rm M_{\mgii,\, ELG}(D_{proj}<r_{500}) \approx (6.1\pm4.5)\times 10^{4}\, M_{\odot}$, implying that $\rm M_{\mgii,\, ELG}/M_{\mgii,\, cluster}\approx 0.19\pm 0.16$ (albeit large error bar), i.e. clusters have $\sim 5$ times more \mgii gas than ELGs, even though both mean $\mean{EW_{\rm 2796}}$ and surface mass density are higher in ELGs. However, as noted above, our column densities are only the lower limits for ELGs, given that majority of \mgii absorbers in their halo are saturated. On the other hand, this also implies that $\rm M_{\mgii,\, ELG}/M_{\mgii,\, LRG}\approx 1.7\pm 1.2$ such that both have quite comparable (error bar is large) \mgii mass at virial scales. Similar trends were found for H\,\textsc{i} mass (traced by \mgii absorbers in the stacked spectra of quasars) around ELGs and LRGs \citep{lan18}. It is clear from above analysis that clusters host more \mgii gas in their haloes compared to LRGs, in part because they reside in larger haloes (see Table~\ref{tab:gal_comp}) and they trace overdense regions.

To test the robustness of the measurement, we also estimated the \mgii mass by simply summing the surface mass densities out to $D_{\rm proj}\lesssim r_{\rm 500}$ and the corresponding masses are $\rm M_{\mgii,\, cluster}(D_{proj}<r_{500})\approx (3.54\pm0.41)\times 10^{5}\, M_{\odot}$ ($(3.68\pm0.41)\times 10^{5}\, \rm M_{\odot}$, for $|\Delta z|\leq 0.003$ case) and $\rm M_{\mgii,\, LRG}(D_{proj}<r_{500})\approx (4.01\pm0.11)\times 10^{4}\, M_{\odot}$ and $\rm M_{\mgii,\, ELG}(D_{proj}<r_{500})\approx (6.2\pm1.1)\times 10^{4}\, M_{\odot}$ for clusters, LRGs and ELGs, respectively, consistent with the previous values.

Finally, we estimate the \mgii mass within $200\, \rm kpc$ of the central galaxy: for clusters we find that $\rm M_{\mgii,\, cluster}(D_{proj}<200\, kpc) \approx (4.23\pm1.70)\times 10^{4}\, M_{\odot}$ and for SDSS LRGs the $\rm M_{\mgii,\, LRG}(D_{proj}<200\, kpc) \approx (1.32\pm0.12)\times 10^{4}\, M_{\odot}$. This gives $\rm M_{\mgii,\, LRG}/M_{\mgii,\, cluster}\approx 0.31\pm 0.12$. Clearly, even on similar scales there is $\approx 3$ times more cool gas in clusters than SDSS LRGs. On the other hand, for ELGs, the corresponding mass is $\rm M_{\mgii,\, ERG}(D_{proj}<200\, kpc) \approx (8.3\pm1.4)\times 10^{4}\, M_{\odot}$, larger than $\rm M_{\mgii,\, ELG}(D_{proj}<r_{500})$, because $D_{\rm proj}=200\, \rm kpc$ is larger than the $r_{\rm 500}$ of ELGs (see Table~\ref{tab:gal_comp}), i.e. we are tracing larger area.

\subsection{Dependence on Cluster Properties}

\subsubsection{Stellar Mass of BCGs}

\begin{figure}
    \includegraphics[width=0.9\linewidth]{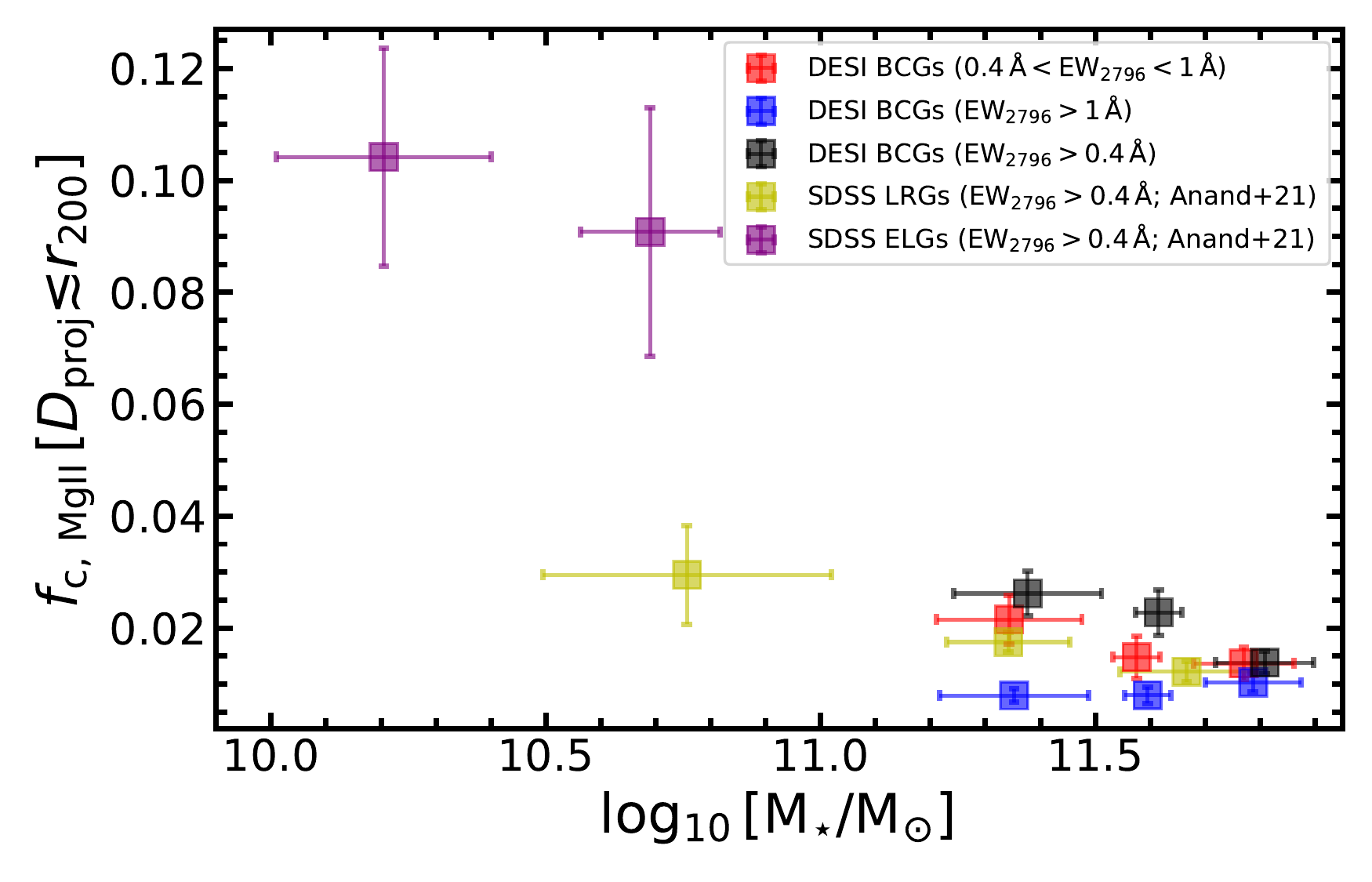}
    \caption{Total covering fraction of weak (red squares) and strong (blue squares) absorbers \mgii absorbers within the $r_{200}$ of cluster haloes, as a function of stellar mass of the BCG. We also show the comparison of DESI BCGs (black squares) with SDSS LRGs (yellow squares) and SDSS ELGs (purple square)} from \citet{anand21}. We clearly see that massive haloes have lower \mgii covering faction for both clusters and galaxies (ELGs and LRGs). Note that we have slightly offset horizontally the points for visual clarity.
    \label{fig:fc_sm_bcg}
\end{figure}

We now study the connection between cluster properties and \mgii properties. We begin with total covering fraction within the cluster halo, $f_{\rm c}\,(D_{\rm proj}\lesssim r_{200})$, as a function of the stellar mass of the cluster BCG. We show the correlations in Figure~\ref{fig:fc_sm_bcg}, where we see that weak absorbers (red squares) have higher covering fractions than strong absorbers (blue squares) at a given BCG stellar mass, implying that weak absorbers are more ubiquitous than strong ones. 

At lower stellar mass ($\rm M_{\star}\approx 10^{11.3}\, M_{\odot}$) the covering fraction of weak absorbers (shown in red) within $r_{\rm 200}$ is $\approx 2.5$ per cent compared to $\approx 1$ per cent for strong absorbers (shown in blue). However, for more massive BCGs ($\rm M_{\star}\approx 10^{11.8}\, M_{\odot}$) the covering fraction for both weak and strong systems are consistent within the error bars. That is, the decreasing trend of \mgii covering fraction as a function of BCG mass is clearly visible for weak systems, but not for strong systems. This suggests that weak and strong systems may have a different physical origin.
 
We also contrast our cluster measurements (\ewGr{0.4}, black squares) with SDSS LRGs (\ewGr{0.4}, yellow squares) and SDSS ELGs (\ewGr{0.4}, purple squares) from \citetalias{anand21}. We combine weak and strong absorbers for a consistent comparison. The boost in covering fraction for BCGs in clusters is approximately the same for the two stellar mass bins. SDSS ELGs trace comparatively smaller haloes and have $3-5$ times higher covering fraction than LRGs and clusters, though BCGs have $5-10$ times larger stellar mass than ELGs.

\subsubsection{\mgii absorbers connection with cluster galaxies?}\label{mgii_gals_connection}

\begin{figure}
	\includegraphics[width=0.9\linewidth]{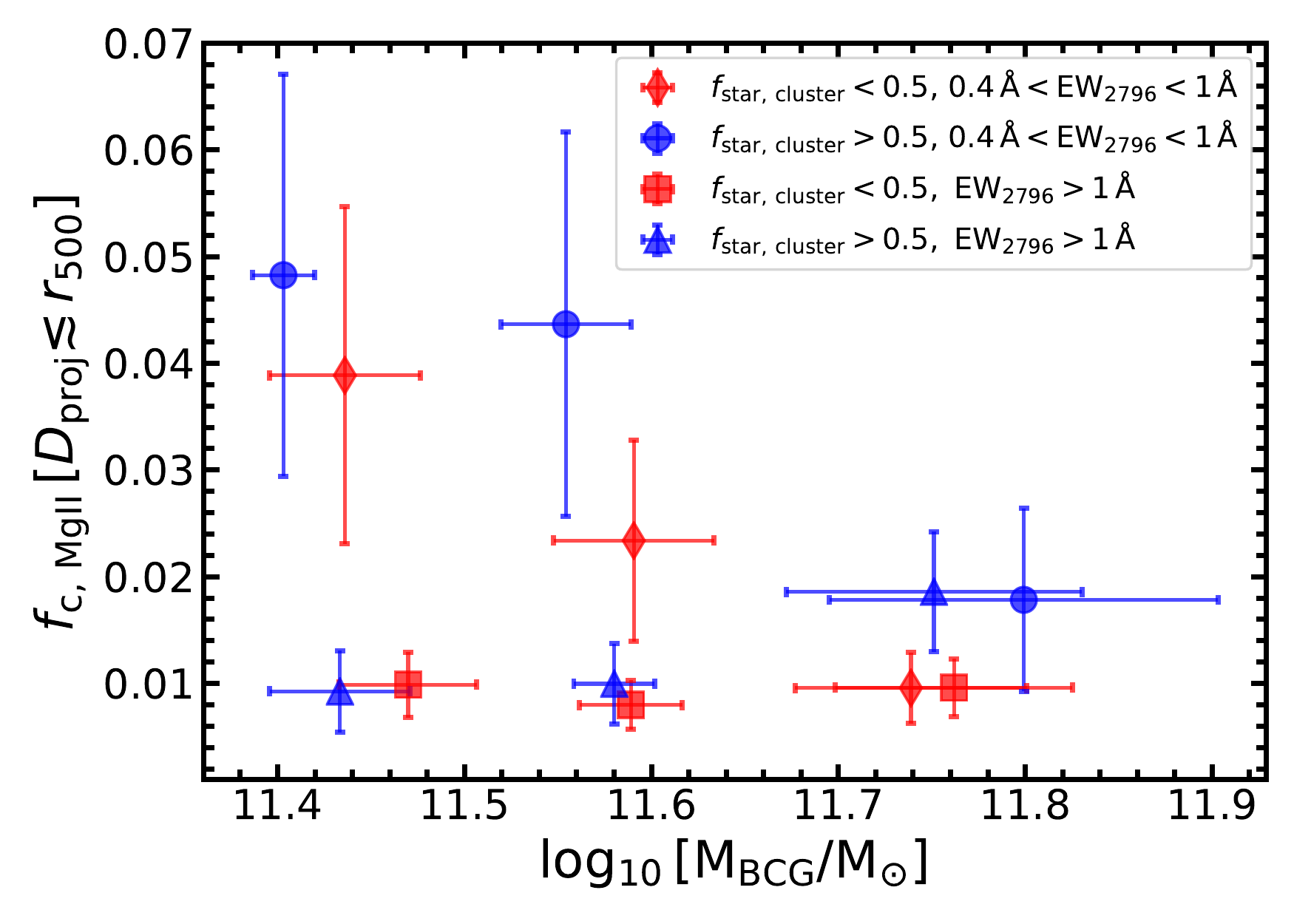}
    \caption{\mgii covering fraction of weak (blue circles and red diamond) and strong (blue triangles and red squares) as a function of fraction of star-forming galaxies in the cluster halo at a given stellar mass of the BCG. Blue shows the total covering fraction of \mgii absorbers in clusters that have more star-forming galaxies within their $r_{\rm 500}, \,\text{i.e.} f_{\rm star}>0.5$, while red shows the measurements for clusters having more passive galaxies inside their halo ($f_{\rm star}<0.5$). For weak absorbers, clusters with more blue galaxies in their haloes have slightly higher covering fraction at a given stellar mass of BCG. Note that we have slightly offset horizontally the points for visual clarity.}
    \label{fig:fstar_sm_bcg}
\end{figure}

Next, as described in Section~\ref{mgii_members}, we divide our clusters into two subsamples based on the fraction of star-forming galaxies $f_{\rm star}$ (>0.5 and <0.5) within $r_{\rm 500}$. For each subsample, we estimate the total covering fraction within $D_{\rm proj}\lesssim r_{\rm 500}$ as a function of the stellar mass of the BCG.

We show the results in Figure~\ref{fig:fstar_sm_bcg}, for clusters with more star-forming galaxies (in blue circles for weak absorbers, blue triangles for strong absorbers) and clusters with more passive galaxies (in red diamonds for weak absorbers and red squares for strong absorbers). The overall anti-correlation of covering fraction with the stellar mass of the BCG is once again visible (Figure~\ref{fig:fc_sm_bcg}). The covering fraction of weak absorbers is higher in clusters with more blue galaxies (compare blue circles and red diamonds), although the offset is small. This hints that some of the \mgii absorbers in the cluster environment could be connected to member galaxies with ongoing star formation. However, the same trend is not apparent for strong absorbers (compare blue triangles and red squares).

\begin{figure*}
	\includegraphics[width=0.45\linewidth]{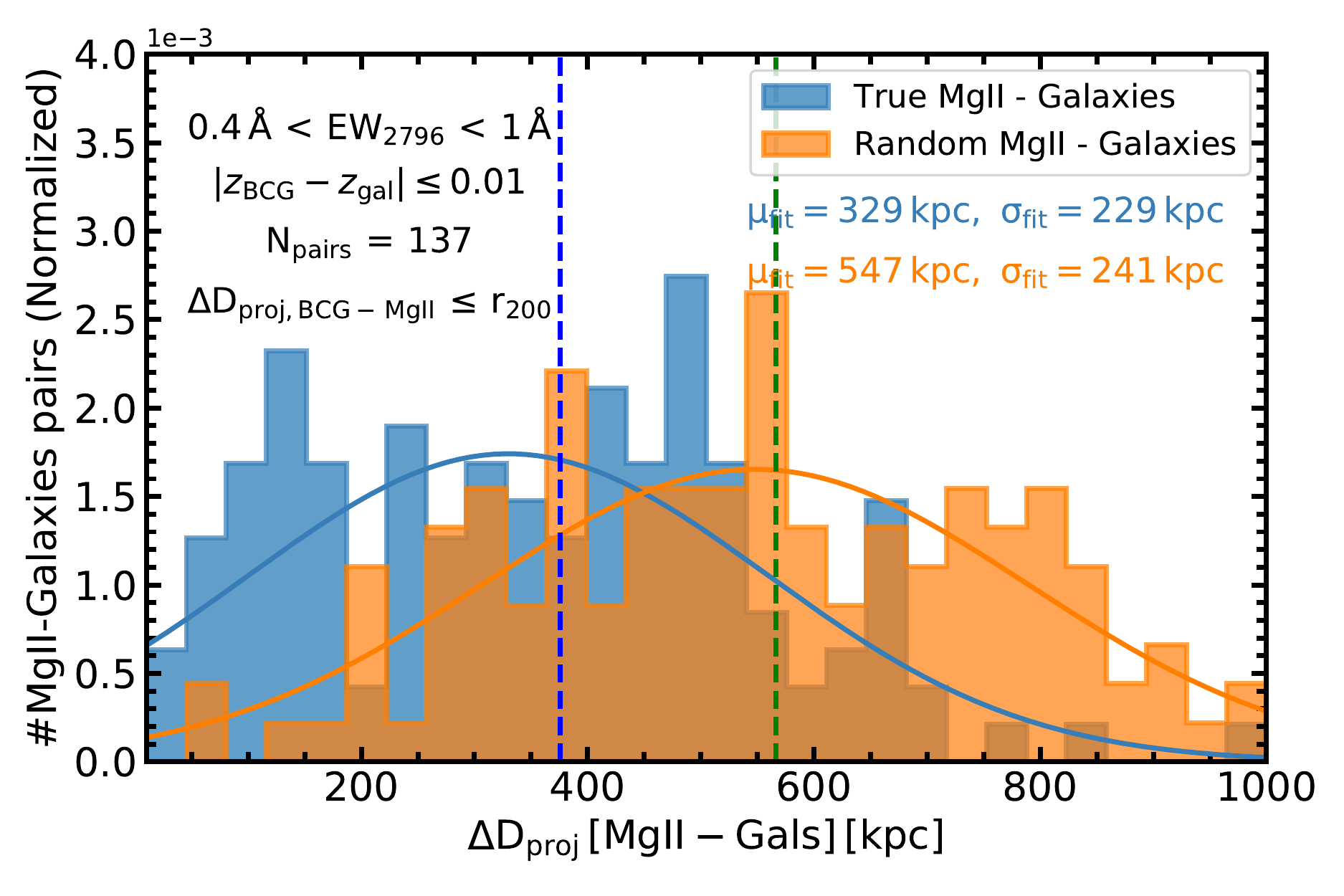}
	\includegraphics[width=0.45\linewidth]{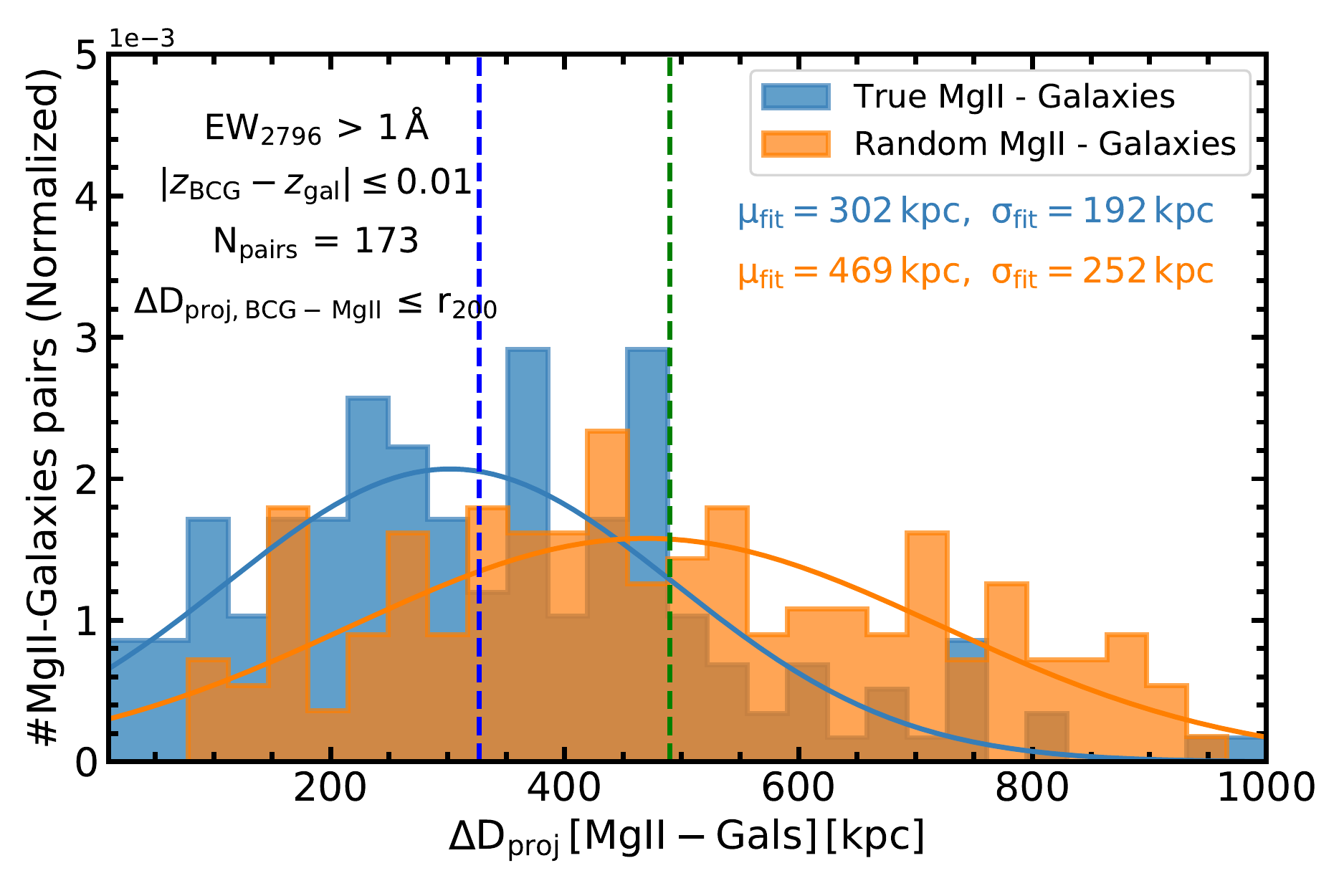}\\
	\includegraphics[width=0.45\linewidth]{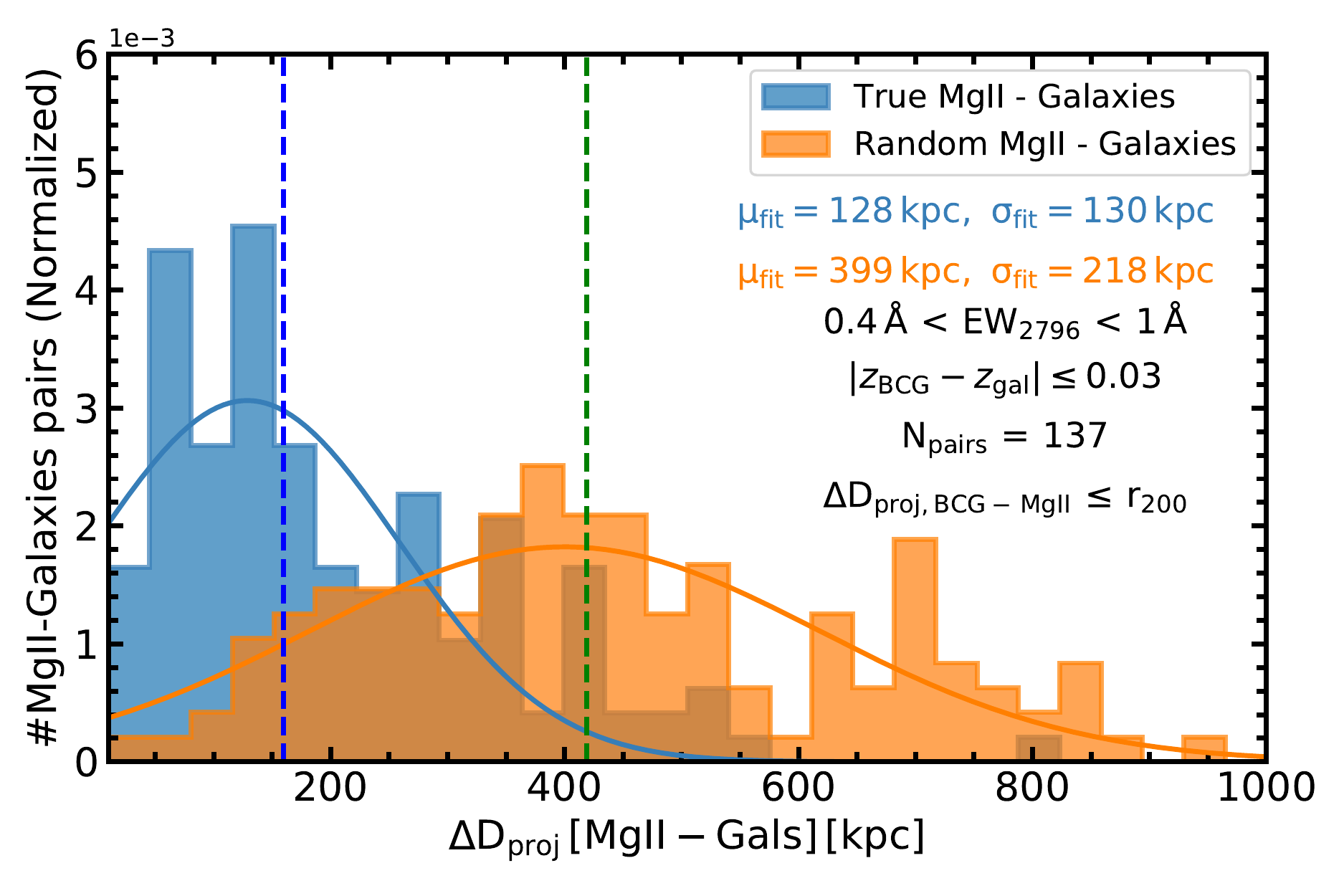}
	\includegraphics[width=0.45\linewidth]{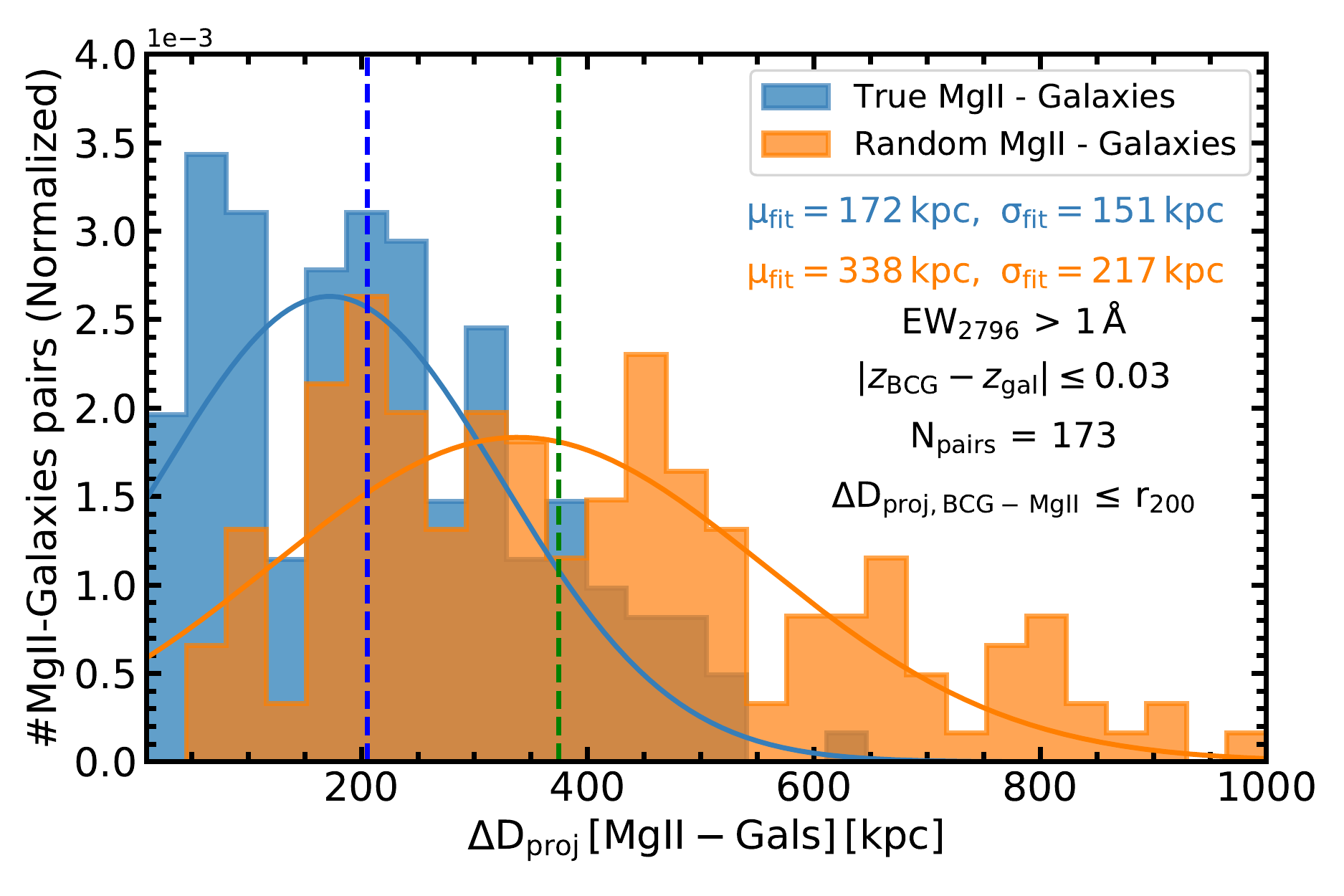}\\
    \caption{1D distribution of projected separation between \mgii absorbers and its nearest member galaxy in clusters. We only take those absorbers that lie within $D_{\rm proj}\lesssim r_{\rm 200}$ of cluster halo. The blue shows the true distribution of absorber-galaxy pair separation and orange shows the distribution when absorbers are distributed randomly in the halo, only in angles, keeping their distances to the BCG fixed. We show the results for two $|\Delta z| = |z_{\rm BCG} - z_{\rm gal}|$ values, namely, a strict cut $|\Delta z|\leq0.01$ (top row) as well as a more relaxed criterion $|\Delta z|\leq0.03$ (bottom row). In each row: \textbf{Left:} Distribution of projected distance between weak absorbers and their nearest cluster galaxies. \textbf{Right:} Same as left panel, for strong absorbers. The corresponding solid blue and green lines represent the best fitting gaussians to the distributions. The dashed vertical lines show the corresponding median values of the distributions.}
    \label{fig:mgii_desi}
\end{figure*}

To understand the connection to satellite galaxies better, we locate the nearest member galaxy (in projected distance) to each absorber within $D_{\rm proj}\lesssim r_{\rm 200}$ of the cluster halo. If the \mgii absorbers are physically associated with the star-forming ISM or extended gas surrounding the member galaxies, the connection should be visible in the distribution of distances between \mgii - galaxy pairs. In this case, a significant fraction of absorbers would be at small distances from their host galaxies.

We show the histogram of absorber-nearest galaxy distance in Figure~\ref{fig:mgii_desi} for both weak (blue, left panels) and strong (blue, right panels) absorbers, for two $|\Delta z| = |z_{\rm BCG} - z_{\rm gal}|$ values namely, $|\Delta z|\leq0.01$ (top row) and a less stringent $|\Delta z|\leq0.03$ (bottom row). The distributions are roughly gaussian, and the resulting fit for the weak absorbers (shown in blue in the top left panel) is an average of $\rm \mu \approx 330\, kpc$ with $\rm \sigma \approx 230\, kpc$. The median separation is about $\approx 350$ kpc. Similarly, for strong absorbers, the distribution (shown in blue in the top right panel) can be described by a gaussian with $\rm \mu \approx 300\, kpc$ and $\sigma \rm \approx 200\, kpc$. The median separation is also about $\approx 325$ kpc.

When we increase the redshift separation ($|z_{\rm BCG} - z_{\rm gal}|$) to $|\Delta z|\leq0.03$, the mean and median separations further decrease (see bottom row). This is of order, or slightly larger than, the uncertainty on the photo$-z$ estimates. Even for this less stringent $|\Delta z|\leq0.03$, the separations between absorbers and the nearest galaxies are larger than the expected size of the star-forming disk and more comparable to the typical halo size ($r_{\rm vir}\approx 200$ kpc) of these galaxies \citep{kravtsov13}.

This suggests that the detected \mgii absorbers are not strongly associated with the ISM of the cluster galaxies. We note that faint, low-mass galaxies below the sensitivity of DESI would be present but not currently detected. It is likely that some \mgii absorbers are associated with the CGM of these fainter galaxies. The completeness of the DESI photo$-z$ galaxy sample is a strong function of redshift and stellar mass. For e.g. at redshift, $z\sim 0.6$, galaxies below $\rm M_{\star}\sim 10^{10.5}\rm M_{\odot}$ have low completeness in the catalogue \citep{zou19}.

\begin{figure*}
 	\includegraphics[width=0.32\linewidth]{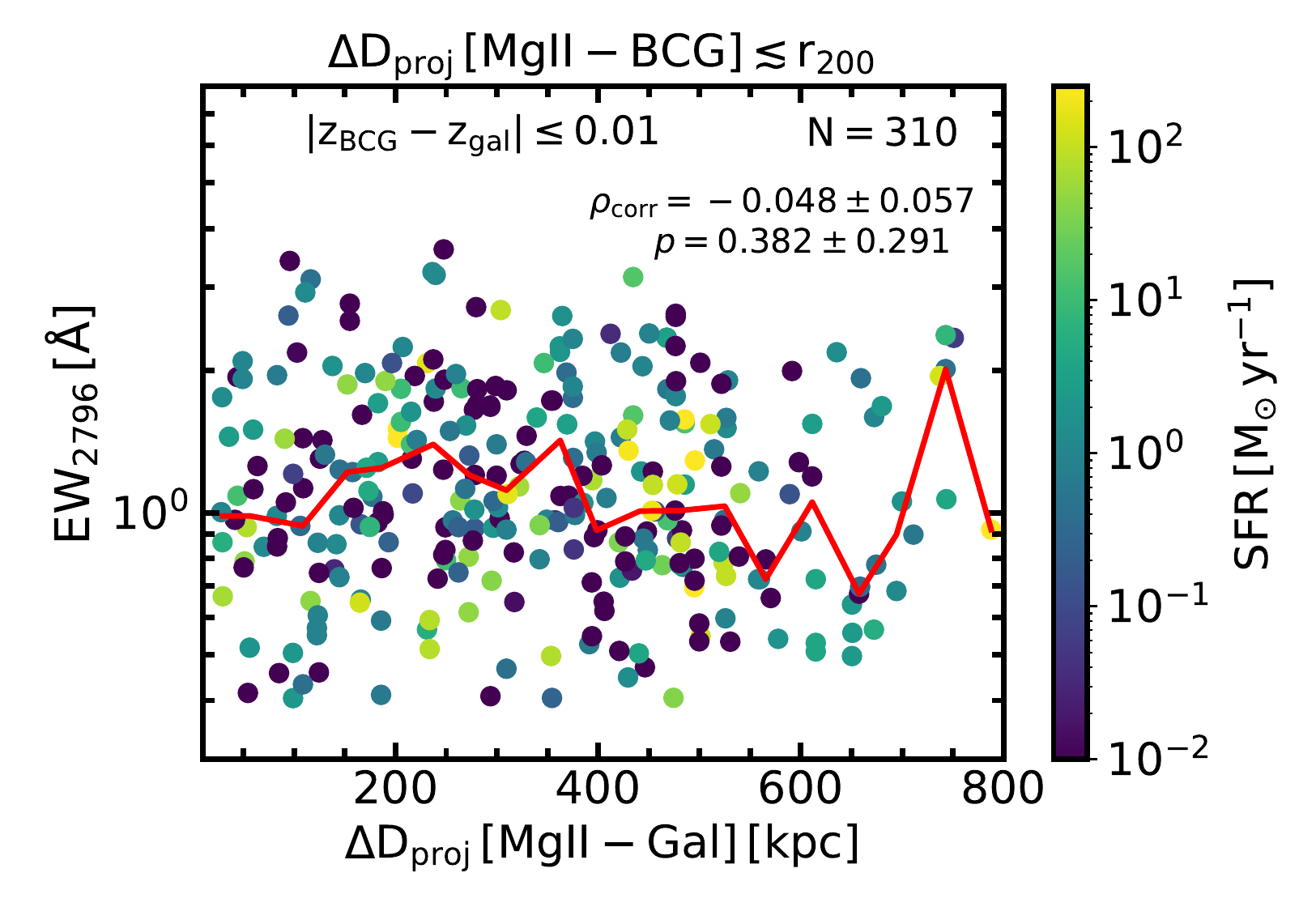}
	\includegraphics[width=0.32\linewidth]{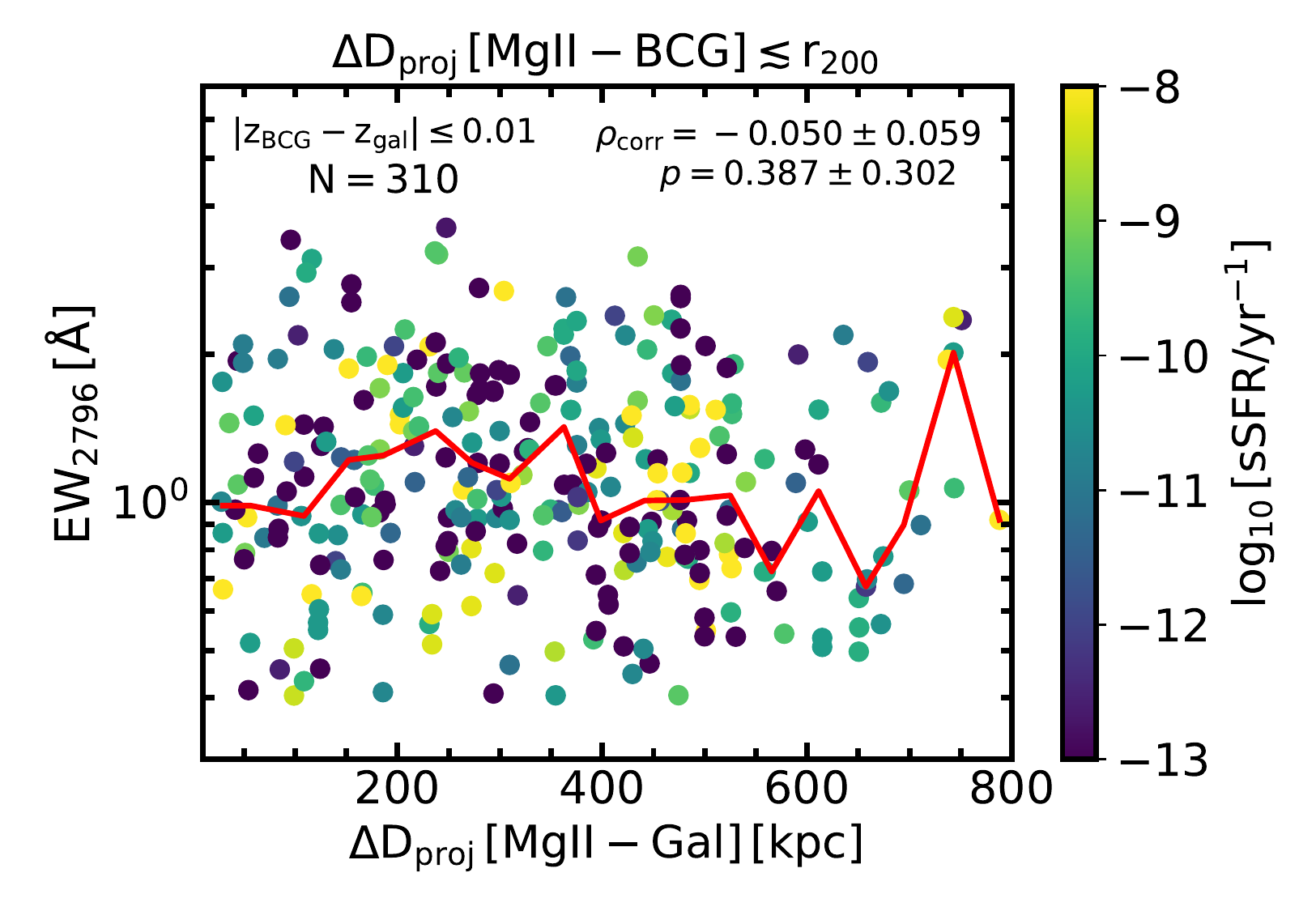}
	\includegraphics[width=0.32\linewidth]{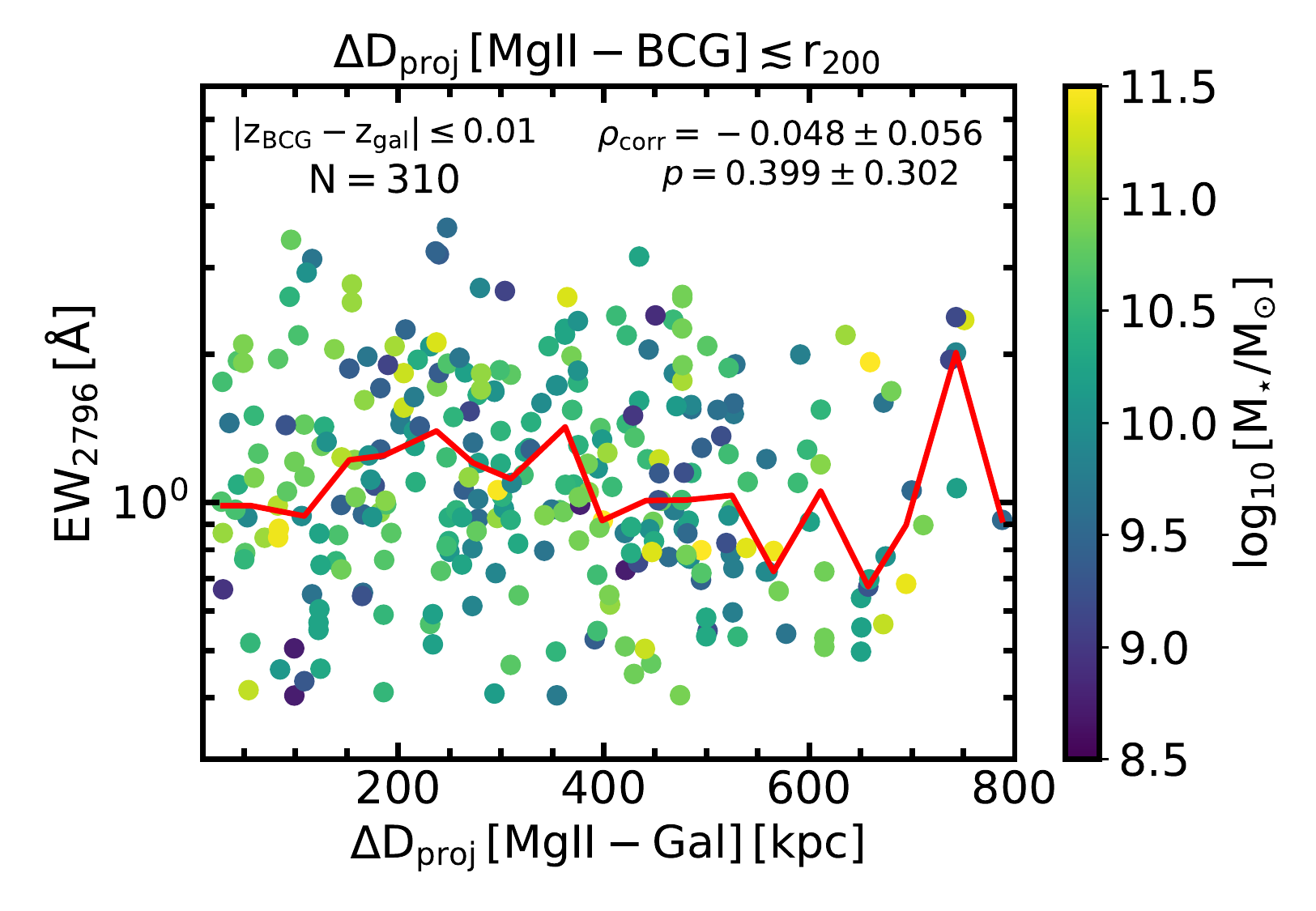}\\
	\includegraphics[width=0.32\linewidth]{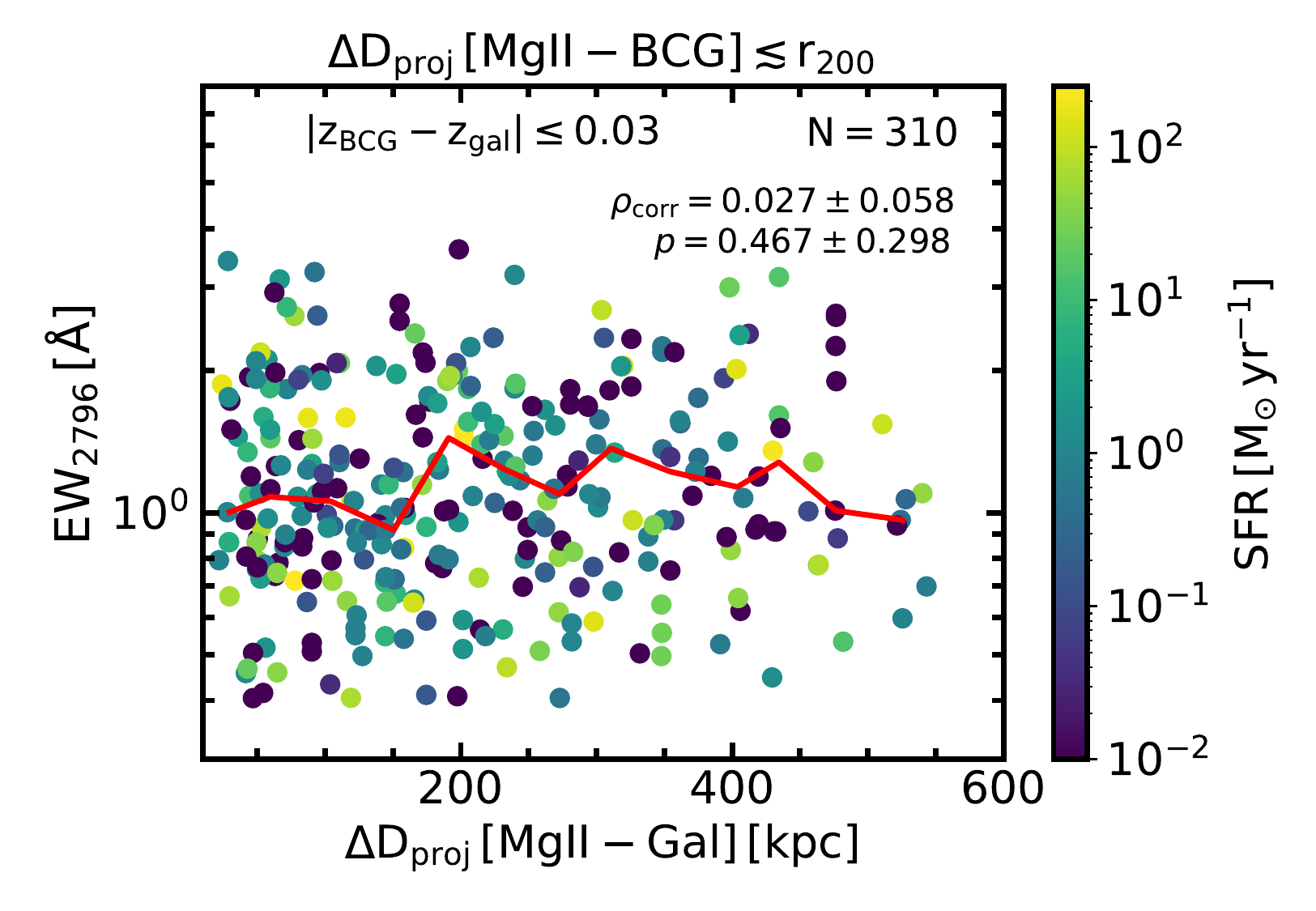}
	\includegraphics[width=0.32\linewidth]{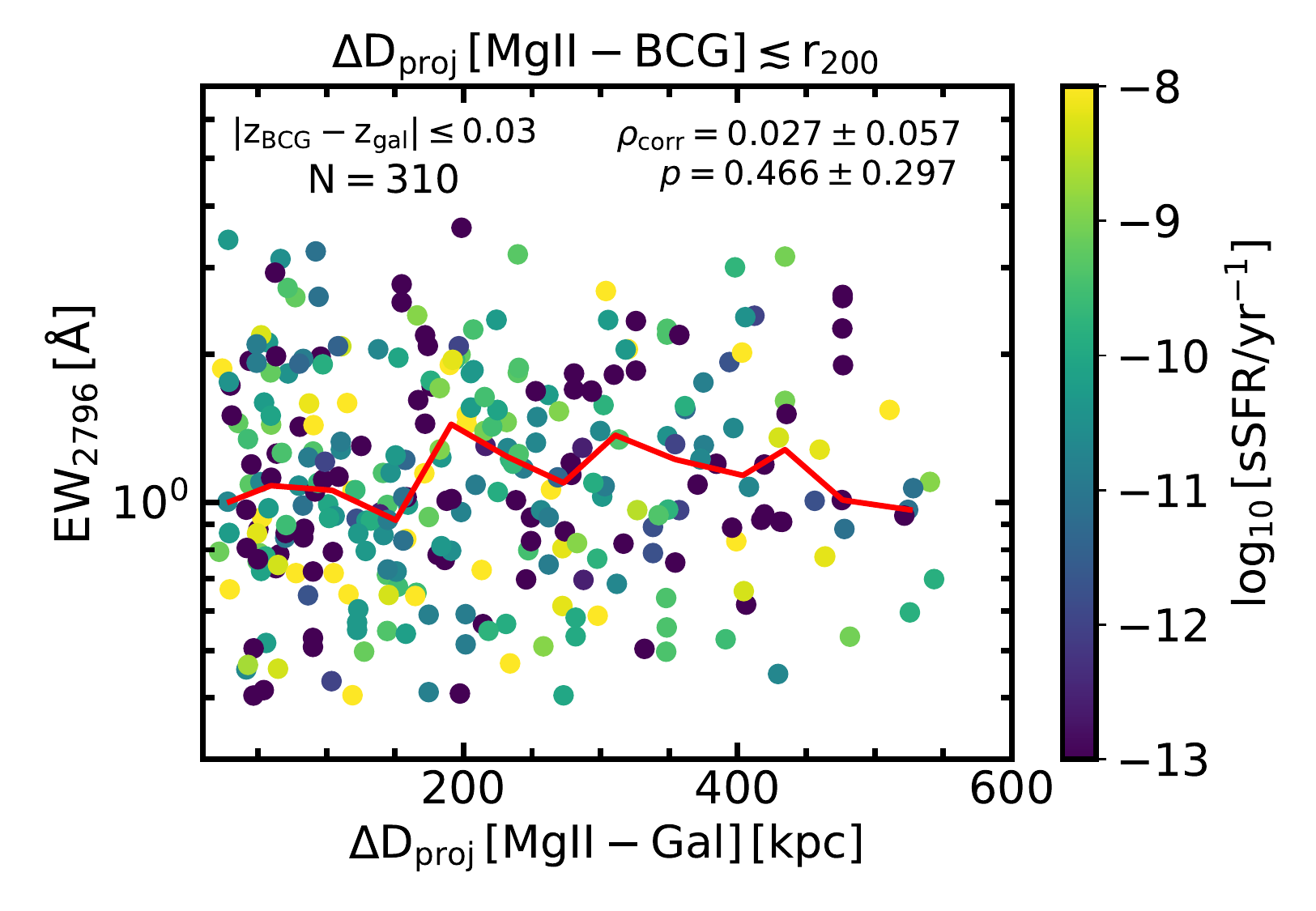}
	\includegraphics[width=0.32\linewidth]{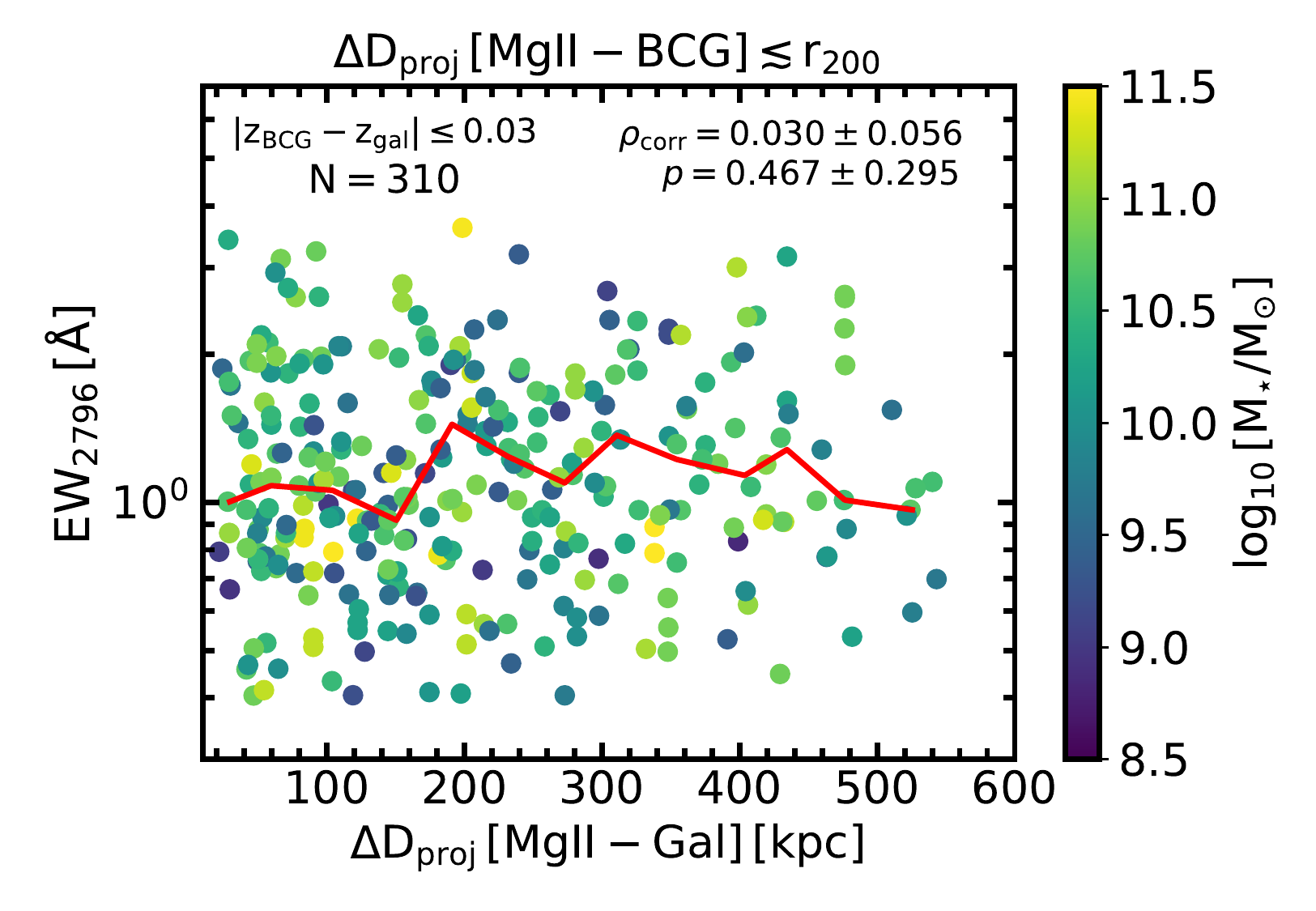}
     \caption{\ewMg of \mgii absorbers (within $r_{\rm 200}$ of the cluster) as a function of impact parameter ($D_{\rm proj}$ from the nearest cluster member). We show the results for a strict $|\Delta z| = |z_{\rm BCG} - z_{\rm gal}|$ value, namely, $|\Delta z|\leq0.01$ (top row), as well as for a more relaxed $|\Delta z|\leq0.03$ (bottom row). In each row: \textbf{Left:} Distribution is colored by the SFR of the nearest galaxy. \textbf{Middle:} Distribution colored by the specific SFR (sSFR) of the nearest galaxy. \textbf{Right:} Distribution colored by the stellar mass ($M_{\rm \star}$) of the nearest galaxy to the absorbers. Note that here, we have taken all the absorbers (\ewGr{0.4}). No significant correlation is visible between \mgii absorption and properties of its nearest galaxy. We also show the Pearson correlation coefficients and the corresponding $p$-values in each panel. The low correlation coefficient and high $p$-values indicate no significant correlation between quantities. The solid red line in each panel shows the median values. Note that, we do not show the error bars on \ewMg for visual clarity.}
     \label{fig:ew_mgii_desi}
 \end{figure*}

To understand the significance of a possible connection between absorbers and satellites we turn to our random catalogues, as described in Section~\ref{mgii_members}. We recall that the randoms place absorbers with the same radial distribution as real absorbers, but redistributed randomly in angle. 

The corresponding distributions for randomly distributed absorbers are shown in Figure~\ref{fig:mgii_desi} in orange, in all panels. A single gaussian function can also describe the random distributions, and in all cases the mean and widths of these distributions are larger than for the true catalogs. In the case of $|\Delta z| \leq 0.01$ (top panel of Figure~\ref{fig:mgii_desi}, orange), for weak absorbers, the best-fitting values, $\rm \mu\approx 550\, kpc, \, \sigma\approx 240\, kpc$, are $\approx 1.5$ higher than true distribution ($\rm \mu\approx 330\, kpc, $). The result is qualitatively similar for the strong absorber case, as well as for larger redshift intervals -- in all cases a statistically significant difference is present between the random mocks and the true data.

To summarize, we find that in cluster environments the \mgii absorbers are preferentially found near cluster galaxies, as opposed to being randomly distributed throughout the halo. Galaxies and absorbers are clearly clustered together to some degree. Because the typical separations are larger than the sizes of the actual galaxy disks, we conclude that the ISM of cluster galaxies (above the detected mass threshold) is not directly responsible for the observed \mgii absorption in clusters. In contrast, the typical separations suggest that  \mgii absorption arises at least in part from cool gas that has been removed from the ISM due to e.g. ram-pressure stripping. Our findings are consistent with a picture where the gas is no longer within the disk, but may be still gravitationally bound to, or recently stripped from, the dark matter subhaloes hosting member galaxies.

\subsubsection{\mgii Equivalent Width vs Cluster Galaxy Properties}\label{ew_vs_gals}

We now study the dependence of \mgii absorption (within $r_{\rm 200}$ of the cluster) on properties of the nearest galaxy, such as its SFR and stellar mass. In Figure~\ref{fig:ew_mgii_desi} we show \ewMg as a function of the projected distance between the absorber and the nearest cluster galaxy. We do not see a clear correlation between \mgii absorption and distance between the absorber-galaxy pair. The equivalent widths have significant scatter at fixed distance, and the median line (red) is largely consistent with being flat. This is true regardless of whether the redshift difference ($\Delta z$) criterion for association is strict (top row) or more relaxed (bottom row).

Previous studies of field galaxies \citep{nielsen13, chen17, lan18, anand21} found that \mgii equivalent width decreases with the impact parameter, though the scatter is large \citep[see][]{dutta21}. We do not see such an anti-correlation here, in fact the Pearson correlation coefficient is consistent with zero ($\rho_{\rm corr}=-0.048\pm0.057$) and the $p$-value is large ($p = 0.4$), clearly indicating that there is no significant correlation in \ewMg and distance between absorber-galaxy pairs in clusters. Even after increasing the $\Delta z$ cuts, we do not see any significant correlation (e.g. for a three times larger $|\Delta z|\leq 0.03$). This suggests that the connection discussed previously is, at best, weak. We also looked at $EW_{\rm 2796}<0.4$ \angstrom\, (not shown here) absorbers, but the statistics are poor, with which no clear additional trends were seen and nothing robust can be said. We discuss more about the limitations and caveats to this analysis in section~\ref{origin}.

We further color the points in each panel according to a given property of the nearest galaxy: star formation rate (left column), specific star formation rate (middle column), and stellar mass (right column). We do not observe any strong dependence of \ewMg on these galactic properties (left: SFR, middle: sSFR and right: $\rm M_{\star}$), even for the closest pairs ($D_{\rm proj}\lesssim 200$ kpc). 

\begin{figure}
	\includegraphics[width=0.9\linewidth]{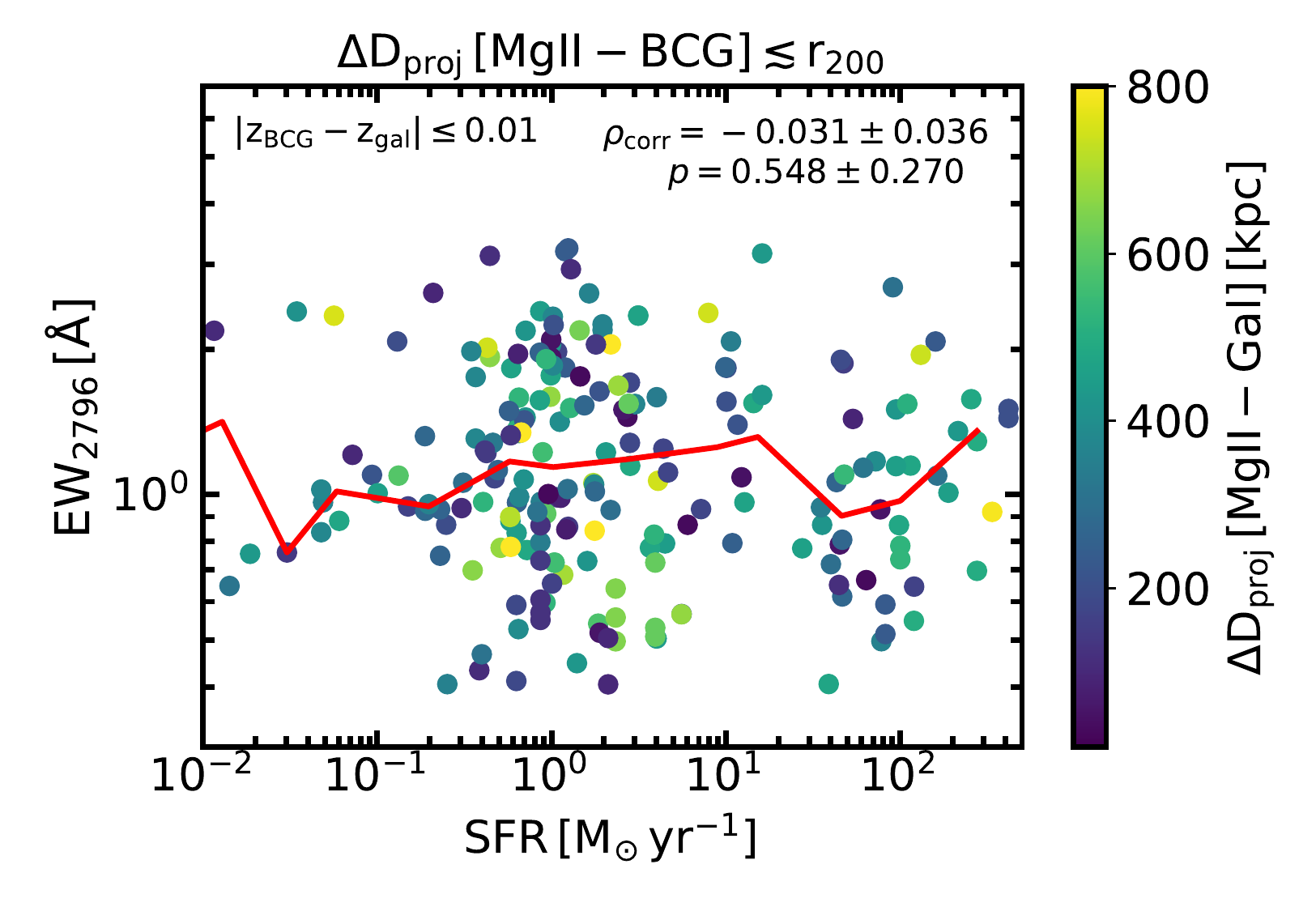}
	\includegraphics[width=0.9\linewidth]{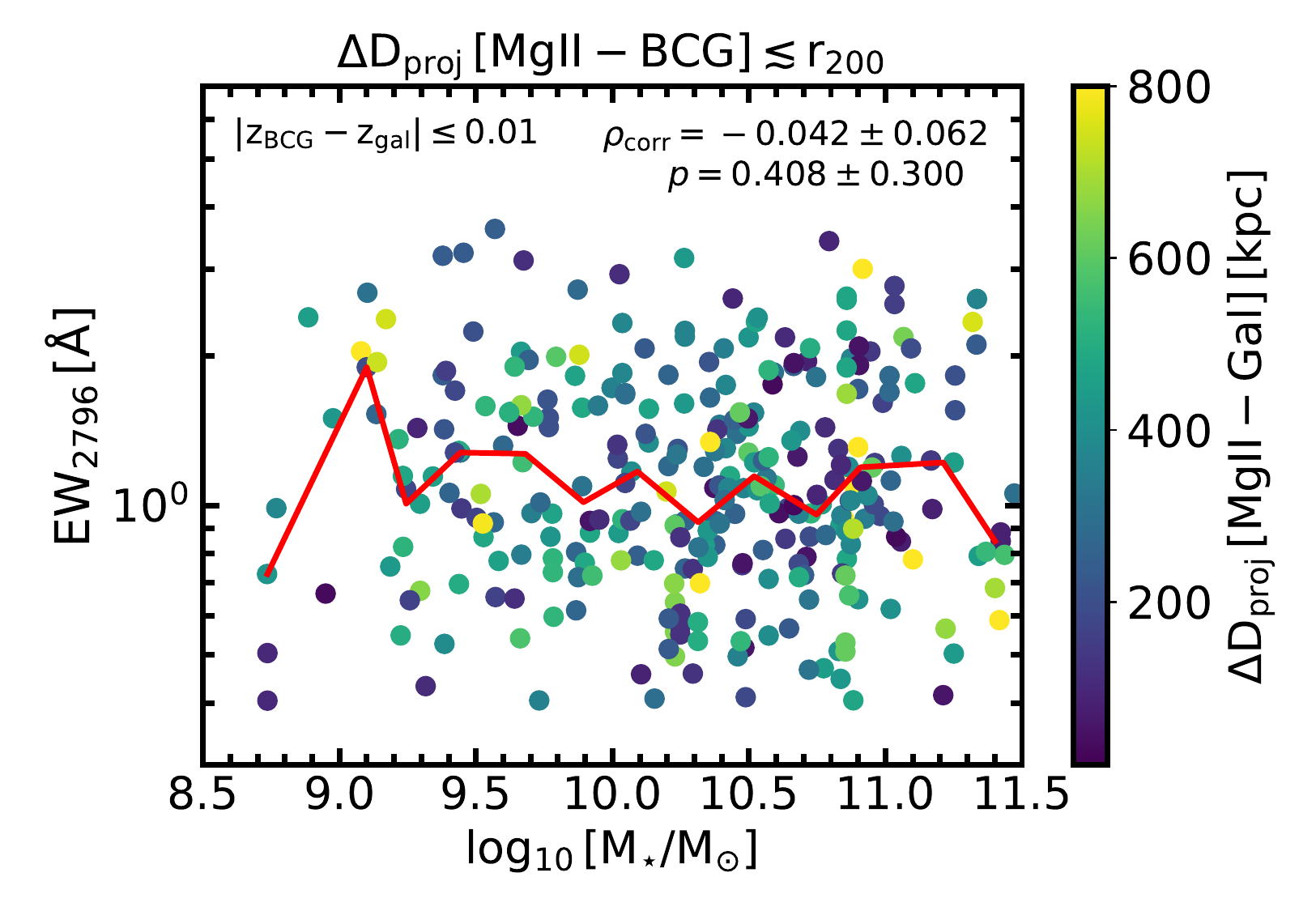}
     \caption{\ewMg of \mgii absorbers (within $r_{\rm 200}$ of the cluster) as a function of SFR (top) and stellar mass (bottom) of cluster galaxies. We show the results for $|\Delta z| = |z_{\rm BCG} - z_{\rm gal}| \leq 0.01$. Points are colored by the distance between absorber and the nearest cluster galaxy. Note that here, we have taken all the absorbers (\ewGr{0.4}). No significant correlation is visible between \mgii absorption and properties of its nearest galaxy -- Pearson correlation coefficients and the corresponding $p$-values are given in each panel. The solid red line in each panel shows the median values.}
     \label{fig:ew_mgii_sfr_desi}
 \end{figure}

To investigate further, we show \ewMg as a function of several properties of the nearest galaxy in Figure~\ref{fig:ew_mgii_sfr_desi}, for $|\Delta z| \leq 0.01$, as the results are similar for other choices. We show \ewMg as a function of SFR (top panel) and stellar mass (bottom panel) of galaxies. We do not observe any strong correlations, even for closest pairs in (see the colour bar, showing distance between absorber-galaxy pairs). The distribution is consistent with random and, the median line is roughly flat, disfavoring any strong connection between \mgii absorption and the SFR or stellar mass of the nearest cluster galaxies. 

Previous studies of field galaxies \citep{bordoloi11, rubin14, lan18, anand21} have shown that \mgii equivalent widths and covering fractions correlate positively with the SFR of the host galaxies, particularly for emission line or star-forming galaxies. This correlation is attributed to the strong outflows due to the stellar activity of the galaxy like stellar winds or supernovae driven winds \citep{bordoloi11, rubin14, tumlinson17}. In the case of quiescent or passive luminous red galaxies, \mgii properties are found to correlate with stellar and halo masses \citep{nielsen15, anand21}. As a result, it appears that \mgii absorbers in clusters are different to \mgii absorbers in the field, in that there is little connection between \ewMg and the SFR or mass of the nearest galaxy. If the \mgii originated from satellite ISM gas, then it is likely that there has been enough time for the properties of the \mgii clouds to have evolved so that the patterns observed in field galaxies are washed out.

\subsection{Relative Kinematics of \mgii absorbers in galaxy clusters}\label{los_dv}

We now turn to the motion of \mgii absorbing gas. Using the BCGs which have robust spectroscopic redshifts, we estimate the line-of-sight (LOS) velocity difference between \mgii absorbers and cluster BCG as, $\Delta v=c \Delta z/(1+z)$, where $z$ is the redshift of the cluster BCG. The distribution of LOS velocity separations allows us to constrain the motion of absorbing clouds in cluster environments.

\begin{figure}
	\includegraphics[width=0.9\linewidth]{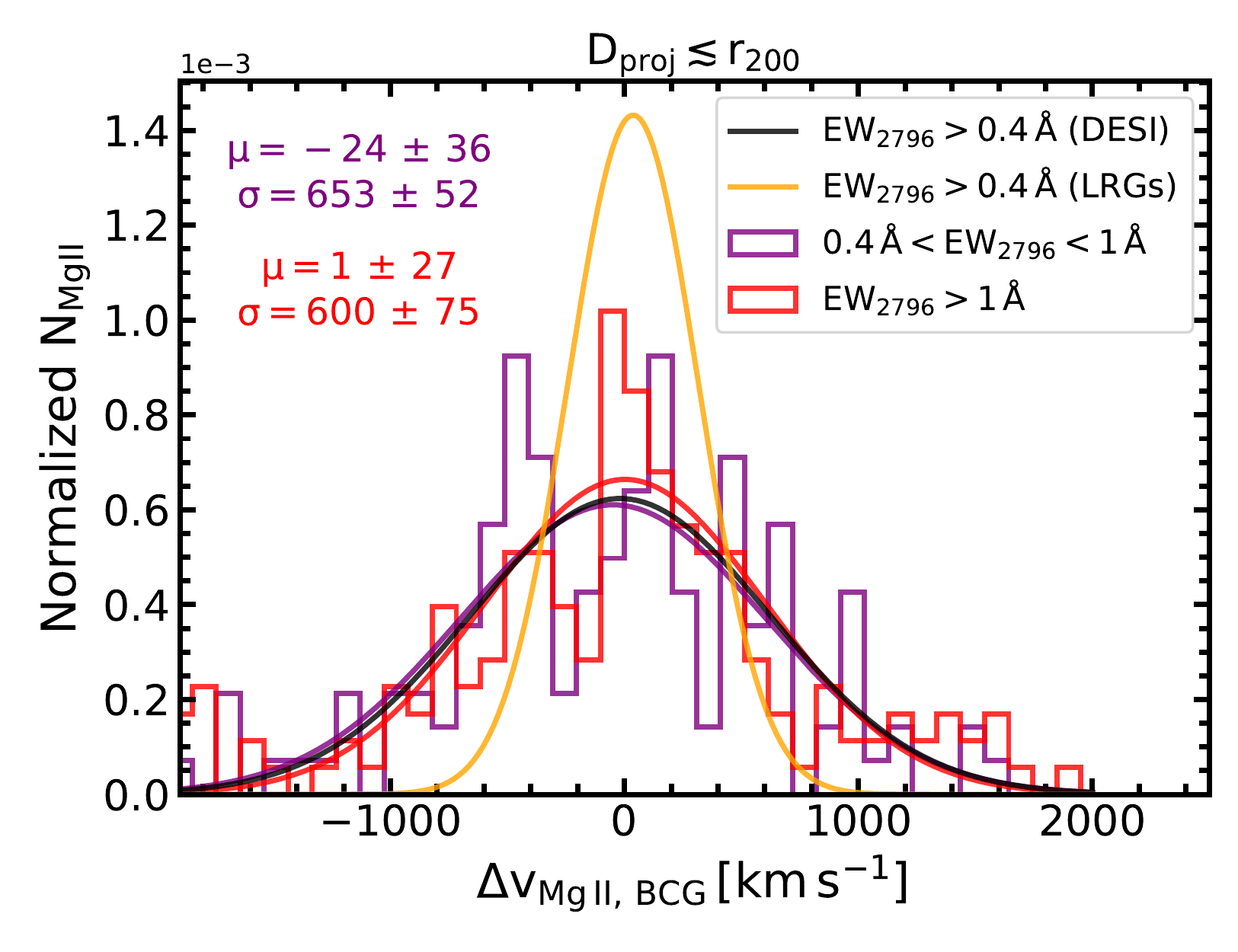}
    \caption{Distribution of LOS velocity difference between cluster BCG and \mgii absorbers pairs within $D_{\rm proj}<r_{\rm 200}$ of the clusters. The purple and red curves show distributions for weak and strong absorbers, respectively. The black curve shows the distribution for all absorbers (weak and strong). We also show the \mgii LOS velocity around LRGs in orange from \citet{anand21}, which is much narrower. The distributions are characterized by single gaussian profiles shown in solid color lines. Best fit parameters for the weak and strong cases are shown in each panel (see text).}
    \label{fig:dv_bcg_mgii}
\end{figure}

Figure~\ref{fig:dv_bcg_mgii} shows results for both weak (blue) and strong (red) absorbers. We see that for weak absorbers, the distribution can be characterized by a single gaussian with mean $\langle\Delta v\rangle\approx -24(\pm36)$ \kms and dispersion, $\sigma_{v}\approx 650$ \kms (shown in the solid purple line). There is weak evidence for an additional peak around $-500$ \kms, which may signpost an inflow scenario where cool gas is flowing toward the centre of the cluster. Similarly, for strong absorbers the best fit gaussian parameters are mean $\langle\Delta v\rangle\approx 1(\pm26)$ \kms and dispersion, $\sigma_{v}\approx 600$ \kms (shown in the solid red line). The mean velocity difference (consistent with zero) is in agreement with results from \mgii-galaxy cross-correlation studies \citep{huang16, chen17, huang21}. We also show the gaussian fit for all absorbers (\ewGr{0.4}, black line) and it is also consistent with the other two distributions. 

The typical dark matter halo velocity dispersion within $r_{\rm 200}$ ($\Delta v_{200}$) for the clusters in our sample is expected to be $\approx 900$ \kms. This implies $\rm \sigma_{weak,\,strong\, \mgii}\approx0.6-0.7 \sigma_{cluster, \, 200}$, i.e. suppressed by $30-40\%$ relative to the velocity dispersion of the cluster dark matter halo. The motion of both weak and strong absorbers in cluster halo is sub-virial. This is very different from the motion of cool CGM around star-forming galaxies (ELGs), where absorbing clouds move with velocities as high as the velocity dispersion of their dark matter halo \citep{nielsen15, lan18, anand21}.

Results for SDSS LRGs are shown for comparison as an orange curve in the figure -- the Gaussian width is $\sim 200$ \kms, i.e. a factor of 3 smaller than for the clusters. Clearly the motion of \mgii absorbers in higher mass halos is significantly larger than in lower mass halos, which is likely tracing the larger gravitationally induced motions in these halos.


\section{Discussion}\label{discuss}

\subsection{Comparison with Previous Studies}\label{previous_study}

\begin{figure*}
	\includegraphics[width=0.48\linewidth]{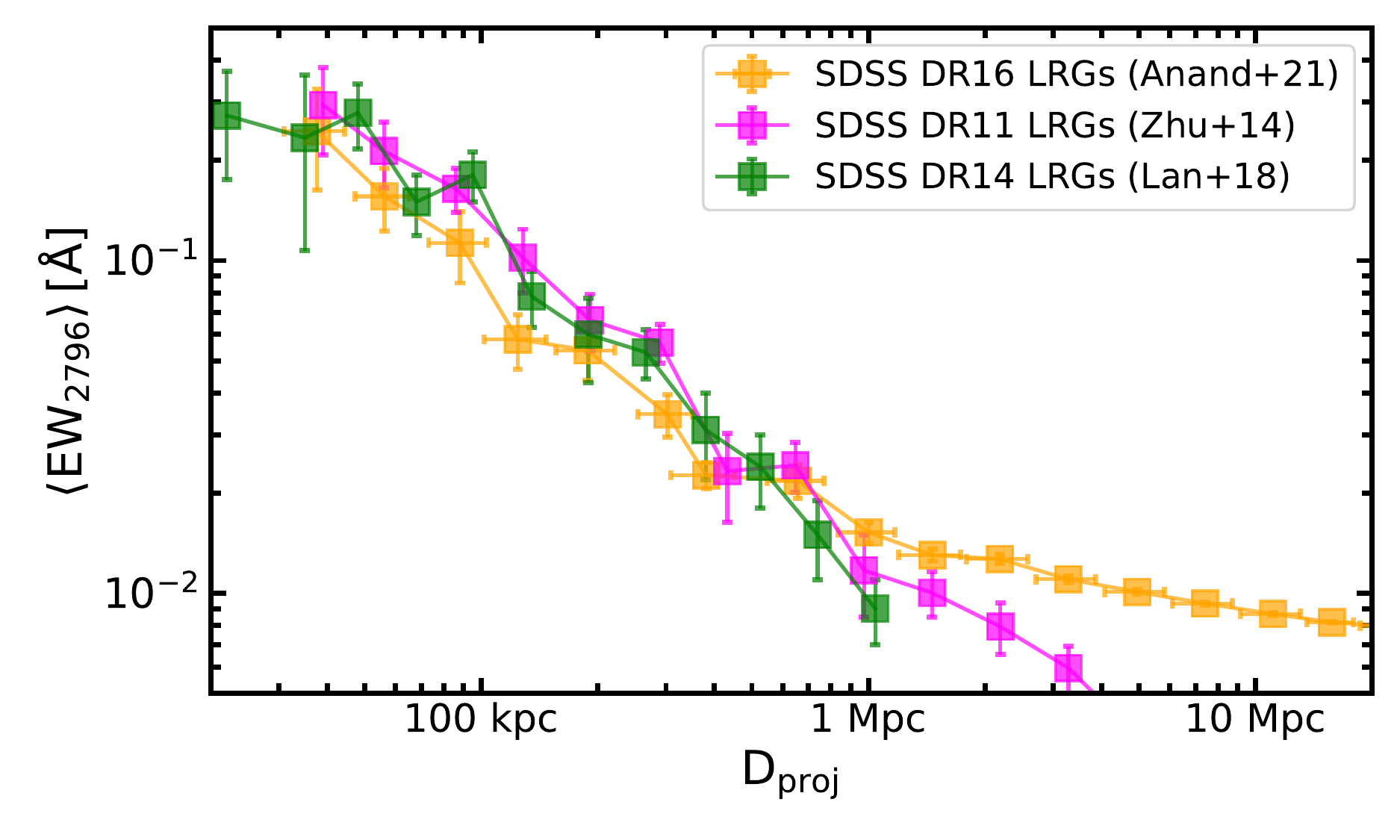}
	\includegraphics[width=0.48\linewidth]{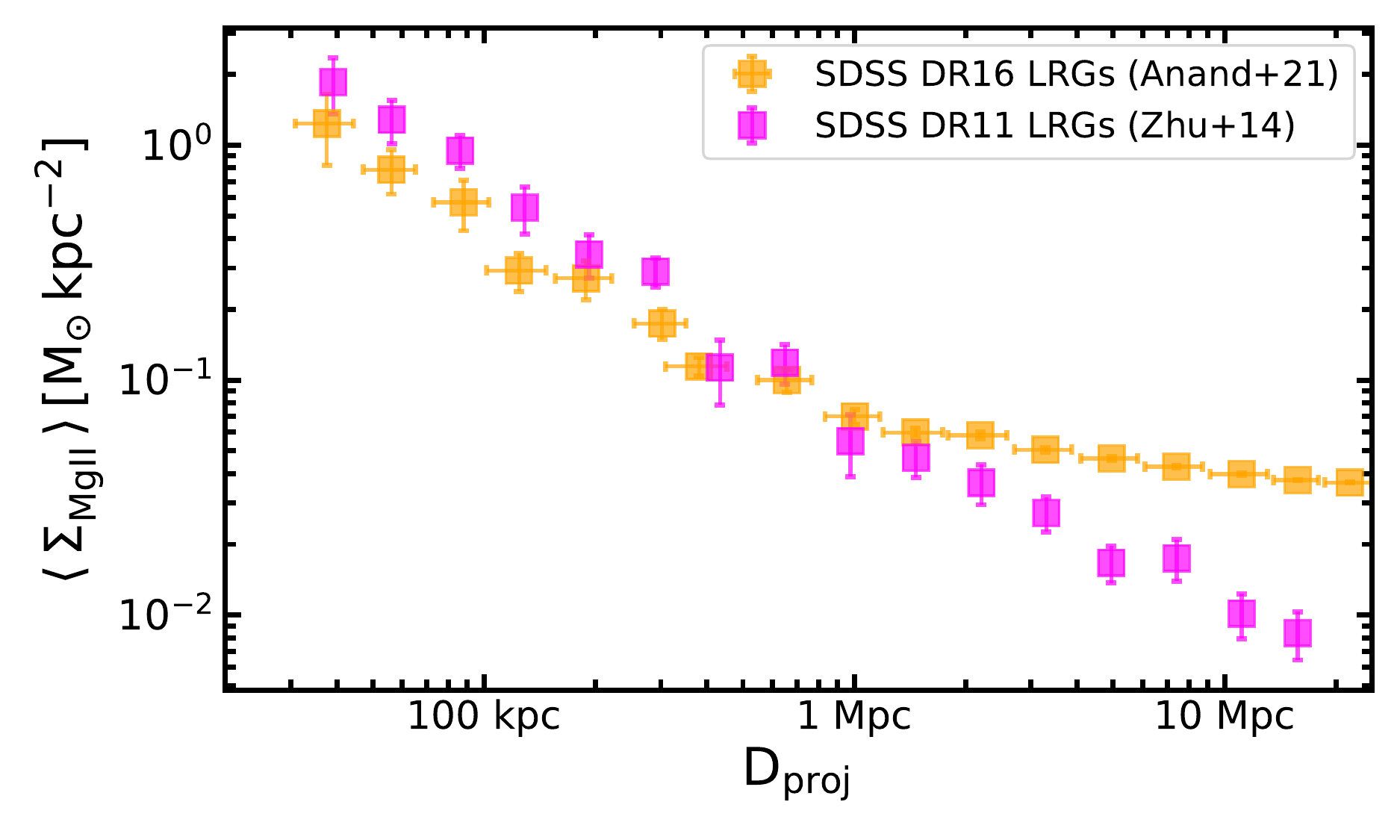}
    \caption{Mean $\mean{EW_{2796}}$ (left) and \mgii surface density (right) as a function of projected distance from the central galaxy. In the first case, we contrast the values around SDSS LRGs from \citet{anand21} (estimated using Eqn~\ref{eqn:mg_ew}, orange) with measurements from \citet{zhu14} (magenta) \citet{lan18} (green) where authors measured the mean absorption by stacking SDSS quasar spectra in the rest-frame of SDSS LRGs. For surface densities (right), we contrast our measurements around SDSS LRGs from \citet{anand21} (estimated using Eqn~\ref{eqn:surf_def}, orange) with \citet{zhu14} (magenta) measurements which was measured by stacking SDSS DR11 quasar spectra in the rest-frame of SDSS LRGs. In both cases the agreement is good at small distances, but methodological differences emerge at large scales (see text). }
    \label{fig:sdss_lrg_anand21_zhu14}
\end{figure*}

We now compare our results with previous studies that performed similar analysis of cool gas in galaxies and clusters. We contrast our mean $\mean{EW_{2796}}$ measurements with results from \citet{zhu14} and \citet{lan18}, where authors estimate the median \ewMg by stacking thousands of quasar spectra from SDSS DR11/DR14 in the rest-frame of LRGs from the same data release. The comparison is shown in Figure~\ref{fig:sdss_lrg_anand21_zhu14} (left panel), where orange squares show the values for SDSS LRGs using our approach (see section~\ref{surf_mass}), green shows the results from \citet{lan18} and magenta represents the measurements from \citet{zhu14}.

Our measurements are in good agreement with previous studies at small distances ($D_{\rm proj}\lesssim 1\, \rm Mpc$), even though the methods and samples are significantly different. We see that at large distances ($D_{\rm proj}\gtrsim 5\, \rm Mpc$) our \ewMg is larger than previous measurements, and this discrepancy is expected given the different samples and methodology. In previous studies, the \ewMg is the mean equivalent width of the \mgii absorbers in the stacked quasar spectra. When we go far from galaxies and start tracing the IGM, the absorption is dominated by weak absorbers. In contrast, we measure the mean equivalent width of absorbers (Section~\ref{surf_mass}) based on individual detections, which is limited by the sensitivity of the detection pipeline. Furthermore, the detection limit is a strong function of absorber strength, redshift and S/N of the spectra \citep[see completeness in][]{anand21} so it is difficult to detect weak systems in the individual spectra. By definition, our method additionally estimates the average equivalent width of absorbers above a threshold, and weak systems are much more ubiquitous than strong systems as we see from a completeness analysis of our \citep{anand21} and JHU-SDSS \citep{zhu13a} \mgii catalogues. 

We also compare our mean \mgii surface mass density (Section~\ref{surf_mass}) in and around SDSS LRGs (orange) with the measurements from \citet{zhu14} (magenta) analysis and this comparison is shown in Figure~\ref{fig:sdss_lrg_anand21_zhu14} (right panel). As above, the agreement is excellent at small distances, but begins to diverge at large scales due to the methodological differences.

Our measurements exhibit a change of slope at $D_{\rm proj}\approx 1\, \rm Mpc$ (see also Section~\ref{best_fit} for a detailed discussion on \mgii scale), which is consistent with previous studies \citep{zhu14, perez15, zu21}. As discussed in \citet{zhu14} and \citet{zu21}, this slope change is indicative of a transition from scales where the measurements are dominated by gas within a single dark matter halo to a regime where the gas properties are determined by contributions from gas in multiple haloes -- the 2-halo term.

In Section~\ref{mgii_ew_cluster}, we found that \mgii equivalent widths and surface mass densities are higher in clusters than SDSS LRGs, at any given impact parameter ($D_{\rm proj}/r_{\rm 500}$). Similar enhancements in \mgii covering fractions \citep{nielsen18, dutta20} in group environments have been noted. On the other hand, \citet{bordoloi11} found extended radial distribution of \mgii equivalent widths in group galaxies in a stacking analysis with $z$COSMOS galaxy survey. In this case, it has been previously assumed that the sum of \ewMg in isolated galaxies can account for total \ewMg in groups. We note that our results on the independence of \mgii equivalent widths on galaxy properties show that this cannot be true in detail.

\subsection{Possible Origin of \mgii absorbers in clusters}\label{origin}

As summarized in Section~\ref{correlation}, our analysis points towards a scenario where \mgii absorption in clusters is higher and more extended than for less massive galaxies. The exact processes that give rise to the observed \mgii absorption in clusters remain unclear, because of the complex nature of galaxy dynamics and gas properties in clusters. A number of different mechanisms have been proposed in the literature to explain the observed properties of \mgii absorbers in overdense regions: (i) the contribution of multiple individual galaxy haloes within the clusters or group \citep{padilla09, bordoloi11, nielsen18, dutta20}; (ii) strong galactic outflows from a single massive galaxy within the cluster; (iii) stripping of cold gas by hot ICM as galaxies move through the ambient hot medium or by tidal interactions between galaxies \citep{fumagalli14, pearson16, johnson18, fossati19}; and (iv) in-situ formed cool gas clumps floating in the intragroup medium or ICM \citep{sharma12, voit15, qiu21}. A combination of satellite interactions/stripping and in-situ thermal instability driven cooling has also been explored \citep{nelson20,dutta21a}.

In our analysis, we find that the median separation between absorbers and the closest cluster galaxy is roughly a few hundred kpc on average. This places a large fraction of absorbers outside the expected radius of the subhaloes that host these systems. The possibility that absorbers may be associated with low-mass galaxies remains open. \citet{tchernyshyov21} recently reported an association of \ovi absorbers with low mass star-forming galaxies in the CGM$^{2}$ survey. However, \citet{dutta20} did a comparative study of \ewMg measurements in groups with predictions from simple superposition models and concluded that such models could not explain the enhanced \mgii absorption in groups. We note that in Section~\ref{mgii_mass} we find that within $r_{\rm 500}$, the cool gas mass in clusters is $\sim10$ times higher than SDSS LRGs, while the \mgii mass is $\sim 3$ times larger in clusters than LRGs within $D_{\rm proj}\lesssim 200\, \rm kpc$ from the central galaxy. This indicates that the \mgii enhancement in clusters is scale dependent, which is not consistent with a simple superposition model.

The observation that \mgii absorbers are spatially correlated with cluster member galaxies (see Figure~\ref{fig:mgii_desi}) compared to the random samples (though the average spatial separation is larger than the size of galaxy discs) points toward the scenario where these absorbers are more likely to be associated with the CGM of cluster galaxies than their ISM. Furthermore, the fact that there is no visible trend (characterized by small Pearson coefficients) between \mgii absorption and stellar activity (SFR) of the nearest galaxy (see Figures~\ref{fig:ew_mgii_desi}, ~\ref{fig:ew_mgii_sfr_desi}) leads us to speculate that current stellar activity could not be the dominant mechanism for the origin of \mgii absorbers in clusters. Increasing the redshift separation between absorbers and cluster galaxies does not change this conclusion (see bottom panels of Figures~\ref{fig:ew_mgii_desi}). 

Our analysis has a number of notable limitations. First, the large errors in the photo$-z$ of the galaxies can introduce large uncertainties in cluster member and \mgii absorbers correlation. Some of the pairs may be spurious and far in redshift space. Hence, the current discrepancy can also arise because previous studies have used spectroscopic samples that allow us to connect galaxies and absorbers more robustly in velocity space. Given these limitations, a direct comparison between previous results and current photo $-z$ galaxies may not be accurate. Second, our analysis is based on a particular \ewMg threshold (\ewGr{0.4}, to retain high completeness) and does not take into account the non-detections or weak systems. It is possible that including such weak systems can show a significant anti-correlation between \ewMg and the impact parameter. We included the absorbers with $EW_{\rm 2796}<0.4$ \angstrom\, (not shown here), but the statistics are poor, and we can not test this hypothesis. Nonetheless, we do not observe any correlation between \ewMg and galactic properties for \ewGr{0.4} absorbers. Hence, the presence of these absorbers perhaps can not be fully explained by the ISM or stellar activity of cluster galaxies. Moreover, while comparing \ewMg or \mgii surface mass density with previous studies, it is important to note that previous studies are based on stacking methods, which also account for non-detections and weak systems. Given our \ewMg threshold, we do not account for such non-detections in the current study. Hence, we conclude that \mgii absorbers (\ewGr{0.4}) in clusters is likely regulated by the interaction between galaxies and the ICM or intragroup medium besides any contribution from stellar outflows of individual galaxies. However, the caveats and limitations should be kept in mind while interpreting the results.

Several studies have shown that galaxies are impacted by environmental effects  before they enter the inner regions ($D_{\rm proj}\lesssim r_{\rm 500}$) of clusters \citep{cen14, marasco16, jung18, ayromlou21}. One important mechanism that can remove the gas from the CGM of an infalling galaxy is ram-pressure stripping, where the hot ambient ICM or intragroup medium strips cold gas from the outskirts of the galactic halo. Tidal disruption of gas from a small satellite galaxy by a massive halo can also play role in removing the gas in dense environments. Both observations and simulations have revealed stripping of gas in groups and clusters \citep{milhos12, marasco16, osullivan18, for19, yun19}. It is plausible that some fraction of the stripped gas may extend to large distances from the host galaxy. In the recent cosmological magnetohydrodynamical simulation TNG50 \citep{pillepich19,nelson19} it was found that cold gas can form due to local density fluctuations driven by ram-pressure stripped gas in massive LRG-like haloes \citep{nelson20}.

As shown in Section~\ref{los_dv}, the motion of absorbing clouds in the cluster is suppressed by $30-40$ per cent compared to the expected dark matter velocity dispersion of the cluster halo. If the absorbing clouds are gravitationally bound to satellite galaxies, the absorbers would follow the dark matter kinematics \citep{more11}. As pointed in several studies \citep{huang16, lan18, anand21}, even though the stripped/accreted gas has initial velocities comparable to cluster velocity dispersion, the drag force exerted by the hot intragroup medium or ICM can decelerate the gas motion. Therefore gas clumps falling towards cluster centre will slow down, as slow-moving clouds will survive longer due to inverse relation between evaporation time scale and cloud velocity \citep{zahedy19}.

Our finding that there is little correlation between absorbers and the properties of the nearest cluster galaxy also supports the gas-stripping scenario. In clusters, the observed abundance and properties of cool, low-ionization gas is undoubtedly a result of the complex interplay between satellite stripping, cloud destruction, cloud formation, and galactic outflows.


\section{Summary}\label{summary}

In this work, we combine $\sim$160,000 \mgii absorbers from SDSS DR16 \mgii catalogue with $72,000$ galaxy clusters from the DESI legacy imaging surveys to characterize the nature and origin of the cool gas in galaxy clusters. Our main findings are:

\begin{itemize}
    \item There is a significant covering fraction of both weak ($0.4\,\angstrom\,<EW_{\rm 2796}\,<\,1\, \angstrom$) and strong (\ewGr{1}) \mgii absorbers in clusters ($D_{\rm proj}\lesssim r_{\rm 500}$). Compared to the random expectation the measured values are $\approx 2-5$ times higher within $D_{\rm proj}\lesssim 2\,r_{\rm 500}$ (slightly depending on our $|\Delta z|$ choice). This implies that the cool gas is readily detected in galaxy clusters, despite the hot ICM. 
    
    \item Characterizing the scale dependence (radial profile) of \mgii absorption, we find that the \mgii covering fraction declines faster for strong absorbers than weak systems, indicating that strong absorbers are preferentially found at small distances.
    
    \item The mean \ewMg decreases with the cluster-centric distance and is higher in clusters compared to SDSS LRGs but lower than SDSS ELGs.
    
    \item Clusters contain $\sim 10$ times more total \mgii gas mass than SDSS LRGs within $r_{\rm 500}$. Within $D_{\rm proj} \lesssim 200$ kpc from the central galaxy, there is $\sim 3$ times more \mgii mass in clusters than SDSS LRGs. Comparison to SDSS ELGs is complicated by the fact that the majority of associated \mgii absorbers are saturated, such that only lower limits on column densities are possible. 
    
    \item The integrated covering fraction of \mgii within $r_{\rm 500}$ of clusters anti-correlates with the stellar mass of the brightest cluster galaxy. This trend is more prominent for weak absorbers.
\end{itemize}
    
We then investigate the correlation between \mgii absorbers and cluster member galaxies, including their physical properties: stellar mass, SFR, and sSFR. Our main results and conclusions are:
    
\begin{itemize}
    \item Clusters with more star-forming galaxies in their haloes have slightly higher \mgii covering fractions, indicating some (weak) level of connection between stellar activity and \mgii absorption.
    
    \item However, there is no significant correlation between \mgii equivalent width (absorption strength) and the stellar activity or stellar mass of the nearest (detected) cluster neighbour. 
    
    \item \mgii absorbers are not randomly distributed within the cluster volume: they are located preferentially closer to satellite galaxies. However, the majority of \mgii absorbers reside at distances greater than 100 kpc from the nearest cluster neighbour.
    
    \item These typical separations suggest that \mgii absorption arises at least in part from cool gas that has been removed from the ISM due to e.g. ram-pressure stripping.
\end{itemize}

Our findings are consistent with the picture whereby the cool \mgii traced gas in clusters is not found within the interstellar medium of satellite galaxies, but may be still gravitationally bound to, or recently stripped from, the dark matter subhaloes hosting member galaxies.

Although we found little direct connection with the stellar activity or the CGM of member galaxies, a more detailed analysis is needed, given the uncertainties in current photo$-z$ measurements, not accounting for non-detection or weak \mgii absorber systems and the stellar mass limits of the DESI legacy survey. With the upcoming DESI spectroscopic survey, which will provide accurate redshifts \citep{besuner21}, a detailed relative velocity analysis and definitive associations will be possible.

Moreover, with deeper surveys such as the Large Synoptic Survey Telescope (LSST) at the Rubin Observatory and  PFS on the Subaru telescope \citep{tamura16}, enormous datasets of low mass galaxies at high redshifts will become available. This will allow us to test our hypotheses in more detail, including the possible association between \mgii absorbers and faint galaxies in groups and clusters. Finally, in addition to optical data, the ongoing \textit{eROSITA} all-sky survey promises direct characterization of the hot gas components in galaxy clusters \citep{bulbul21, liu21}. In the future, it will be possible to compare the gas contents of clusters as a function of temperature and phase \citep[see][]{truong20, oppenheimer20, truong21, das21} through multi-wavelength and multi-dataset analyses.

\section*{Data Availability}

Data directly related to this publication and its figures is available on request from the corresponding author. This work is based on data which is publicly available in its entirety as part of DR16 of the Sloan Digital Sky Survey (SDSS). Our \mgii absorber catalogue is available at \url{www.mpa-garching.mpg.de/SDSS/MgII/}.

The public data of DESI legacy imaging surveys is available at \url{www.legacysurvey.org}. The galaxy cluster catalogue compiled by \citet{zou21} is  \href{https://cdsarc.cds.unistra.fr/viz-bin/cat/J/ApJS/253/56}{publicly available}. The DESI photo$-z$ galaxy catalogue \citep{zou19} is also \href{http://cdsarc.u-strasbg.fr/viz-bin/cat/J/ApJS/242/8}{publicly available}.


\section*{Acknowledgements}

We thank the referee for providing a very constructive and insightful report that helped us improve our paper. AA thanks Hu Zou for discussions and assistance with various aspects of the cluster and galaxy catalogue. AA would also like to thank his colleagues Mohammadreza Ayromlou, Periklis Okalidis and Ilkham Galiullin for insightful discussions. DN acknowledges funding from the Deutsche Forschungsgemeinschaft (DFG) through an Emmy Noether Research Group (grant number NE 2441/1-1).

We make use of Python libraries such as \textsc{numpy} \citep{harris20}, \textsc{scipy} \citep{scipy20}, \textsc{astropy} \citep{astropy18}, Jupyter Notebook \citep{jupyter16} and \textsc{matplotlib} \citep{hunter2007} for data analysis and visualization. The analysis herein was
carried out on the Freya cluster of the Max Planck Computing and Data Facility (MPCDF).

Funding for the Sloan Digital Sky Survey IV has been provided by the Alfred P. Sloan Foundation, the U.S. Department of Energy Office of Science, and the Participating Institutions. SDSS acknowledges support and resources from the Center for High Performance Computing at the University of Utah. The SDSS web site is \url{www.sdss.org}. SDSS is managed by the Astrophysical Research Consortium for the Participating Institutions of the SDSS Collaboration including the Brazilian Participation Group, the Carnegie Institution for Science, Carnegie Mellon University, Center for Astrophysics | Harvard \& Smithsonian (CfA), the Chilean Participation Group, the French Participation Group, Instituto de Astrof\'isica de Canarias, The Johns Hopkins University, Kavli Institute for the Physics and Mathematics of the Universe (IPMU) / University of Tokyo, the Korean Participation Group, Lawrence Berkeley National Laboratory, Leibniz Institut f\"ur Astrophysik Potsdam (AIP), Max- Planck-Institut f\"ur Astronomie (MPIA Heidelberg), Max-Planck- Institut für Astrophysik (MPA Garching), Max-Planck-Institut f\"ur Extraterrestrische Physik (MPE), National Astronomical Observatories of China, New Mexico State University, New York University, University of Notre Dame, Observat\'orio Nacional MCTI, The Ohio State University, Pennsylvania State University, Shanghai Astronomical Observatory, United Kingdom Participation Group, Universidad Nacional Aut\'onoma de M\'exico, University of Arizona, University of Colorado Boulder, University of Oxford, University of Portsmouth, University of Utah, University of Virginia, University of Washington, University of Wisconsin, Vanderbilt University, and Yale University.

This paper uses data that were obtained by The Legacy Surveys: the Dark Energy Camera Legacy Survey (DECaLS; NOAO Proposal ID \# 2014B-0404; PIs: David Schlegel and Arjun Dey), the Beijing-Arizona Sky Survey (BASS; NOAO Proposal ID \# 2015A-0801; PIs: Zhou Xu and Xiaohui Fan), and the Mayall $z-$ band Legacy Survey (MzLS; NOAO Proposal ID \# 2016A-0453; PI: Arjun Dey). DECaLS, BASS and MzLS together include data obtained, respectively, at the Blanco telescope, Cerro Tololo Inter-American Observatory, National Optical Astronomy Observatory (NOAO); the Bok telescope, Steward Observatory, University of Arizona; and the Mayall telescope, Kitt Peak National Observatory, NOAO. NOAO is operated by the Association of Universities for Research in Astronomy (AURA) under a cooperative agreement with the National Science Foundation. Please see \url{http://legacysurvey.org} for details regarding the Legacy Surveys. BASS is a key project of the Telescope Access Program (TAP), which has been funded by the National Astronomical Observatories of China, the Chinese Academy of Sciences (the Strategic Priority Research Program "The Emergence of Cosmological Structures" Grant No. XDB09000000), and the Special Fund for Astronomy from the Ministry of Finance. The BASS is also supported by the External Cooperation Program of Chinese Academy of Sciences (Grant No. 114A11KYSB20160057) and Chinese National Natural Science Foundation (Grant No. 11433005). The Legacy Surveys imaging of the DESI footprint is supported by the Director, Office of Science, Office of High Energy Physics of the U.S. Department of Energy under Contract No. DE-AC02-05CH1123, and by the National Energy Research Scientific Computing Center, a DOE Office of Science User Facility under the same contract; and by the U.S. National Science Foundation, Division of Astronomical Sciences under Contract No.AST-0950945 to the National Optical Astronomy Observatory.

\bibliographystyle{mnras}
\bibliography{refs}

\begin{thebibliography}{}
\makeatletter
\relax
\def\mn@urlcharsother{\let\do\@makeother \do\$\do\&\do\#\do\^\do\_\do\%\do\~}
\def\mn@doi{\begingroup\mn@urlcharsother \@ifnextchar [ {\mn@doi@}
  {\mn@doi@[]}}
\def\mn@doi@[#1]#2{\def\@tempa{#1}\ifx\@tempa\@empty \href
  {http://dx.doi.org/#2} {doi:#2}\else \href {http://dx.doi.org/#2} {#1}\fi
  \endgroup}
\def\mn@eprint#1#2{\mn@eprint@#1:#2::\@nil}
\def\mn@eprint@arXiv#1{\href {http://arxiv.org/abs/#1} {{\tt arXiv:#1}}}
\def\mn@eprint@dblp#1{\href {http://dblp.uni-trier.de/rec/bibtex/#1.xml}
  {dblp:#1}}
\def\mn@eprint@#1:#2:#3:#4\@nil{\def\@tempa {#1}\def\@tempb {#2}\def\@tempc
  {#3}\ifx \@tempc \@empty \let \@tempc \@tempb \let \@tempb \@tempa \fi \ifx
  \@tempb \@empty \def\@tempb {arXiv}\fi \@ifundefined
  {mn@eprint@\@tempb}{\@tempb:\@tempc}{\expandafter \expandafter \csname
  mn@eprint@\@tempb\endcsname \expandafter{\@tempc}}}

\bibitem[\protect\citeauthoryear{{Ahumada} et~al.,}{{Ahumada}
  et~al.}{2020}]{ahumada20}
{Ahumada} R.,  et~al., 2020, \mn@doi [\apjs] {10.3847/1538-4365/ab929e}, \href
  {https://ui.adsabs.harvard.edu/abs/2020ApJS..249....3A} {249, 3}

\bibitem[\protect\citeauthoryear{{Anand}, {Nelson}  \& {Kauffmann}}{{Anand}
  et~al.}{2021}]{anand21}
{Anand} A.,  {Nelson} D.,   {Kauffmann} G.,  2021, \mn@doi [\mnras]
  {10.1093/mnras/stab871}, \href
  {https://ui.adsabs.harvard.edu/abs/2021MNRAS.504...65A} {504, 65 (Paper I)}

\bibitem[\protect\citeauthoryear{{Astropy Collaboration} et~al.,}{{Astropy
  Collaboration} et~al.}{2018}]{astropy18}
{Astropy Collaboration} et~al., 2018, \mn@doi [\aj] {10.3847/1538-3881/aabc4f},
  \href {https://ui.adsabs.harvard.edu/abs/2018AJ....156..123A} {156, 123}

\bibitem[\protect\citeauthoryear{{Ayromlou}, {Kauffmann}, {Yates}, {Nelson}  \&
  {White}}{{Ayromlou} et~al.}{2021}]{ayromlou21}
{Ayromlou} M.,  {Kauffmann} G.,  {Yates} R.~M.,  {Nelson} D.,   {White} S.
  D.~M.,  2021, \mn@doi [\mnras] {10.1093/mnras/stab1245}, \href
  {https://ui.adsabs.harvard.edu/abs/2021MNRAS.505..492A} {505, 492}

\bibitem[\protect\citeauthoryear{{Bah{\'e}}, {McCarthy}, {Balogh}  \&
  {Font}}{{Bah{\'e}} et~al.}{2013}]{bahe13}
{Bah{\'e}} Y.~M.,  {McCarthy} I.~G.,  {Balogh} M.~L.,   {Font} A.~S.,  2013,
  \mn@doi [\mnras] {10.1093/mnras/stt109}, \href
  {https://ui.adsabs.harvard.edu/abs/2013MNRAS.430.3017B} {430, 3017}

\bibitem[\protect\citeauthoryear{{Beck}, {Dobos}, {Budav{\'a}ri}, {Szalay}  \&
  {Csabai}}{{Beck} et~al.}{2016}]{beck16}
{Beck} R.,  {Dobos} L.,  {Budav{\'a}ri} T.,  {Szalay} A.~S.,   {Csabai} I.,
  2016, \mn@doi [\mnras] {10.1093/mnras/stw1009}, \href
  {https://ui.adsabs.harvard.edu/abs/2016MNRAS.460.1371B} {460, 1371}

\bibitem[\protect\citeauthoryear{{Behroozi}, {Wechsler}, {Hearin}  \&
  {Conroy}}{{Behroozi} et~al.}{2019}]{behroozi19}
{Behroozi} P.,  {Wechsler} R.~H.,  {Hearin} A.~P.,   {Conroy} C.,  2019,
  \mn@doi [\mnras] {10.1093/mnras/stz1182}, \href
  {https://ui.adsabs.harvard.edu/abs/2019MNRAS.488.3143B} {488, 3143}

\bibitem[\protect\citeauthoryear{{Besuner} et~al.,}{{Besuner}
  et~al.}{2021}]{besuner21}
{Besuner} R.,  et~al., 2021, arXiv e-prints, \href
  {https://ui.adsabs.harvard.edu/abs/2021arXiv210111794B} {p. arXiv:2101.11794}

\bibitem[\protect\citeauthoryear{{B{\"o}hringer} \& {Werner}}{{B{\"o}hringer}
  \& {Werner}}{2010}]{bohringer10}
{B{\"o}hringer} H.,  {Werner} N.,  2010, \mn@doi [\aapr]
  {10.1007/s00159-009-0023-3}, \href
  {https://ui.adsabs.harvard.edu/abs/2010A&ARv..18..127B} {18, 127}

\bibitem[\protect\citeauthoryear{{Bordoloi} et~al.,}{{Bordoloi}
  et~al.}{2011}]{bordoloi11}
{Bordoloi} R.,  et~al., 2011, \mn@doi [\apj] {10.1088/0004-637X/743/1/10},
  \href {https://ui.adsabs.harvard.edu/abs/2011ApJ...743...10B} {743, 10}

\bibitem[\protect\citeauthoryear{{Bruzual} \& {Charlot}}{{Bruzual} \&
  {Charlot}}{2003}]{bruzal03}
{Bruzual} G.,  {Charlot} S.,  2003, \mn@doi [\mnras]
  {10.1046/j.1365-8711.2003.06897.x}, \href
  {https://ui.adsabs.harvard.edu/abs/2003MNRAS.344.1000B} {344, 1000}

\bibitem[\protect\citeauthoryear{{Bulbul} et~al.,}{{Bulbul}
  et~al.}{2021}]{bulbul21}
{Bulbul} E.,  et~al., 2021, arXiv e-prints, \href
  {https://ui.adsabs.harvard.edu/abs/2021arXiv211009544B} {p. arXiv:2110.09544}

\bibitem[\protect\citeauthoryear{{Burchett}, {Tripp}, {Wang}, {Willmer},
  {Bowen}  \& {Jenkins}}{{Burchett} et~al.}{2018}]{burchett18}
{Burchett} J.~N.,  {Tripp} T.~M.,  {Wang} Q.~D.,  {Willmer} C. N.~A.,  {Bowen}
  D.~V.,   {Jenkins} E.~B.,  2018, \mn@doi [\mnras] {10.1093/mnras/stx3170},
  \href {https://ui.adsabs.harvard.edu/abs/2018MNRAS.475.2067B} {475, 2067}

\bibitem[\protect\citeauthoryear{{Burchett}, {Rubin}, {Prochaska}, {Coil},
  {Vaught}  \& {Hennawi}}{{Burchett} et~al.}{2021}]{burchett21}
{Burchett} J.~N.,  {Rubin} K. H.~R.,  {Prochaska} J.~X.,  {Coil} A.~L.,
  {Vaught} R.~R.,   {Hennawi} J.~F.,  2021, \mn@doi [\apj]
  {10.3847/1538-4357/abd4e0}, \href
  {https://ui.adsabs.harvard.edu/abs/2021ApJ...909..151B} {909, 151}

\bibitem[\protect\citeauthoryear{{Butsky}, {Burchett}, {Nagai}, {Tremmel},
  {Quinn}  \& {Werk}}{{Butsky} et~al.}{2019}]{butsky19}
{Butsky} I.~S.,  {Burchett} J.~N.,  {Nagai} D.,  {Tremmel} M.,  {Quinn} T.~R.,
   {Werk} J.~K.,  2019, \mn@doi [\mnras] {10.1093/mnras/stz2859}, \href
  {https://ui.adsabs.harvard.edu/abs/2019MNRAS.490.4292B} {490, 4292}

\bibitem[\protect\citeauthoryear{{Byrohl} et~al.,}{{Byrohl}
  et~al.}{2021}]{byrohl21}
{Byrohl} C.,  et~al., 2021, \mn@doi [\mnras] {10.1093/mnras/stab1958}, \href
  {https://ui.adsabs.harvard.edu/abs/2021MNRAS.506.5129B} {506, 5129}

\bibitem[\protect\citeauthoryear{{Cen}, {Roxana Pop}  \& {Bahcall}}{{Cen}
  et~al.}{2014}]{cen14}
{Cen} R.,  {Roxana Pop} A.,   {Bahcall} N.~A.,  2014, \mn@doi [Proceedings of
  the National Academy of Science] {10.1073/pnas.1407300111}, \href
  {https://ui.adsabs.harvard.edu/abs/2014PNAS..111.7914C} {111, 7914}

\bibitem[\protect\citeauthoryear{{Chabrier}}{{Chabrier}}{2003}]{chabrier03}
{Chabrier} G.,  2003, \mn@doi [\pasp] {10.1086/376392}, \href
  {https://ui.adsabs.harvard.edu/abs/2003PASP..115..763C} {115, 763}

\bibitem[\protect\citeauthoryear{Chen}{Chen}{2017}]{chen17}
Chen H.-W.,  2017, The Circumgalactic Medium in Massive Halos.
Springer International Publishing, Cham, pp 167--194,
  \mn@doi{10.1007/978-3-319-52512-9_8}, \url
  {https://doi.org/10.1007/978-3-319-52512-9_8}

\bibitem[\protect\citeauthoryear{{Chen} et~al.,}{{Chen} et~al.}{2012}]{chen12}
{Chen} Y.-M.,  et~al., 2012, \mn@doi [\mnras]
  {10.1111/j.1365-2966.2011.20306.x}, \href
  {https://ui.adsabs.harvard.edu/abs/2012MNRAS.421..314C} {421, 314}

\bibitem[\protect\citeauthoryear{{Churchill}, {Mellon}, {Charlton}, {Jannuzi},
  {Kirhakos}, {Steidel}  \& {Schneider}}{{Churchill}
  et~al.}{2000}]{churchill2000}
{Churchill} C.~W.,  {Mellon} R.~R.,  {Charlton} J.~C.,  {Jannuzi} B.~T.,
  {Kirhakos} S.,  {Steidel} C.~C.,   {Schneider} D.~P.,  2000, \mn@doi [\apj]
  {10.1086/317120}, \href
  {https://ui.adsabs.harvard.edu/abs/2000ApJ...543..577C} {543, 577}

\bibitem[\protect\citeauthoryear{{Churchill}, {Vogt}  \&
  {Charlton}}{{Churchill} et~al.}{2003}]{churchill03}
{Churchill} C.~W.,  {Vogt} S.~S.,   {Charlton} J.~C.,  2003, \mn@doi [\aj]
  {10.1086/345513}, \href
  {https://ui.adsabs.harvard.edu/abs/2003AJ....125...98C} {125, 98}

\bibitem[\protect\citeauthoryear{{Circosta} et~al.,}{{Circosta}
  et~al.}{2018}]{circosta19}
{Circosta} C.,  et~al., 2018, \mn@doi [\aap] {10.1051/0004-6361/201833520},
  \href {https://ui.adsabs.harvard.edu/abs/2018A&A...620A..82C} {620, A82}

\bibitem[\protect\citeauthoryear{{Cortese} et~al.,}{{Cortese}
  et~al.}{2007}]{cortese07}
{Cortese} L.,  et~al., 2007, \mn@doi [\mnras]
  {10.1111/j.1365-2966.2006.11369.x}, \href
  {https://ui.adsabs.harvard.edu/abs/2007MNRAS.376..157C} {376, 157}

\bibitem[\protect\citeauthoryear{{Das}, {Mathur}, {Gupta}  \& {Krongold}}{{Das}
  et~al.}{2021}]{das21}
{Das} S.,  {Mathur} S.,  {Gupta} A.,   {Krongold} Y.,  2021, \mn@doi [\apj]
  {10.3847/1538-4357/ac0e8e}, \href
  {https://ui.adsabs.harvard.edu/abs/2021ApJ...918...83D} {918, 83}

\bibitem[\protect\citeauthoryear{{Dav{\'e}}, {Angl{\'e}s-Alc{\'a}zar},
  {Narayanan}, {Li}, {Rafieferantsoa}  \& {Appleby}}{{Dav{\'e}}
  et~al.}{2019}]{dave19}
{Dav{\'e}} R.,  {Angl{\'e}s-Alc{\'a}zar} D.,  {Narayanan} D.,  {Li} Q.,
  {Rafieferantsoa} M.~H.,   {Appleby} S.,  2019, \mn@doi [\mnras]
  {10.1093/mnras/stz937}, \href
  {https://ui.adsabs.harvard.edu/abs/2019MNRAS.486.2827D} {486, 2827}

\bibitem[\protect\citeauthoryear{{Dey} et~al.,}{{Dey} et~al.}{2019}]{dey19}
{Dey} A.,  et~al., 2019, \mn@doi [\aj] {10.3847/1538-3881/ab089d}, \href
  {https://ui.adsabs.harvard.edu/abs/2019AJ....157..168D} {157, 168}

\bibitem[\protect\citeauthoryear{{Donnari} et~al.,}{{Donnari}
  et~al.}{2019}]{donnari19}
{Donnari} M.,  et~al., 2019, \mn@doi [\mnras] {10.1093/mnras/stz712}, \href
  {https://ui.adsabs.harvard.edu/abs/2019MNRAS.485.4817D} {485, 4817}

\bibitem[\protect\citeauthoryear{{Draine}}{{Draine}}{2011}]{draine11}
{Draine} B.~T.,  2011, {Physics of the Interstellar and Intergalactic Medium}.
Princeton University Press

\bibitem[\protect\citeauthoryear{{Dutta} et~al.,}{{Dutta}
  et~al.}{2020}]{dutta20}
{Dutta} R.,  et~al., 2020, \mn@doi [\mnras] {10.1093/mnras/staa3147}, \href
  {https://ui.adsabs.harvard.edu/abs/2020MNRAS.499.5022D} {499, 5022}

\bibitem[\protect\citeauthoryear{{Dutta} et~al.,}{{Dutta}
  et~al.}{2021}]{dutta21}
{Dutta} R.,  et~al., 2021, \mn@doi [\mnras] {10.1093/mnras/stab2752}, \href
  {https://ui.adsabs.harvard.edu/abs/2021MNRAS.508.4573D} {508, 4573}

\bibitem[\protect\citeauthoryear{{Dutta}, {Sharma}  \& {Nelson}}{{Dutta}
  et~al.}{2022}]{dutta21a}
{Dutta} A.,  {Sharma} P.,   {Nelson} D.,  2022, \mn@doi [\mnras]
  {10.1093/mnras/stab3653}, \href
  {https://ui.adsabs.harvard.edu/abs/2022MNRAS.510.3561D} {510, 3561}

\bibitem[\protect\citeauthoryear{{Edge}}{{Edge}}{2001}]{edge01}
{Edge} A.~C.,  2001, \mn@doi [\mnras] {10.1046/j.1365-8711.2001.04802.x}, \href
  {https://ui.adsabs.harvard.edu/abs/2001MNRAS.328..762E} {328, 762}

\bibitem[\protect\citeauthoryear{{Fielding} et~al.,}{{Fielding}
  et~al.}{2020}]{fielding20}
{Fielding} D.~B.,  et~al., 2020, \mn@doi [\apj] {10.3847/1538-4357/abbc6d},
  \href {https://ui.adsabs.harvard.edu/abs/2020ApJ...903...32F} {903, 32}

\bibitem[\protect\citeauthoryear{{For} et~al.,}{{For} et~al.}{2019}]{for19}
{For} B.~Q.,  et~al., 2019, \mn@doi [\mnras] {10.1093/mnras/stz2501}, \href
  {https://ui.adsabs.harvard.edu/abs/2019MNRAS.489.5723F} {489, 5723}

\bibitem[\protect\citeauthoryear{{Fossati}, {Gavazzi}, {Boselli}  \&
  {Fumagalli}}{{Fossati} et~al.}{2012}]{fossati12}
{Fossati} M.,  {Gavazzi} G.,  {Boselli} A.,   {Fumagalli} M.,  2012, \mn@doi
  [\aap] {10.1051/0004-6361/201219933}, \href
  {https://ui.adsabs.harvard.edu/abs/2012A&A...544A.128F} {544, A128}

\bibitem[\protect\citeauthoryear{{Fossati} et~al.,}{{Fossati}
  et~al.}{2019}]{fossati19}
{Fossati} M.,  et~al., 2019, \mn@doi [\mnras] {10.1093/mnras/stz2693}, \href
  {https://ui.adsabs.harvard.edu/abs/2019MNRAS.490.1451F} {490, 1451}

\bibitem[\protect\citeauthoryear{{Fumagalli}, {Fossati}, {Hau}, {Gavazzi},
  {Bower}, {Sun}  \& {Boselli}}{{Fumagalli} et~al.}{2014}]{fumagalli14}
{Fumagalli} M.,  {Fossati} M.,  {Hau} G. K.~T.,  {Gavazzi} G.,  {Bower} R.,
  {Sun} M.,   {Boselli} A.,  2014, \mn@doi [\mnras] {10.1093/mnras/stu2092},
  \href {https://ui.adsabs.harvard.edu/abs/2014MNRAS.445.4335F} {445, 4335}

\bibitem[\protect\citeauthoryear{{Gao}, {Zou}, {Zhou}  \& {Kong}}{{Gao}
  et~al.}{2020}]{gao20}
{Gao} J.,  {Zou} H.,  {Zhou} X.,   {Kong} X.,  2020, \mn@doi [\pasp]
  {10.1088/1538-3873/ab6151}, \href
  {https://ui.adsabs.harvard.edu/abs/2020PASP..132b4101G} {132, 024101}

\bibitem[\protect\citeauthoryear{{Gladders} \& {Yee}}{{Gladders} \&
  {Yee}}{2005}]{gladders05}
{Gladders} M.~D.,  {Yee} H.~K.~C.,  2005, \mn@doi [\apjs] {10.1086/427327},
  \href {https://ui.adsabs.harvard.edu/abs/2005ApJS..157....1G} {157, 1}

\bibitem[\protect\citeauthoryear{{Gunn} \& {Gott}}{{Gunn} \&
  {Gott}}{1972}]{gunn72}
{Gunn} J.~E.,  {Gott} J.~Richard I.,  1972, \mn@doi [\apj] {10.1086/151605},
  \href {https://ui.adsabs.harvard.edu/abs/1972ApJ...176....1G} {176, 1}

\bibitem[\protect\citeauthoryear{{Hamanowicz} et~al.,}{{Hamanowicz}
  et~al.}{2020}]{hamanowicz20}
{Hamanowicz} A.,  et~al., 2020, \mn@doi [\mnras] {10.1093/mnras/stz3590}, \href
  {https://ui.adsabs.harvard.edu/abs/2020MNRAS.492.2347H} {492, 2347}

\bibitem[\protect\citeauthoryear{Harris et~al.,}{Harris
  et~al.}{2020}]{harris20}
Harris C.~R.,  et~al., 2020, \mn@doi [Nature] {10.1038/s41586-020-2649-2}, 585,
  357

\bibitem[\protect\citeauthoryear{{Hasan}, {Churchill}, {Stemock}, {Nielsen},
  {Kacprzak}, {Croom}  \& {Murphy}}{{Hasan} et~al.}{2021}]{hasan21}
{Hasan} F.,  {Churchill} C.~W.,  {Stemock} B.,  {Nielsen} N.~M.,  {Kacprzak}
  G.~G.,  {Croom} M.,   {Murphy} M.~T.,  2021, arXiv e-prints, \href
  {https://ui.adsabs.harvard.edu/abs/2021arXiv210804924H} {p. arXiv:2108.04924}

\bibitem[\protect\citeauthoryear{{Huang}, {Chen}, {Johnson}  \&
  {Weiner}}{{Huang} et~al.}{2016}]{huang16}
{Huang} Y.-H.,  {Chen} H.-W.,  {Johnson} S.~D.,   {Weiner} B.~J.,  2016,
  \mn@doi [\mnras] {10.1093/mnras/stv2327}, \href
  {https://ui.adsabs.harvard.edu/abs/2016MNRAS.455.1713H} {455, 1713}

\bibitem[\protect\citeauthoryear{{Huang}, {Chen}, {Shectman}, {Johnson},
  {Zahedy}, {Helsby}, {Gauthier}  \& {Thompson}}{{Huang}
  et~al.}{2021}]{huang21}
{Huang} Y.-H.,  {Chen} H.-W.,  {Shectman} S.~A.,  {Johnson} S.~D.,  {Zahedy}
  F.~S.,  {Helsby} J.~E.,  {Gauthier} J.-R.,   {Thompson} I.~B.,  2021, \mn@doi
  [\mnras] {10.1093/mnras/stab360}, \href
  {https://ui.adsabs.harvard.edu/abs/2021MNRAS.502.4743H} {502, 4743}

\bibitem[\protect\citeauthoryear{Hunter}{Hunter}{2007}]{hunter2007}
Hunter J.~D.,  2007, \mn@doi [Computing in Science \& Engineering]
  {10.1109/MCSE.2007.55}, 9, 90

\bibitem[\protect\citeauthoryear{{Ilbert} et~al.,}{{Ilbert}
  et~al.}{2009}]{ilbert09}
{Ilbert} O.,  et~al., 2009, \mn@doi [\apj] {10.1088/0004-637X/690/2/1236},
  \href {https://ui.adsabs.harvard.edu/abs/2009ApJ...690.1236I} {690, 1236}

\bibitem[\protect\citeauthoryear{{Johnson} et~al.,}{{Johnson}
  et~al.}{2018}]{johnson18}
{Johnson} S.~D.,  et~al., 2018, \mn@doi [\apjl] {10.3847/2041-8213/aaf1cf},
  \href {https://ui.adsabs.harvard.edu/abs/2018ApJ...869L...1J} {869, L1}

\bibitem[\protect\citeauthoryear{{Jung}, {Choi}, {Wong}, {Kimm}, {Chung}  \&
  {Yi}}{{Jung} et~al.}{2018}]{jung18}
{Jung} S.~L.,  {Choi} H.,  {Wong} O.~I.,  {Kimm} T.,  {Chung} A.,   {Yi} S.~K.,
   2018, \mn@doi [\apj] {10.3847/1538-4357/aadda2}, \href
  {https://ui.adsabs.harvard.edu/abs/2018ApJ...865..156J} {865, 156}

\bibitem[\protect\citeauthoryear{Kluyver et~al.,}{Kluyver
  et~al.}{2016}]{jupyter16}
Kluyver T.,  et~al., 2016, in Loizides F.,  Schmidt B.,  eds, Positioning and
  Power in Academic Publishing: Players, Agents and Agendas. pp 87 -- 90

\bibitem[\protect\citeauthoryear{{Kravtsov}}{{Kravtsov}}{2013}]{kravtsov13}
{Kravtsov} A.~V.,  2013, \mn@doi [\apjl] {10.1088/2041-8205/764/2/L31}, \href
  {https://ui.adsabs.harvard.edu/abs/2013ApJ...764L..31K} {764, L31}

\bibitem[\protect\citeauthoryear{{Kravtsov} \& {Borgani}}{{Kravtsov} \&
  {Borgani}}{2012}]{kravtsov12}
{Kravtsov} A.~V.,  {Borgani} S.,  2012, \mn@doi [\araa]
  {10.1146/annurev-astro-081811-125502}, \href
  {https://ui.adsabs.harvard.edu/abs/2012ARA&A..50..353K} {50, 353}

\bibitem[\protect\citeauthoryear{{Lan}}{{Lan}}{2020}]{lan20}
{Lan} T.-W.,  2020, \mn@doi [\apj] {10.3847/1538-4357/ab989a}, \href
  {https://ui.adsabs.harvard.edu/abs/2020ApJ...897...97L} {897, 97}

\bibitem[\protect\citeauthoryear{{Lan} \& {Mo}}{{Lan} \& {Mo}}{2018}]{lan18}
{Lan} T.-W.,  {Mo} H.,  2018, \mn@doi [\apj] {10.3847/1538-4357/aadc08}, \href
  {https://ui.adsabs.harvard.edu/abs/2018ApJ...866...36L} {866, 36}

\bibitem[\protect\citeauthoryear{{Lan}, {M{\'e}nard}  \& {Zhu}}{{Lan}
  et~al.}{2014}]{lan14}
{Lan} T.-W.,  {M{\'e}nard} B.,   {Zhu} G.,  2014, \mn@doi [\apj]
  {10.1088/0004-637X/795/1/31}, \href
  {https://ui.adsabs.harvard.edu/abs/2014ApJ...795...31L} {795, 31}

\bibitem[\protect\citeauthoryear{{Lee} \& {Seung}}{{Lee} \&
  {Seung}}{1999}]{lee99}
{Lee} D.~D.,  {Seung} H.~S.,  1999, \mn@doi [\nat] {10.1038/44565}, \href
  {https://ui.adsabs.harvard.edu/abs/1999Natur.401..788L} {401, 788}

\bibitem[\protect\citeauthoryear{{Lee}, {Hwang}  \& {Song}}{{Lee}
  et~al.}{2021}]{lee21}
{Lee} J.~C.,  {Hwang} H.~S.,   {Song} H.,  2021, \mn@doi [\mnras]
  {10.1093/mnras/stab637}, \href
  {https://ui.adsabs.harvard.edu/abs/2021MNRAS.503.4309L} {503, 4309}

\bibitem[\protect\citeauthoryear{{Liu} et~al.,}{{Liu} et~al.}{2021}]{liu21}
{Liu} A.,  et~al., 2021, arXiv e-prints, \href
  {https://ui.adsabs.harvard.edu/abs/2021arXiv210614518L} {p. arXiv:2106.14518}

\bibitem[\protect\citeauthoryear{{Lopez} et~al.,}{{Lopez}
  et~al.}{2008}]{lopez08}
{Lopez} S.,  et~al., 2008, \mn@doi [ApJ] {10.1086/587678}, \href
  {https://ui.adsabs.harvard.edu/abs/2008ApJ...679.1144L} {679, 1144}

\bibitem[\protect\citeauthoryear{{Lyke} et~al.,}{{Lyke}
  et~al.}{2020}]{lyke2020}
{Lyke} B.~W.,  et~al., 2020, \mn@doi [\apjs] {10.3847/1538-4365/aba623}, \href
  {https://ui.adsabs.harvard.edu/abs/2020ApJS..250....8L} {250, 8}

\bibitem[\protect\citeauthoryear{{Marasco}, {Crain}, {Schaye}, {Bah{\'e}}, {van
  der Hulst}, {Theuns}  \& {Bower}}{{Marasco} et~al.}{2016}]{marasco16}
{Marasco} A.,  {Crain} R.~A.,  {Schaye} J.,  {Bah{\'e}} Y.~M.,  {van der Hulst}
  T.,  {Theuns} T.,   {Bower} R.~G.,  2016, \mn@doi [\mnras]
  {10.1093/mnras/stw1498}, \href
  {https://ui.adsabs.harvard.edu/abs/2016MNRAS.461.2630M} {461, 2630}

\bibitem[\protect\citeauthoryear{{McCarthy}, {Frenk}, {Font}, {Lacey}, {Bower},
  {Mitchell}, {Balogh}  \& {Theuns}}{{McCarthy} et~al.}{2008}]{mccarthy08}
{McCarthy} I.~G.,  {Frenk} C.~S.,  {Font} A.~S.,  {Lacey} C.~G.,  {Bower}
  R.~G.,  {Mitchell} N.~L.,  {Balogh} M.~L.,   {Theuns} T.,  2008, \mn@doi
  [\mnras] {10.1111/j.1365-2966.2007.12577.x}, \href
  {https://ui.adsabs.harvard.edu/abs/2008MNRAS.383..593M} {383, 593}

\bibitem[\protect\citeauthoryear{{McCourt}, {Sharma}, {Quataert}  \&
  {Parrish}}{{McCourt} et~al.}{2012}]{mccourt12}
{McCourt} M.,  {Sharma} P.,  {Quataert} E.,   {Parrish} I.~J.,  2012, \mn@doi
  [\mnras] {10.1111/j.1365-2966.2011.19972.x}, \href
  {https://ui.adsabs.harvard.edu/abs/2012MNRAS.419.3319M} {419, 3319}

\bibitem[\protect\citeauthoryear{{Mihos}, {Keating}, {Holley-Bockelmann},
  {Pisano}  \& {Kassim}}{{Mihos} et~al.}{2012}]{milhos12}
{Mihos} J.~C.,  {Keating} K.~M.,  {Holley-Bockelmann} K.,  {Pisano} D.~J.,
  {Kassim} N.~E.,  2012, \mn@doi [\apj] {10.1088/0004-637X/761/2/186}, \href
  {https://ui.adsabs.harvard.edu/abs/2012ApJ...761..186M} {761, 186}

\bibitem[\protect\citeauthoryear{{Mishra} \& {Muzahid}}{{Mishra} \&
  {Muzahid}}{2022}]{mishra22}
{Mishra} S.,  {Muzahid} S.,  2022, arXiv e-prints, \href
  {https://ui.adsabs.harvard.edu/abs/2022arXiv220108545M} {p. arXiv:2201.08545}

\bibitem[\protect\citeauthoryear{{More}, {van den Bosch}, {Cacciato}, {Skibba},
  {Mo}  \& {Yang}}{{More} et~al.}{2011}]{more11}
{More} S.,  {van den Bosch} F.~C.,  {Cacciato} M.,  {Skibba} R.,  {Mo} H.~J.,
  {Yang} X.,  2011, \mn@doi [\mnras] {10.1111/j.1365-2966.2010.17436.x}, \href
  {https://ui.adsabs.harvard.edu/abs/2011MNRAS.410..210M} {410, 210}

\bibitem[\protect\citeauthoryear{{Muzahid}, {Charlton}, {Nagai}, {Schaye}  \&
  {Srianand}}{{Muzahid} et~al.}{2017}]{muzahid17}
{Muzahid} S.,  {Charlton} J.,  {Nagai} D.,  {Schaye} J.,   {Srianand} R.,
  2017, \mn@doi [\apjl] {10.3847/2041-8213/aa8559}, \href
  {https://ui.adsabs.harvard.edu/abs/2017ApJ...846L...8M} {846, L8}

\bibitem[\protect\citeauthoryear{{Naab} \& {Ostriker}}{{Naab} \&
  {Ostriker}}{2017}]{naab17}
{Naab} T.,  {Ostriker} J.~P.,  2017, \mn@doi [\araa]
  {10.1146/annurev-astro-081913-040019}, \href
  {https://ui.adsabs.harvard.edu/abs/2017ARA&A..55...59N} {55, 59}

\bibitem[\protect\citeauthoryear{{Nelson} et~al.,}{{Nelson}
  et~al.}{2019}]{nelson19}
{Nelson} D.,  et~al., 2019, \mn@doi [\mnras] {10.1093/mnras/stz2306}, \href
  {https://ui.adsabs.harvard.edu/abs/2019MNRAS.490.3234N} {490, 3234}

\bibitem[\protect\citeauthoryear{{Nelson} et~al.,}{{Nelson}
  et~al.}{2020}]{nelson20}
{Nelson} D.,  et~al., 2020, \mn@doi [\mnras] {10.1093/mnras/staa2419}, \href
  {https://ui.adsabs.harvard.edu/abs/2020MNRAS.498.2391N} {498, 2391}

\bibitem[\protect\citeauthoryear{{Nelson}, {Byrohl}, {Peroux}, {Rubin}  \&
  {Burchett}}{{Nelson} et~al.}{2021}]{nelson21}
{Nelson} D.,  {Byrohl} C.,  {Peroux} C.,  {Rubin} K. H.~R.,   {Burchett} J.~N.,
   2021, \mn@doi [\mnras] {10.1093/mnras/stab2177}, \href
  {https://ui.adsabs.harvard.edu/abs/2021MNRAS.507.4445N} {507, 4445}

\bibitem[\protect\citeauthoryear{{Nielsen}, {Churchill}, {Kacprzak}  \&
  {Murphy}}{{Nielsen} et~al.}{2013}]{nielsen13}
{Nielsen} N.~M.,  {Churchill} C.~W.,  {Kacprzak} G.~G.,   {Murphy} M.~T.,
  2013, \mn@doi [\apj] {10.1088/0004-637X/776/2/114}, \href
  {https://ui.adsabs.harvard.edu/abs/2013ApJ...776..114N} {776, 114}

\bibitem[\protect\citeauthoryear{{Nielsen}, {Churchill}, {Kacprzak}, {Murphy}
  \& {Evans}}{{Nielsen} et~al.}{2015}]{nielsen15}
{Nielsen} N.~M.,  {Churchill} C.~W.,  {Kacprzak} G.~G.,  {Murphy} M.~T.,
  {Evans} J.~L.,  2015, \mn@doi [\apj] {10.1088/0004-637X/812/1/83}, \href
  {https://ui.adsabs.harvard.edu/abs/2015ApJ...812...83N} {812, 83}

\bibitem[\protect\citeauthoryear{{Nielsen}, {Kacprzak}, {Pointon}, {Churchill}
  \& {Murphy}}{{Nielsen} et~al.}{2018}]{nielsen18}
{Nielsen} N.~M.,  {Kacprzak} G.~G.,  {Pointon} S.~K.,  {Churchill} C.~W.,
  {Murphy} M.~T.,  2018, \mn@doi [\apj] {10.3847/1538-4357/aaedbd}, \href
  {https://ui.adsabs.harvard.edu/abs/2018ApJ...869..153N} {869, 153}

\bibitem[\protect\citeauthoryear{{O'Sullivan}, {Kolokythas}, {Kantharia},
  {Raychaudhury}, {David}  \& {Vrtilek}}{{O'Sullivan}
  et~al.}{2018}]{osullivan18}
{O'Sullivan} E.,  {Kolokythas} K.,  {Kantharia} N.~G.,  {Raychaudhury} S.,
  {David} L.~P.,   {Vrtilek} J.~M.,  2018, \mn@doi [\mnras]
  {10.1093/mnras/stx2702}, \href
  {https://ui.adsabs.harvard.edu/abs/2018MNRAS.473.5248O} {473, 5248}

\bibitem[\protect\citeauthoryear{{Olivares} et~al.,}{{Olivares}
  et~al.}{2019}]{olivares19}
{Olivares} V.,  et~al., 2019, \mn@doi [\aap] {10.1051/0004-6361/201935350},
  \href {https://ui.adsabs.harvard.edu/abs/2019A&A...631A..22O} {631, A22}

\bibitem[\protect\citeauthoryear{{Oppenheimer}, {Schaye}, {Crain}, {Werk}  \&
  {Richings}}{{Oppenheimer} et~al.}{2018}]{oppenheimer18}
{Oppenheimer} B.~D.,  {Schaye} J.,  {Crain} R.~A.,  {Werk} J.~K.,   {Richings}
  A.~J.,  2018, \mn@doi [\mnras] {10.1093/mnras/sty2281}, \href
  {https://ui.adsabs.harvard.edu/abs/2018MNRAS.481..835O} {481, 835}

\bibitem[\protect\citeauthoryear{{Oppenheimer} et~al.,}{{Oppenheimer}
  et~al.}{2020}]{oppenheimer20}
{Oppenheimer} B.~D.,  et~al., 2020, \mn@doi [\apjl] {10.3847/2041-8213/ab846f},
  \href {https://ui.adsabs.harvard.edu/abs/2020ApJ...893L..24O} {893, L24}

\bibitem[\protect\citeauthoryear{{Padilla}, {Lacerna}, {Lopez}, {Barrientos},
  {Lira}, {Andrews}  \& {Tejos}}{{Padilla} et~al.}{2009}]{padilla09}
{Padilla} N.,  {Lacerna} I.,  {Lopez} S.,  {Barrientos} L.~F.,  {Lira} P.,
  {Andrews} H.,   {Tejos} N.,  2009, \mn@doi [MNRAS]
  {10.1111/j.1365-2966.2009.14621.x}, \href
  {https://ui.adsabs.harvard.edu/abs/2009MNRAS.395.1135P} {395, 1135}

\bibitem[\protect\citeauthoryear{{P{\^a}ris} et~al.,}{{P{\^a}ris}
  et~al.}{2018}]{paris18}
{P{\^a}ris} I.,  et~al., 2018, \mn@doi [\aap] {10.1051/0004-6361/201732445},
  \href {https://ui.adsabs.harvard.edu/abs/2018A&A...613A..51P} {613, A51}

\bibitem[\protect\citeauthoryear{{Pearson} et~al.,}{{Pearson}
  et~al.}{2016}]{pearson16}
{Pearson} S.,  et~al., 2016, \mn@doi [\mnras] {10.1093/mnras/stw757}, \href
  {https://ui.adsabs.harvard.edu/abs/2016MNRAS.459.1827P} {459, 1827}

\bibitem[\protect\citeauthoryear{{Peeples} et~al.,}{{Peeples}
  et~al.}{2019}]{peeples19}
{Peeples} M.~S.,  et~al., 2019, \mn@doi [\apj] {10.3847/1538-4357/ab0654},
  \href {https://ui.adsabs.harvard.edu/abs/2019ApJ...873..129P} {873, 129}

\bibitem[\protect\citeauthoryear{{P{\'e}rez-R{\`a}fols}, {Miralda-Escud{\'e}},
  {Lundgren}, {Ge}, {Petitjean}, {Schneider}, {York}  \&
  {Weaver}}{{P{\'e}rez-R{\`a}fols} et~al.}{2015}]{perez15}
{P{\'e}rez-R{\`a}fols} I.,  {Miralda-Escud{\'e}} J.,  {Lundgren} B.,  {Ge} J.,
  {Petitjean} P.,  {Schneider} D.~P.,  {York} D.~G.,   {Weaver} B.~A.,  2015,
  \mn@doi [\mnras] {10.1093/mnras/stu2645}, \href
  {https://ui.adsabs.harvard.edu/abs/2015MNRAS.447.2784P} {447, 2784}

\bibitem[\protect\citeauthoryear{{P{\'e}roux} \& {Howk}}{{P{\'e}roux} \&
  {Howk}}{2020}]{peroux20a}
{P{\'e}roux} C.,  {Howk} J.~C.,  2020, \mn@doi [\araa]
  {10.1146/annurev-astro-021820-120014}, \href
  {https://ui.adsabs.harvard.edu/abs/2020ARA&A..58..363P} {58, 363}

\bibitem[\protect\citeauthoryear{{P{\'e}roux}, {Nelson}, {van de Voort},
  {Pillepich}, {Marinacci}, {Vogelsberger}  \& {Hernquist}}{{P{\'e}roux}
  et~al.}{2020}]{peroux20b}
{P{\'e}roux} C.,  {Nelson} D.,  {van de Voort} F.,  {Pillepich} A.,
  {Marinacci} F.,  {Vogelsberger} M.,   {Hernquist} L.,  2020, \mn@doi [\mnras]
  {10.1093/mnras/staa2888}, \href
  {https://ui.adsabs.harvard.edu/abs/2020MNRAS.499.2462P} {499, 2462}

\bibitem[\protect\citeauthoryear{{Pillepich} et~al.,}{{Pillepich}
  et~al.}{2018}]{pillepich18}
{Pillepich} A.,  et~al., 2018, \mn@doi [\mnras] {10.1093/mnras/stx2656}, \href
  {https://ui.adsabs.harvard.edu/abs/2018MNRAS.473.4077P} {473, 4077}

\bibitem[\protect\citeauthoryear{{Pillepich} et~al.,}{{Pillepich}
  et~al.}{2019}]{pillepich19}
{Pillepich} A.,  et~al., 2019, \mn@doi [\mnras] {10.1093/mnras/stz2338}, \href
  {https://ui.adsabs.harvard.edu/abs/2019MNRAS.490.3196P} {490, 3196}

\bibitem[\protect\citeauthoryear{{Planck Collaboration} et~al.,}{{Planck
  Collaboration} et~al.}{2020}]{planck20}
{Planck Collaboration} et~al., 2020, \mn@doi [\aap]
  {10.1051/0004-6361/201833910}, \href
  {https://ui.adsabs.harvard.edu/abs/2020A&A...641A...6P} {641, A6}

\bibitem[\protect\citeauthoryear{{Pradeep}, {Narayanan}, {Muzahid}, {Nagai},
  {Charlton}  \& {Srianand}}{{Pradeep} et~al.}{2019}]{pradeep19}
{Pradeep} J.,  {Narayanan} A.,  {Muzahid} S.,  {Nagai} D.,  {Charlton} J.~C.,
  {Srianand} R.,  2019, \mn@doi [\mnras] {10.1093/mnras/stz2059}, \href
  {https://ui.adsabs.harvard.edu/abs/2019MNRAS.488.5327P} {488, 5327}

\bibitem[\protect\citeauthoryear{{Prochaska}, {Kasen}  \& {Rubin}}{{Prochaska}
  et~al.}{2011}]{prochaska11}
{Prochaska} J.~X.,  {Kasen} D.,   {Rubin} K.,  2011, \mn@doi [\apj]
  {10.1088/0004-637X/734/1/24}, \href
  {https://ui.adsabs.harvard.edu/abs/2011ApJ...734...24P} {734, 24}

\bibitem[\protect\citeauthoryear{{Qiu}, {Bogdanovi{\'c}}, {Li}, {McDonald}  \&
  {McNamara}}{{Qiu} et~al.}{2020}]{qiu21}
{Qiu} Y.,  {Bogdanovi{\'c}} T.,  {Li} Y.,  {McDonald} M.,   {McNamara} B.~R.,
  2020, \mn@doi [Nature Astronomy] {10.1038/s41550-020-1090-7}, \href
  {https://ui.adsabs.harvard.edu/abs/2020NatAs...4..900Q} {4, 900}

\bibitem[\protect\citeauthoryear{{Raichoor} et~al.,}{{Raichoor}
  et~al.}{2021}]{raichoor21}
{Raichoor} A.,  et~al., 2021, \mn@doi [\mnras] {10.1093/mnras/staa3336}, \href
  {https://ui.adsabs.harvard.edu/abs/2021MNRAS.500.3254R} {500, 3254}

\bibitem[\protect\citeauthoryear{{Reiprich}, {Basu}, {Ettori}, {Israel},
  {Lovisari}, {Molendi}, {Pointecouteau}  \& {Roncarelli}}{{Reiprich}
  et~al.}{2013}]{reiprich13}
{Reiprich} T.~H.,  {Basu} K.,  {Ettori} S.,  {Israel} H.,  {Lovisari} L.,
  {Molendi} S.,  {Pointecouteau} E.,   {Roncarelli} M.,  2013, \mn@doi [\ssr]
  {10.1007/s11214-013-9983-8}, \href
  {https://ui.adsabs.harvard.edu/abs/2013SSRv..177..195R} {177, 195}

\bibitem[\protect\citeauthoryear{{Rubin}, {Prochaska}, {Koo}, {Phillips},
  {Martin}  \& {Winstrom}}{{Rubin} et~al.}{2014}]{rubin14}
{Rubin} K. H.~R.,  {Prochaska} J.~X.,  {Koo} D.~C.,  {Phillips} A.~C.,
  {Martin} C.~L.,   {Winstrom} L.~O.,  2014, \mn@doi [\apj]
  {10.1088/0004-637X/794/2/156}, \href
  {https://ui.adsabs.harvard.edu/abs/2014ApJ...794..156R} {794, 156}

\bibitem[\protect\citeauthoryear{{Rudie} et~al.,}{{Rudie}
  et~al.}{2012}]{rudie12}
{Rudie} G.~C.,  et~al., 2012, \mn@doi [\apj] {10.1088/0004-637X/750/1/67},
  \href {https://ui.adsabs.harvard.edu/abs/2012ApJ...750...67R} {750, 67}

\bibitem[\protect\citeauthoryear{{Rykoff} et~al.,}{{Rykoff}
  et~al.}{2014}]{rykoff14}
{Rykoff} E.~S.,  et~al., 2014, \mn@doi [\apj] {10.1088/0004-637X/785/2/104},
  \href {https://ui.adsabs.harvard.edu/abs/2014ApJ...785..104R} {785, 104}

\bibitem[\protect\citeauthoryear{{Rykoff} et~al.,}{{Rykoff}
  et~al.}{2016}]{rykoff16}
{Rykoff} E.~S.,  et~al., 2016, \mn@doi [\apjs] {10.3847/0067-0049/224/1/1},
  \href {https://ui.adsabs.harvard.edu/abs/2016ApJS..224....1R} {224, 1}

\bibitem[\protect\citeauthoryear{{Schneider} et~al.,}{{Schneider}
  et~al.}{2005}]{schneider05}
{Schneider} D.~P.,  et~al., 2005, \mn@doi [\aj] {10.1086/431156}, \href
  {https://ui.adsabs.harvard.edu/abs/2005AJ....130..367S} {130, 367}

\bibitem[\protect\citeauthoryear{{Schroetter} et~al.,}{{Schroetter}
  et~al.}{2021}]{schroetter21}
{Schroetter} I.,  et~al., 2021, \mn@doi [\mnras] {10.1093/mnras/stab1447},
  \href {https://ui.adsabs.harvard.edu/abs/2021MNRAS.506.1355S} {506, 1355}

\bibitem[\protect\citeauthoryear{{Sharma}, {McCourt}, {Quataert}  \&
  {Parrish}}{{Sharma} et~al.}{2012}]{sharma12}
{Sharma} P.,  {McCourt} M.,  {Quataert} E.,   {Parrish} I.~J.,  2012, \mn@doi
  [\mnras] {10.1111/j.1365-2966.2011.20246.x}, \href
  {https://ui.adsabs.harvard.edu/abs/2012MNRAS.420.3174S} {420, 3174}

\bibitem[\protect\citeauthoryear{{Stocke}, {Keeney}, {Danforth}, {Shull},
  {Froning}, {Green}, {Penton}  \& {Savage}}{{Stocke} et~al.}{2013}]{stocke13}
{Stocke} J.~T.,  {Keeney} B.~A.,  {Danforth} C.~W.,  {Shull} J.~M.,  {Froning}
  C.~S.,  {Green} J.~C.,  {Penton} S.~V.,   {Savage} B.~D.,  2013, \mn@doi
  [\apj] {10.1088/0004-637X/763/2/148}, \href
  {https://ui.adsabs.harvard.edu/abs/2013ApJ...763..148S} {763, 148}

\bibitem[\protect\citeauthoryear{{Tamura} et~al.,}{{Tamura}
  et~al.}{2016}]{tamura16}
{Tamura} N.,  et~al., 2016, in {Evans} C.~J.,  {Simard} L.,   {Takami} H.,
  eds,  Society of Photo-Optical Instrumentation Engineers (SPIE) Conference
  Series Vol. 9908, Ground-based and Airborne Instrumentation for Astronomy VI.
  p. 99081M (\mn@eprint {arXiv} {1608.01075}), \mn@doi{10.1117/12.2232103}

\bibitem[\protect\citeauthoryear{{Tchernyshyov} et~al.,}{{Tchernyshyov}
  et~al.}{2021}]{tchernyshyov21}
{Tchernyshyov} K.,  et~al., 2021, arXiv e-prints, \href
  {https://ui.adsabs.harvard.edu/abs/2021arXiv211013167T} {p. arXiv:2110.13167}

\bibitem[\protect\citeauthoryear{{Truong} et~al.,}{{Truong}
  et~al.}{2020}]{truong20}
{Truong} N.,  et~al., 2020, \mn@doi [\mnras] {10.1093/mnras/staa685}, \href
  {https://ui.adsabs.harvard.edu/abs/2020MNRAS.494..549T} {494, 549}

\bibitem[\protect\citeauthoryear{{Truong}, {Pillepich}, {Nelson}, {Werner}  \&
  {Hernquist}}{{Truong} et~al.}{2021}]{truong21}
{Truong} N.,  {Pillepich} A.,  {Nelson} D.,  {Werner} N.,   {Hernquist} L.,
  2021, \mn@doi [\mnras] {10.1093/mnras/stab2638}, \href
  {https://ui.adsabs.harvard.edu/abs/2021MNRAS.508.1563T} {508, 1563}

\bibitem[\protect\citeauthoryear{{Tumlinson}, {Peeples}  \& {Werk}}{{Tumlinson}
  et~al.}{2017}]{tumlinson17}
{Tumlinson} J.,  {Peeples} M.~S.,   {Werk} J.~K.,  2017, \mn@doi [\araa]
  {10.1146/annurev-astro-091916-055240}, \href
  {https://ui.adsabs.harvard.edu/abs/2017ARA&A..55..389T} {55, 389}

\bibitem[\protect\citeauthoryear{Virtanen et~al.,}{Virtanen
  et~al.}{2020}]{scipy20}
Virtanen P.,  et~al., 2020, \mn@doi [Nature Methods]
  {10.1038/s41592-019-0686-2}, \href {https://rdcu.be/b08Wh} {17, 261}

\bibitem[\protect\citeauthoryear{{Voit} \& {Donahue}}{{Voit} \&
  {Donahue}}{2015}]{voit15}
{Voit} G.~M.,  {Donahue} M.,  2015, \mn@doi [\apjl]
  {10.1088/2041-8205/799/1/L1}, \href
  {https://ui.adsabs.harvard.edu/abs/2015ApJ...799L...1V} {799, L1}

\bibitem[\protect\citeauthoryear{{Voit}, {Meece}, {Li}, {O'Shea}, {Bryan}  \&
  {Donahue}}{{Voit} et~al.}{2017}]{voit17}
{Voit} G.~M.,  {Meece} G.,  {Li} Y.,  {O'Shea} B.~W.,  {Bryan} G.~L.,
  {Donahue} M.,  2017, \mn@doi [\apj] {10.3847/1538-4357/aa7d04}, \href
  {https://ui.adsabs.harvard.edu/abs/2017ApJ...845...80V} {845, 80}

\bibitem[\protect\citeauthoryear{{Vollmer} et~al.,}{{Vollmer}
  et~al.}{2012}]{vollmer12}
{Vollmer} B.,  et~al., 2012, \mn@doi [\aap] {10.1051/0004-6361/201117680},
  \href {https://ui.adsabs.harvard.edu/abs/2012A&A...537A.143V} {537, A143}

\bibitem[\protect\citeauthoryear{{Wang}}{{Wang}}{1993}]{wang93}
{Wang} B.,  1993, \mn@doi [\apj] {10.1086/173153}, \href
  {https://ui.adsabs.harvard.edu/abs/1993ApJ...415..174W} {415, 174}

\bibitem[\protect\citeauthoryear{{Wen} \& {Han}}{{Wen} \& {Han}}{2015}]{wen15}
{Wen} Z.~L.,  {Han} J.~L.,  2015, \mn@doi [\apj] {10.1088/0004-637X/807/2/178},
  \href {https://ui.adsabs.harvard.edu/abs/2015ApJ...807..178W} {807, 178}

\bibitem[\protect\citeauthoryear{{Werk}, {Prochaska}, {Thom}, {Tumlinson},
  {Tripp}, {O'Meara}  \& {Meiring}}{{Werk} et~al.}{2012}]{werk12}
{Werk} J.~K.,  {Prochaska} J.~X.,  {Thom} C.,  {Tumlinson} J.,  {Tripp} T.~M.,
  {O'Meara} J.~M.,   {Meiring} J.~D.,  2012, \mn@doi [\apjs]
  {10.1088/0067-0049/198/1/3}, \href
  {https://ui.adsabs.harvard.edu/abs/2012ApJS..198....3W} {198, 3}

\bibitem[\protect\citeauthoryear{{Yang} \& {Reynolds}}{{Yang} \&
  {Reynolds}}{2016}]{yang16}
{Yang} H. Y.~K.,  {Reynolds} C.~S.,  2016, \mn@doi [\apj]
  {10.3847/0004-637X/829/2/90}, \href
  {https://ui.adsabs.harvard.edu/abs/2016ApJ...829...90Y} {829, 90}

\bibitem[\protect\citeauthoryear{{Yoon} \& {Putman}}{{Yoon} \&
  {Putman}}{2017}]{yoon17}
{Yoon} J.~H.,  {Putman} M.~E.,  2017, \mn@doi [\apj]
  {10.3847/1538-4357/aa697b}, \href
  {https://ui.adsabs.harvard.edu/abs/2017ApJ...839..117Y} {839, 117}

\bibitem[\protect\citeauthoryear{{Yun} et~al.,}{{Yun} et~al.}{2019}]{yun19}
{Yun} K.,  et~al., 2019, \mn@doi [\mnras] {10.1093/mnras/sty3156}, \href
  {https://ui.adsabs.harvard.edu/abs/2019MNRAS.483.1042Y} {483, 1042}

\bibitem[\protect\citeauthoryear{{Zabl} et~al.,}{{Zabl} et~al.}{2021}]{zabl21}
{Zabl} J.,  et~al., 2021, \mn@doi [\mnras] {10.1093/mnras/stab2165}, \href
  {https://ui.adsabs.harvard.edu/abs/2021MNRAS.507.4294Z} {507, 4294}

\bibitem[\protect\citeauthoryear{{Zahedy}, {Chen}, {Johnson}, {Pierce},
  {Rauch}, {Huang}, {Weiner}  \& {Gauthier}}{{Zahedy} et~al.}{2019}]{zahedy19}
{Zahedy} F.~S.,  {Chen} H.-W.,  {Johnson} S.~D.,  {Pierce} R.~M.,  {Rauch} M.,
  {Huang} Y.-H.,  {Weiner} B.~J.,   {Gauthier} J.-R.,  2019, \mn@doi [\mnras]
  {10.1093/mnras/sty3482}, \href
  {https://ui.adsabs.harvard.edu/abs/2019MNRAS.484.2257Z} {484, 2257}

\bibitem[\protect\citeauthoryear{{Zhu} \& {M{\'e}nard}}{{Zhu} \&
  {M{\'e}nard}}{2013}]{zhu13a}
{Zhu} G.,  {M{\'e}nard} B.,  2013, \mn@doi [\apj]
  {10.1088/0004-637X/770/2/130}, \href
  {https://ui.adsabs.harvard.edu/abs/2013ApJ...770..130Z} {770, 130}

\bibitem[\protect\citeauthoryear{{Zhu} et~al.,}{{Zhu} et~al.}{2014}]{zhu14}
{Zhu} G.,  et~al., 2014, \mn@doi [\mnras] {10.1093/mnras/stu186}, \href
  {https://ui.adsabs.harvard.edu/abs/2014MNRAS.439.3139Z} {439, 3139}

\bibitem[\protect\citeauthoryear{{Zou}, {Gao}, {Zhou}  \& {Kong}}{{Zou}
  et~al.}{2019}]{zou19}
{Zou} H.,  {Gao} J.,  {Zhou} X.,   {Kong} X.,  2019, \mn@doi [\apjs]
  {10.3847/1538-4365/ab1847}, \href
  {https://ui.adsabs.harvard.edu/abs/2019ApJS..242....8Z} {242, 8}

\bibitem[\protect\citeauthoryear{{Zou} et~al.,}{{Zou} et~al.}{2021}]{zou21}
{Zou} H.,  et~al., 2021, \mn@doi [\apjs] {10.3847/1538-4365/abe5b0}, \href
  {https://ui.adsabs.harvard.edu/abs/2021ApJS..253...56Z} {253, 56}

\bibitem[\protect\citeauthoryear{{Zu}}{{Zu}}{2021}]{zu21}
{Zu} Y.,  2021, \mn@doi [\mnras] {10.1093/mnras/stab1752}, \href
  {https://ui.adsabs.harvard.edu/abs/2021MNRAS.506..115Z} {506, 115}

\bibitem[\protect\citeauthoryear{{Zuhone} \& {Markevitch}}{{Zuhone} \&
  {Markevitch}}{2009}]{zuhone09}
{Zuhone} J.,  {Markevitch} M.,  2009, in {Heinz} S.,  {Wilcots} E.,  eds,
  American Institute of Physics Conference Series Vol. 1201, The Monster's
  Fiery Breath: Feedback in Galaxies, Groups, and Clusters. pp 383--386
  (\mn@eprint {arXiv} {0909.0560}), \mn@doi{10.1063/1.3293082}

\makeatother
\end{thebibliography}

\appendix
\section{Effect of Redshift Difference between BCG and \mgii Absorbers}\label{dz_compare}

\begin{figure*}
	\includegraphics[width=0.45\linewidth]{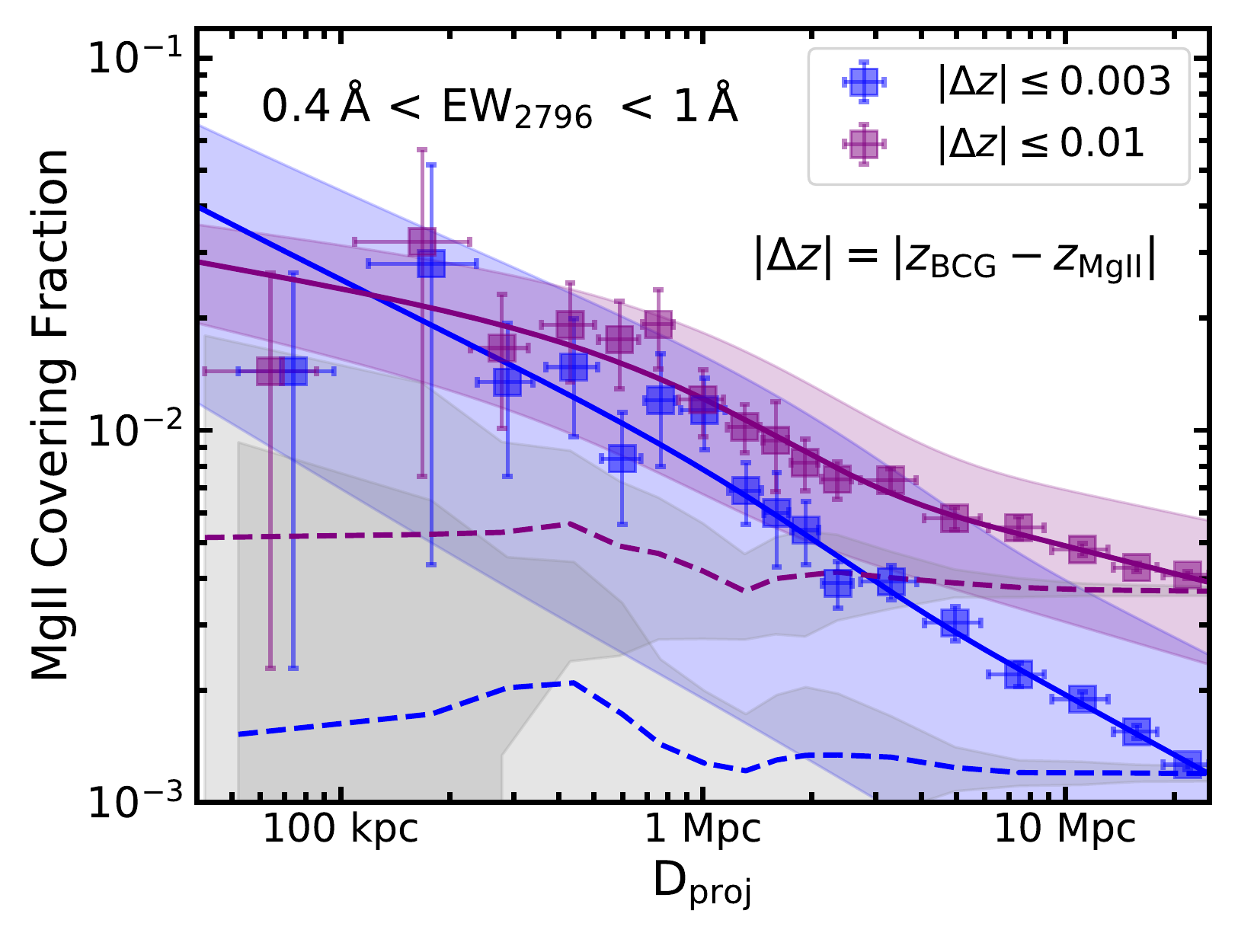}
	\includegraphics[width=0.45\linewidth]{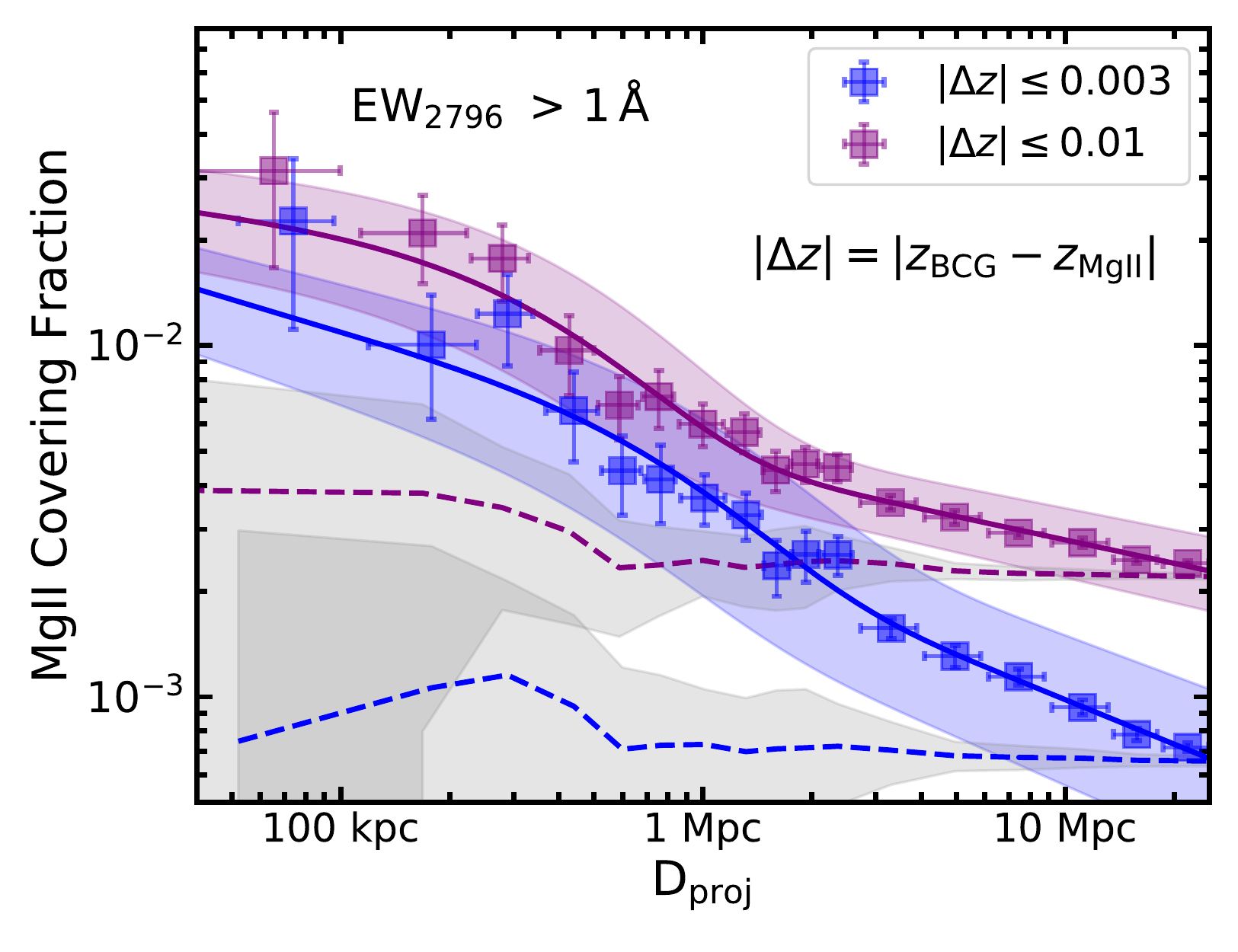}\\
	\includegraphics[width=0.45\linewidth]{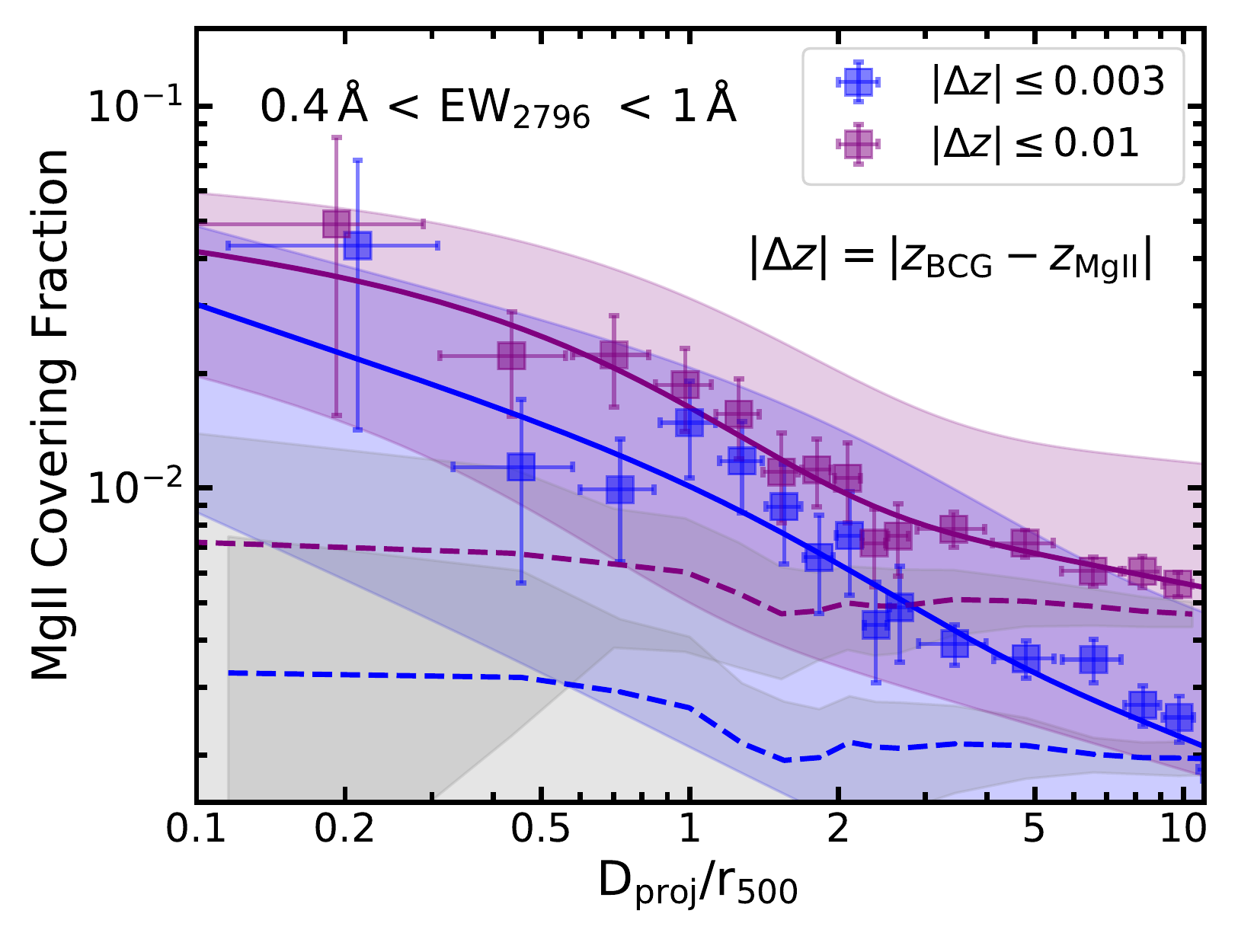}
    \includegraphics[width=0.45\linewidth]{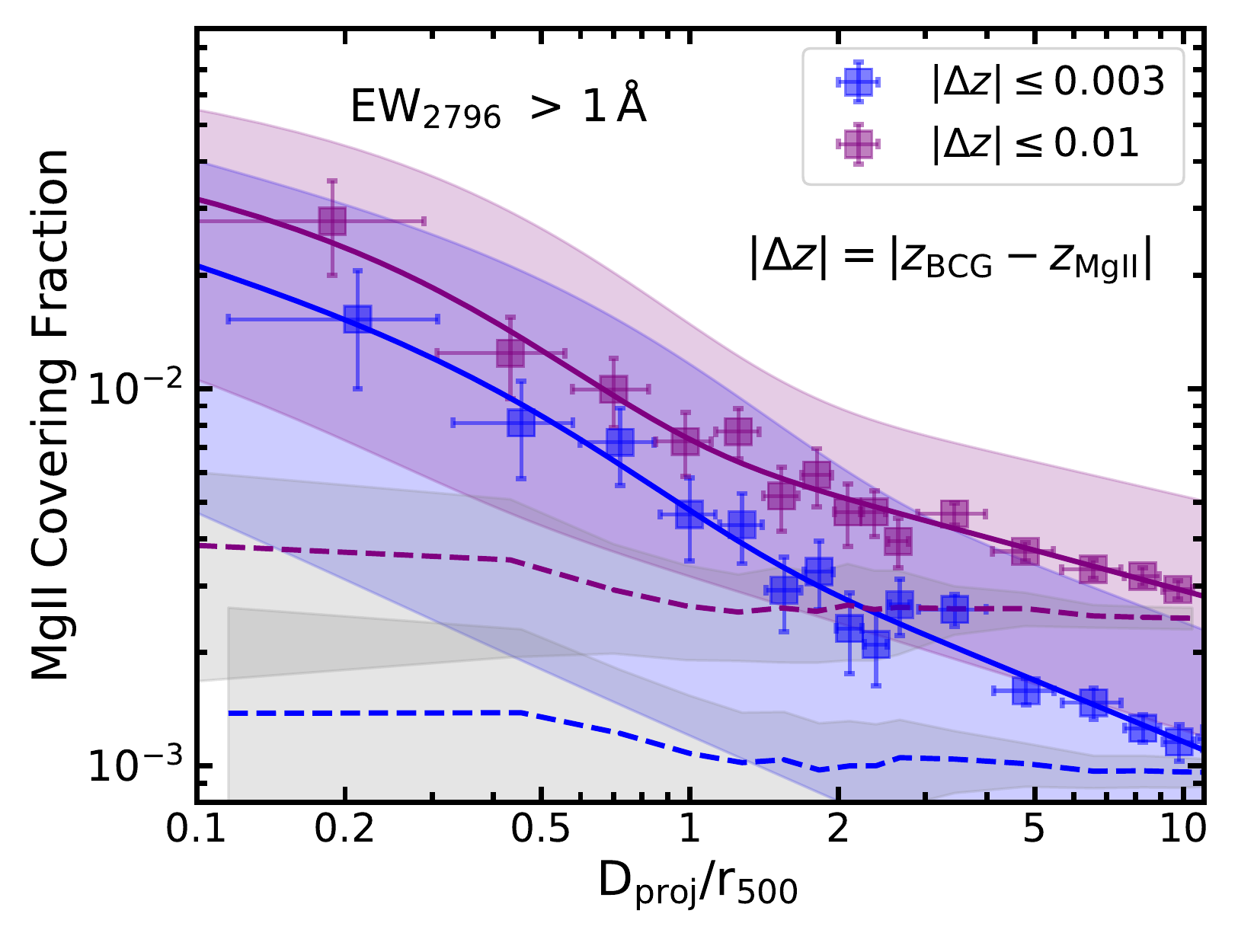}\\
    
    \caption{Effect of changing redshift interval on \mgii covering fraction. Comparing results for $|\Delta z|\leq 0.01$ (purple squares) and $|\Delta z|\leq 0.003$ (blue squares). \textbf{Top:} Differential covering fraction of weak and strong \mgii absorbers around DESI galaxy clusters as a function of projected distance from the BCG. \textbf{Bottom:} \mgii covering fractions of weak and strong systems as a function of projected distance normalized by $r_{\rm 500}$ of cluster. The corresponding dashed lines show the measurements around random quasar sightlines. We compile the fitting parameters for these measurements in Table~\ref{tab:params_fc} and the shaded regions show the corresponding $1\sigma$ uncertainty intervals.}
    \label{fig:dz003_diff_fc}
\end{figure*}

\subsection{\mgii Covering Fraction}

To investigate the effect of $|\Delta z| = |z_{\rm BCG}-z_{\rm \mgii}|$ choice on the measurements (particularly the \mgii covering fractions such that there are not many spurious absorber-cluster pairs), we also estimate the \mgii covering fraction for absorbers that reside within $|\Delta z|\leq 0.003$ ($|\Delta v|\leq 600$ \kms). As we understand, this is smaller than the typical $v_{\rm 500}\sim 900$ \kms,\, of the clusters hence, all the absorbers can be assumed to be gravitationally bound to the cluster halo. In Figure~\ref{fig:dz003_diff_fc}, we compare the \mgii covering fraction (for both weak and strong absorbers) as a function of $D_{\rm proj}/r_{\rm 500}$ and $D_{\rm proj}$ for both $|\Delta z|\leq 0.003$ (blue squares) and $|\Delta z|\leq 0.01$ (purple squares) choices. The best-fitting parameters (solid lines) are compiled in Table~\ref{tab:params_fc}. We find that even after decreasing the $|\Delta z|$, \mgii covering fractions (compare blue squares with purple squares) do not differ significantly, particularly at small distances on both projected and normalized scales though the difference is significant at large distances. At large scales, a large $|\Delta z|$ choice will include many absorber-BCG pairs, implying a higher covering fraction in that case. The analysis confirms that a significant fraction of the absorbers at small distances indeed have small $\Delta z$, relative to the cluster BCG. Hence, our results do not vary significantly at small distances even after decreasing the $\Delta z$ to small values. Besides this, we also see that the slope (parametrized by $\alpha$) is also larger, implying a faster decline in covering fractions on larger scales. On the other hand, the characteristic scales (parametrized by $x_{\rm o}$) of \mgii covering fraction are slightly larger (see Table~\ref{tab:params_fc}), than the $|\Delta z|\leq 0.01$, though the error bars are also larger as S/N of the measurements also goes down.

\subsection{\mgii Mean Equivalent Width and Surface Mass Density}
\begin{figure*}
	\includegraphics[width=0.475\linewidth]{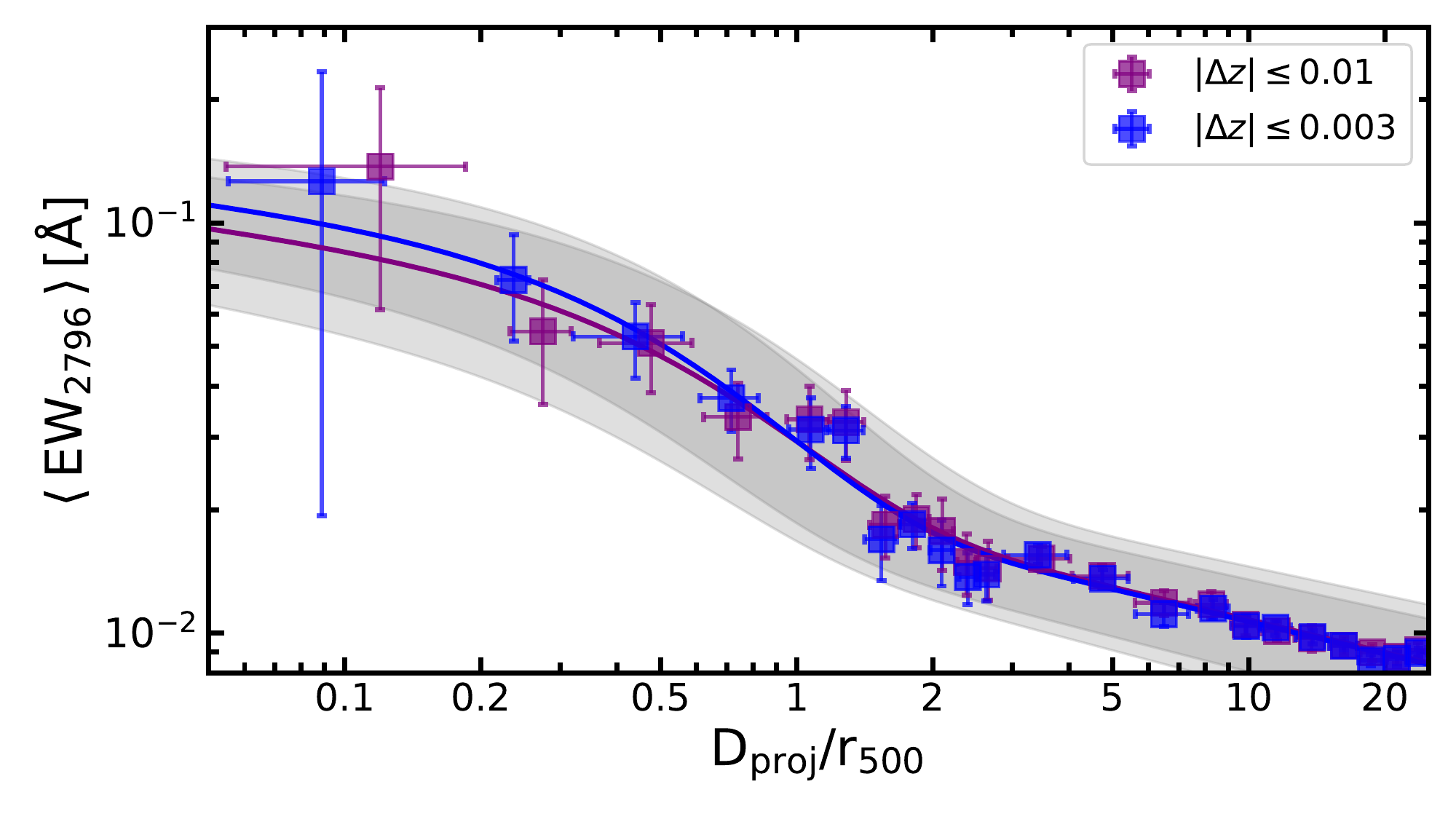}
	\includegraphics[width=0.475\linewidth]{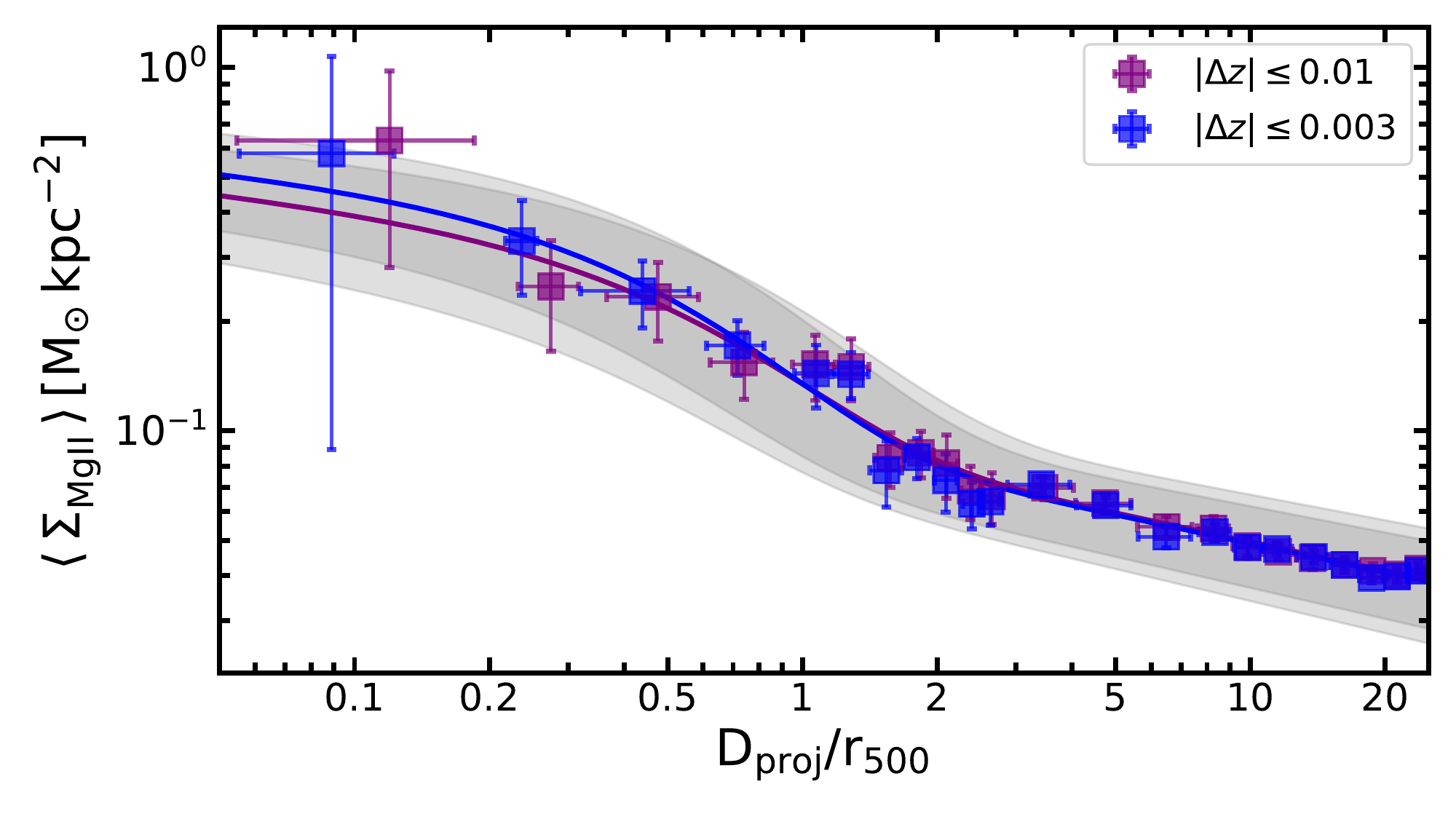}
    \caption{\textbf{Left:} The mean $\mean{EW_{2796}}$ in and around DESI legacy galaxy clusters (purple squares for $|\Delta z|\leq0.01$ and blue squares for $|\Delta z|\leq0.003$) as a function of projected distance normalized by $r_{\rm 500}$ of the corresponding haloes. \textbf{Right:} Differential average surface mass density of \mgii absorbers in and around DESI legacy galaxy for both $|\Delta z|$ cuts. The solid lines are the best-fitting curves described in section~\ref{best_fit} and the best-fitting parameters are summarized in Table~\ref{tab:params_fc}. The shaded regions show the corresponding $1\sigma$ uncertainty intervals.}
    \label{fig:dz003_ew_mgii}
\end{figure*}

Similarly, we do not see any significant variation in mean equivalent widths and surface mass densities of \mgii absorbers in clusters with $|\Delta z|$ choice, particularly at small scales. This analysis shows that our results and conclusions are not dependent on this choice. We show the measurements for $|\Delta z|\leq 0.003$ case in Figure~\ref{fig:dz003_ew_mgii}. The best-fitting parameters for these measurements are also shown in Table~\ref{tab:params_fc}.

\end{document}